\documentclass{cernyrep}
\usepackage{texnames}
\usepackage[T1]{fontenc}
\usepackage{mcite}
\usepackage{units}
\begin{document}

\input{./MacrosJakobs/atlasphysics.sty}

 
\newcommand{\ppbar}{\mbox{$p\overline{p}$}}
\newcommand{\dzero}{\mbox{D\O}}
\newcommand{\Zg}{\mbox{$Z /\gamma^*$}}

\newcommand{\stotWp}{\mbox{$\sigma_{W^+}^{\rm{tot}}$}}
\newcommand{\stotWm}{\mbox{$\sigma_{W^-}^{\rm{tot}}$}}
\newcommand{\stotW}{\mbox{$\sigma_{W}^{\rm{tot}}$}}
\newcommand{\stotZ}{\mbox{$\sigma_{Z}^{\rm{tot}}$}}
\newcommand{\stotZg}{\mbox{$\sigma_{Z/\gamma^*}^{\rm{tot}}$}}
\newcommand{\stotWZ}{\mbox{$\sigma_{W(Z)}^{\rm{tot}}$}}
\newcommand{\sfidWp}{\mbox{$\sigma_{W^+}^{\rm{fid}}$}}
\newcommand{\sfidWm}{\mbox{$\sigma_{W^-}^{\rm{fid}}$}}
\newcommand{\sfidW}{\mbox{$\sigma_{W}^{\rm{fid}}$}}
\newcommand{\sfidZ}{\mbox{$\sigma_{Z}^{\rm{fid}}$}}
\newcommand{\sfidZg}{\mbox{$\sigma_{Z/\gamma^*}^{\rm{fid}}$}}
\newcommand{\sfidWZ}{\mbox{$\sigma_{W(Z)}^{\rm{fid}}$}}
\newcommand{\cw}{\mbox{$C_W$}}
\newcommand{\cz}{\mbox{$C_Z$}}

\title{Physics at the LHC \\ -From Standard Model measurements to Searches for New Physics-}

\author{Karl Jakobs}

\institute{Physikalisches Institut, University of Freiburg, Germany}

\maketitle 

\begin{abstract}
The successful operation of the {\em Large Hadron Collider} (LHC) during the past two years
allowed to explore particle interaction in a new energy regime.
Measurements of important Standard Model processes like the production
of  high-\pt\ jets, $W$ and $Z$ bosons and top and $b$-quarks were performed by the
LHC experiments. In addition, the high collision energy allowed to search for
new particles in so far unexplored mass regions. Important constraints
on the existence of new particles predicted in many models of physics beyond the Standard
Model could be established.
With integrated luminosities reaching values around 5~\ifb\ in 2011, the experiments reached
as well sensitivity to probe the existence of the Standard Model Higgs boson
over a large mass range.
In the present report the major physics results obtained
by the two general-purpose experiments ATLAS and CMS are summarized.
\end{abstract}

\section{Introduction}
In March 2010 the Large Hadron Collider started its operation at the highest
centre-of-mass energy ever reached and delivered first proton-proton collisions at
7~\TeV. The years 2010 and 2011 showed a very successful operation of both the collider and
the associated experiments. During the start-up year 2010 data corresponding to an
integrated luminosity of about 48~\ipb\ could be delivered. This successful operation was
followed by an even more successful year 2011 where the collider delivered data corresponding
to an integrated luminosity of 5.5~\ifb\ and exceeded the original design goal of 1~\ifb\
 by far.
In April 2011 the world record on the instantaneous luminosity was reached with a luminosity
of 4.7~$\cdot$~10$^{32}$~cm$^{-2}$sec$^{-1}$. Meanwhile luminosities beyond 3$\cdot$10$^{33}$~cm$^{-2}$sec$^{-1}$ have been reached.

However, not only the accelerator but also the experiments
showed an extremely successful operation. They were able to record the delivered luminosity
with efficiencies of the order of 94\%. All detector subsystems worked well with a high
number of functioning channels, typically exceeding 99\%.

The data were used to test the Standard Model \cite{SM1,*SM2,*SM3,qcd1,*qcd2,*qcd3} of particle physics in the new energy regime.
At the LHC as a hadron
collider the production of particles via the strong interaction is dominating.
Therefore, the test of Quantum Chromodynamics (QCD) \cite{qcd1,*qcd2,*qcd3},
the theory of strong interactions, was in the focus during the early phase.
Tests of QCD can be performed at small distances or for
processes with large momentum transfer. Among them the production of jets with
large transverse momenta ($\pt$) has the largest cross section.
The investigation of the production of $W$ and $Z$ bosons, their
associated production with jets and the production of
top quarks constitute other important tests of QCD in the new energy regime.

Due to the high centre-of-mass energy of 7~\TeV\ the LHC has a large discovery potential
for new particles with masses beyond the limits set by the Tevatron experiments, even already in the initial
phase of operation. Due to the dominating strong production, this holds in particular for
particles that carry colour charge, like e.g. the supersymmetric partners of quarks and gluons.
Due to the excellent luminosity performance of the LHC in 2011 the sensitivity
for many models of new physics were pushed far beyond the  mass range explored so far.

In this review the physics motivation for the LHC is briefly recalled in Section 2.
The phenomenology of proton-proton collisions and the calculation of cross sections
is briefly reviewed in Section 3.
The measurement of important Standard Model processes at the LHC is discussed in Section 4.
The status of the search for
the Standard Model Higgs boson is summarized in Section 5. It should be noted that in this paper
the status of the Higgs boson search as of March 2012 is presented. Given the
large increase in the integrated luminosity during the second half of 2011, these results
supersede by far those presented at the school in September 2011.
In the remaining sections of the paper the searches for physics beyond the Standard
Model are discussed. The search for supersymmetric particles is described in Section~6,
the search for other scenarios is summarized in Section~7.

\section{The Physics Questions at the LHC}

The Standard Model is a very successful description of the interactions
of particles
at the smallest scales (10$^{-18}$m) and highest energies accessible to current experiments.
It is a quantum field theory which describes the interactions of spin-$\nicefrac{1}{2}$ pointlike
fermions whose interactions are mediated by spin-1 gauge bosons. The bosons
are a consequence of local gauge invariance of the underlying Lagrangian under the
symmetry group $SU(3)$x$SU(2)$x$U(1)$ \cite{SM1,*SM2,*SM3}.

The $SU(2)$ x $U(1)$ symmetry group, which describes the electroweak interactions, is
spontaneously broken by the existence of a postulated scalar field, the so-called
Higgs field, with a non-zero vacuum expectation value \cite{Higgs1,Higgs3,Higgs4,PhysRevLett.13.508,PhysRev.145.1156,Kibble:1967sv}.
This leads to the emergence of massive vector bosons, the $W$ and $Z$ bosons, which
mediate the weak interaction, while the photon of electromagnetism remains massless.
One physical degree of freedom remains in the Higgs sector which should manifest
as a neutral scalar boson $H$ which is so far unobserved.
The description of the strong interaction (Quantum Chromodynamics or QCD) is based
on the gauge group $SU(3)$ \cite{qcd1,*qcd2,*qcd3}. Eight massless gluons mediate this interaction.

All experimental particle physics measurements performed up to date are in
excellent agreement
with the predictions of the Standard Model. The only noticeable exception is the
evidence for non-zero neutrino masses observed in neutrino-oscillation
experiments \cite{neutrinomasses}.
There remain, however, many key questions open and it is generally believed that the
Standard Model can only be a low energy effective theory of a more fundamental underlying
theory. One of the strongest arguments for an extension of the Standard Model is the
existence of Dark Matter \cite{dark-matter} in the universe. There is no explanation for
such a type of matter in the Standard Model.

The key questions can be classified to be linked to mass, unification and flavour:

\begin{itemize}
\item Mass:  What is the origin of mass? \\
How is the electroweak symmetry broken? Is the solution, as implemented in the
Standard Model, realized in Nature, and linked to this, does the Higgs boson exist? \\

\item Unification: What is the underlying fundamental theory? \\
Can the three interactions which are relevant for particle physics be unified at larger
energy and are there new symmetries found towards unification?
Are there new particles, e.g. supersymmetric particles, at higher energy scales?
And finally, it must also be answered how gravity can eventually be incorporated. \\

\item Flavour: Why are there three generations of matter particles? What is the
origin of CP violation in the weak interaction? What is the origin of neutrino masses and
mixings?

\end{itemize}

The high energy and luminosity of the LHC offers a large range of physics opportunities.
The primary role of the LHC is to explore the \TeV-energy range where answers to at least
some of the aforementioned questions are expected to be found. In the focus is certainly the
search for the Higgs boson. The LHC experiments have the potential to explore the full
relevant mass range and either to discover the Standard Model Higgs boson or to exclude its existence.

Another focus area constitutes the search for supersymmetric particles which can be
carried out at the LHC up the masses of a few \TeV. If such particles are
discovered, their link to the dark matter in the universe must be investigated. This can only be done in conjunction
with experiments on direct dark matter detection \cite{direct_dark_matter}.

However, it is important to stress that the remit of the LHC is not only to look for the
{\em expected} or theoretically favoured models, but to carry out a thorough investigation
of as many final states as possible. It is important to search for any deviation from
the Standard Model predictions. This implies that the Standard Model predictions must
be reliably tested in the new energy domain. In particular during the early phase
of experimentation at the LHC, detailed measurements of Standard Model
processes must be carried out. Some of these processes can as well be used for the understanding of the detector
response  and its calibration.

Finally, with increasing precision of the Standard Model measurements, it is also
important to test the consistency of the model via quantum corrections. Important contributions
from the LHC in this area will be  precise measurements of the $W$ mass and of the
top-quark mass, which can be used to constrain the Higgs boson mass. A direct confrontation
of this prediction to a direct Higgs boson mass measurement may constitute the
ultimate test of the Standard Model at the LHC.

\section{Phenomenology of proton-proton collisions}

Scattering processes at high-energy hadron colliders can be classified as either hard or soft.
Quantum Chromodynamics is the underlying theory for all such processes, but the
approach and level of understanding is very different for the two cases. For hard processes,
e.g. high-\pT\ jet production or $W$ and $Z$ production, the rates and event properties
can be predicted with good precision using perturbation theory.
For soft processes, e.g. the total cross section, the underlying event etc., the rates
and properties are dominated by non-perturbative QCD effects,
which are less well understood. An understanding of the rates and characteristics
of predictions for hard processes, both signals and backgrounds, using perturbative QCD
(pQCD) is crucial for tests of the theory and for searches for new physics.

\begin{figure}[hbtn]
\begin{center}
\includegraphics[width=0.30\textwidth,angle=0]{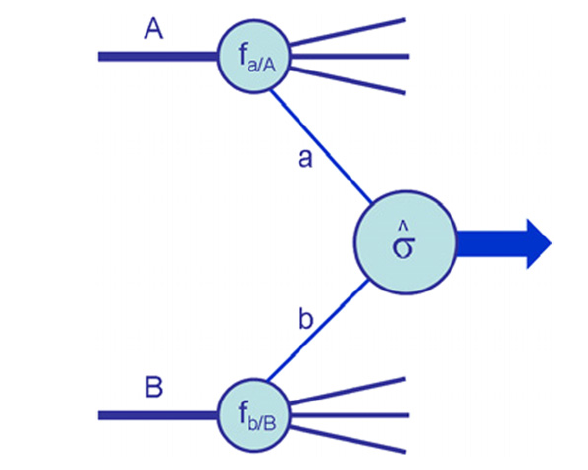}
\end{center}
\caption{\small \it
Diagrammatic structure of a generic hard-scattering process (from
Ref.~\protect\cite{stirling_QCD}).
}
\label{f:pp-collision}
\end{figure}

The calculation of a hard-scattering process for two hadrons $A$ and $B$ can be
illustrated as displayed in Fig.~\ref{f:pp-collision}. Two partons of the incoming
hadrons undergo a hard scattering process characterized by the cross section
$\hat{\sigma}$. The structure of the incoming hadrons is described by the parton
density functions (PDFs) $ f_{a/A}(x_a,\mu_F^2)$ (see Section~\ref{s:PDFs}),
i.e. the probability to find a parton
$a$ in hadron $A$ with a momentum fraction $x_a$ at the energy scale $\mu_F^2$. To obtain the
hadron-hadron cross section, a summation over all possible parton-parton
scattering processes and an integration over the momentum fractions has
to be performed \cite{stirling_QCD}:
\begin{equation}
\sigma_{AB} = \sum\limits_{a,b} \int \rm{d}x_a \cdot \rm{d}x_b \ f_{a/A} (x_a,\mu_F^2) \ f_{b/B}(x_b,\mu_F^2) \ \hat{\sigma}_{ab}(x_a,x_b,\alpha_s(\mu_R^2))~.
\end{equation}
The calculations of the hard
scattering process $\hat{\sigma}_{ab}$ are performed in
perturbative QCD and the results depend on the strong coupling constant $\alpha_s$ and
its renormalization scale $\mu_R$. The scale $\mu_F$ that appears in the parton
density functions is the so-called factorization scale, which can be thought of as
the scale that separates long- and short-distance physics \cite{stirling_QCD}.
Large logarithms related to gluons emitted collinear with incoming quarks can be absorbed
in the definition of the parton densities, giving rise to logarithmic
scaling violations which can be described via the DGLAP\footnote{Dokshitzer-Gribov-Lipatov-Altarelli-Parisi} evolution equations \cite{DGLAP}.
The perturbative calculation can be written as
\begin{equation}\label{eq1}
\hat{\sigma}_{ab}^{[n]} =  \hat{\sigma}_{ab}^{[0]} +
\sum \limits_{j=k+1}^{k+n} c_j \cdot \alpha_s^{j} ~,
\end{equation}
where $\hat{\sigma}_{ab}^{[0]}$ denotes the leading order (LO) cross section and
$n$ denotes the perturbative order of the calculation. The index $k$ denotes the
order of $\alpha_s$ appearing in the leading order calculation, which might as well be 0,
like for the Drell-Yan production of $W$ and $Z$ bosons, as discussed below.
The cross sections at higher orders, which are usually denoted as next-to-leading order (NLO)
and next-to-next-to-leading order (NNLO) etc., are often parametrized in terms
of total $K$ factors, defined at each perturbative order [$n$] as the ratio of the cross
section computed to that order normalized to the Born level cross section:
\begin{equation}
\sigma^{[n]} \ =  \ \sigma^{[0]}  \cdot K_{\rm tot}^{[n]} .
\end{equation}
As discussed above, the scale $\mu$ is an arbitrary parameter, which in
general is, however, chosen to be of the order of the energy characterizing
the parton-parton interaction,
like, for example, the mass of the vector bosons or the transverse
momenta of outgoing jets. The more orders are included in the perturbative
expansion, the weaker the dependence on $\mu$. As an example, the production of $W$ and $Z$
bosons is discussed in Section~\ref{s:WZ_prod}.

Those partons which do not take part in the hard scattering process will produce
what is generally called the `underlying event'. Finally, it should be stressed that
Eq.~\ref{eq1} does not describe the bulk of the events which occur at a hadron collider.
It can only be used to describe the most interesting classes of events which involve
a hard interaction. Most events result from elastic and soft inelastic interactions
generally called `minimum bias' events. In the following a few specific examples of
hard scattering processes are discussed.

\subsection{Parton Distribution Functions \label{s:PDFs}}
\label{sec:PartonDistributionFunctions}

The parton distribution functions (PDFs) $f(x,Q^2)$ for a hadron
provide the probability density of finding a parton with momentum fraction $x$
at momentum transfer $Q^2$ which defines the energy scale of the
process.
The $Q^2$ dependence is induced by the usage
of perturbation theory and the resulting higher order corrections.
It is described by the DGLAP
evolution equations~\cite{DGLAP}.
However, the functional form of the PDFs is not predicted by perturbative QCD
and has to be measured experimentally.

Various classes of experiments are sensitive to the proton PDFs,
such as deep inelastic scattering at fixed target experiments
with electron, muon or neutrino beams, and electron-proton scattering
at the HERA collider. Also experiments at pure hadronic colliders
such as the Tevatron ($p\bar{p}$) and the LHC ($pp$)
can yield valuable information.

In order to determine the parton distributions from the measurements,
a parametrization is assumed to be valid at some starting value $Q^2=Q^2_0$.
The DGLAP evolution functions are used to evolve the PDFs to a different
$Q^2$ where predictions of the measured quantities (e.g. structure functions)
are obtained. The predictions are then fitted to the measured datasets, thus
constraining the parameters (typically $10$ to $20$) of the
parametrisation~\cite{DevenishCooperSarkar}.
Various collaborations performed fits to the available datasets
and provided PDF sets for the proton, for instance the groups
ABKM~\cite{abkm}, CTEQ~\cite{cteq66}, CT10~\cite{ct10}, HERAPDF~\cite{herapdf10,herapdf15},
JR~\cite{jr09}, MSTW~\cite{mstw2008} and NNPDF~\cite{nnpdf20,nnpdf21}.
The NNPDF collaboration has already included the first lepton charge
asymmetry measurements in the $W$ boson production by the ATLAS~\cite{Atlas:WAsymmetry2011}
and CMS~\cite{CMS:WAsymmetry2011} experiments in their fit~\cite{nnpdf21WithLHCAsymmetry}.
The PDFs determined by MSTW are shown in Fig.~\ref{fig:MSTW2008_PDFs} at
two different $Q^2$ scales. For example, it can be observed that gluons dominate the low $x$ region
and the contributions from sea quarks become more dominant at higher $Q^2$.

\begin{figure}
  \begin{center}
    \includegraphics[width=0.4\textwidth]{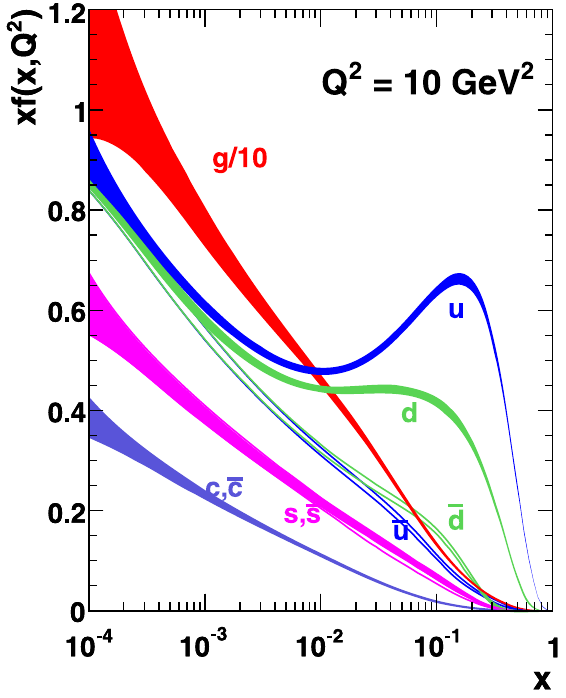}\qquad\quad
    \includegraphics[width=0.4\textwidth]{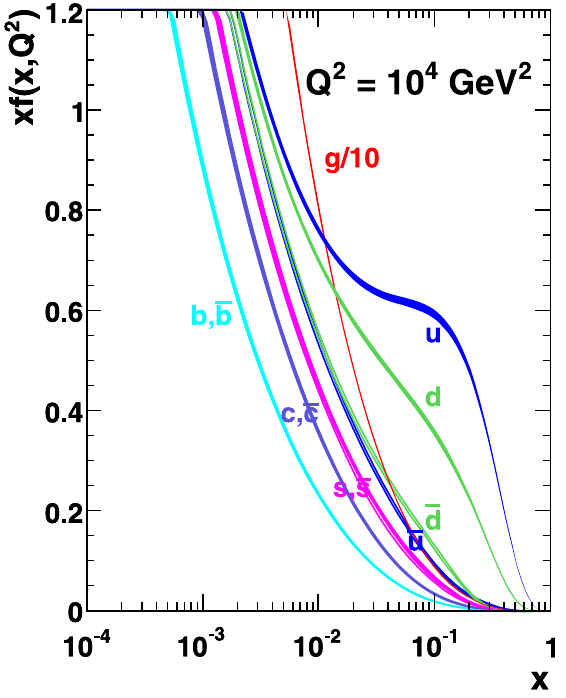}
    \caption{\it Parton distribution functions of the proton as determined for the
      MSTW08 PDF set for (left) $Q^2=10\,\GeV^2$ and (right) $Q^2=10^4\,\GeV^2$.
      The bands reflect the uncertainties at the $68$\% confidence level
(from Ref.~\protect\cite{mstw2008}).}
    \label{fig:MSTW2008_PDFs}
  \end{center}
\end{figure}

\subsection{Jet Production via QCD scattering processes}
Two-jet events result when an incoming parton from one hadron scatters off an
incoming parton from the other hadron to produce two high transverse momentum
partons which are observed as jets. The parton processes that contribute at leading
order are shown in Fig.~\ref{f:feynman_qq-qg-gg}. The matrix elements have been
calculated at leading order \cite{LO_jets} and next-to-leading order \cite{NLO_jets,NLO_jets2}.
At the LHC, terms involving gluons in the initial state are dominant at low \pT.
\begin{figure}[hbtn]
\begin{center}
\includegraphics[width=0.50\textwidth,angle=0]{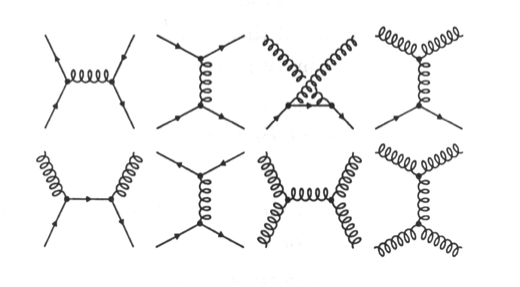}
\end{center}
\vspace*{-0.9cm}
\caption{\small \it
Leading order diagrams for the production of high-\pT\ jets.
}
\label{f:feynman_qq-qg-gg}
\end{figure}
Unlike in lowest order, where a direct correspondence between
a jet cross section and the parton cross section can be made, a prescription is
needed to derive jet cross sections in next-to-leading order. When such prescriptions
are applied, the next-to-leading order cross sections show substantially smaller
sensitivities to variations of the renormalization scale than at lowest order.

\subsection{$W$ and $Z$ Production \label{s:WZ_prod}}
In leading order the production of the vector bosons $W$ and $Z$ is
described by the Drell-Yan process, where a quark and an antiquark
from the incoming hadrons annihilate. This process has been calculated
up to next-to-next-to-leading order
in the strong coupling constant \alphas~\cite{zwprod1991,zwprod1992,DrellYanNNLO_3,fewz1}.
Some of the relevant Feynman diagrams are given in Fig.~\ref{f:feynman_WZ}.
\begin{figure}[htbn]
\begin{center}
\includegraphics[width=0.19\textwidth,angle=0]{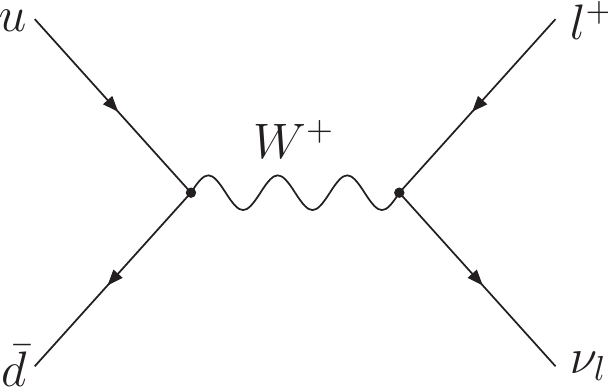}
\includegraphics[width=0.19\textwidth,angle=0]{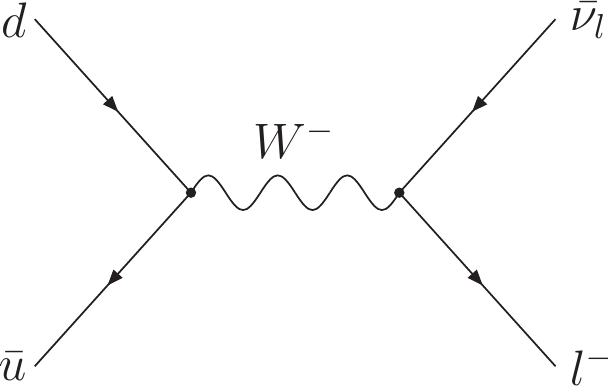}
\includegraphics[width=0.19\textwidth,angle=0]{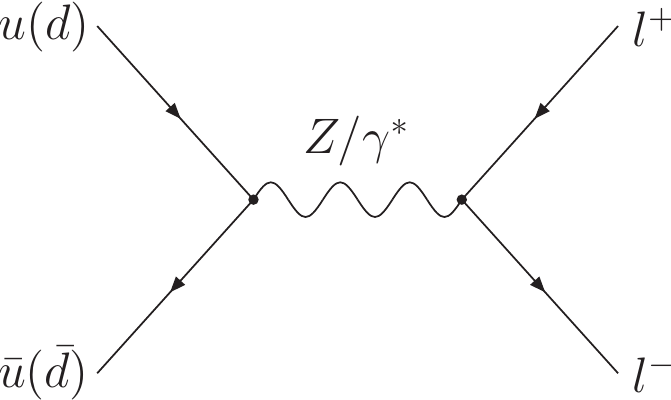} \\
\ \\
\includegraphics[width=0.15\textwidth,angle=0]{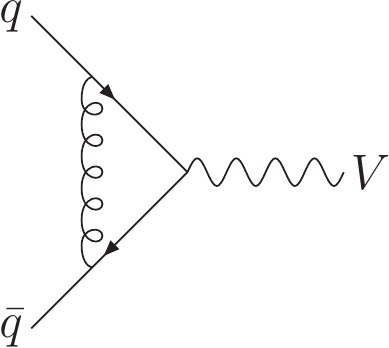}
\includegraphics[width=0.15\textwidth,angle=0]{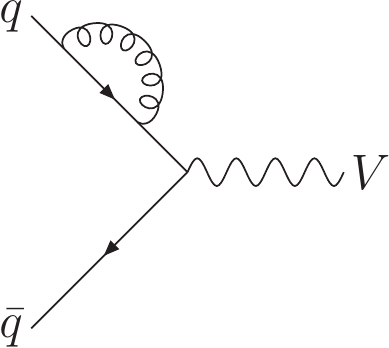}
\includegraphics[width=0.15\textwidth,angle=0]{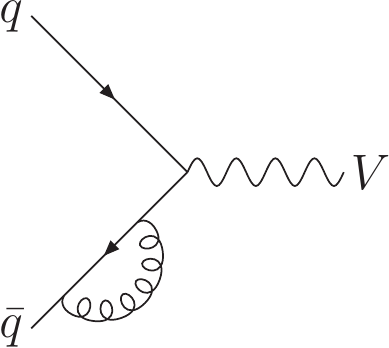}
\includegraphics[width=0.15\textwidth,angle=0]{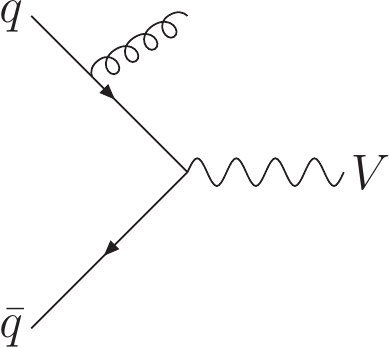}
\includegraphics[width=0.15\textwidth,angle=0]{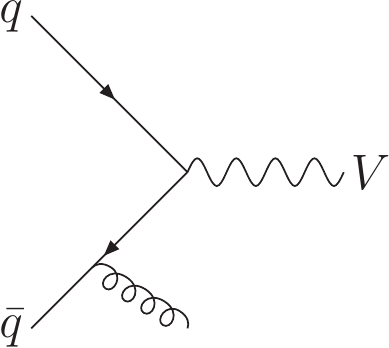}
\end{center}
\caption{\small \it
Leading order (top) and some next-to-leading order diagrams (bottom)
for the production of $W$ and $Z$ bosons.}
\label{f:feynman_WZ}
\end{figure}
When going from LO to NLO the cross sections increase by about 20\% and the factorization
and renormalization scale uncertainties decrease. This is nicely shown in
Fig.~\ref{f:WZ_renormalization}.
\begin{figure}[h!]
  \begin{center}
    \includegraphics[width=0.49\textwidth]{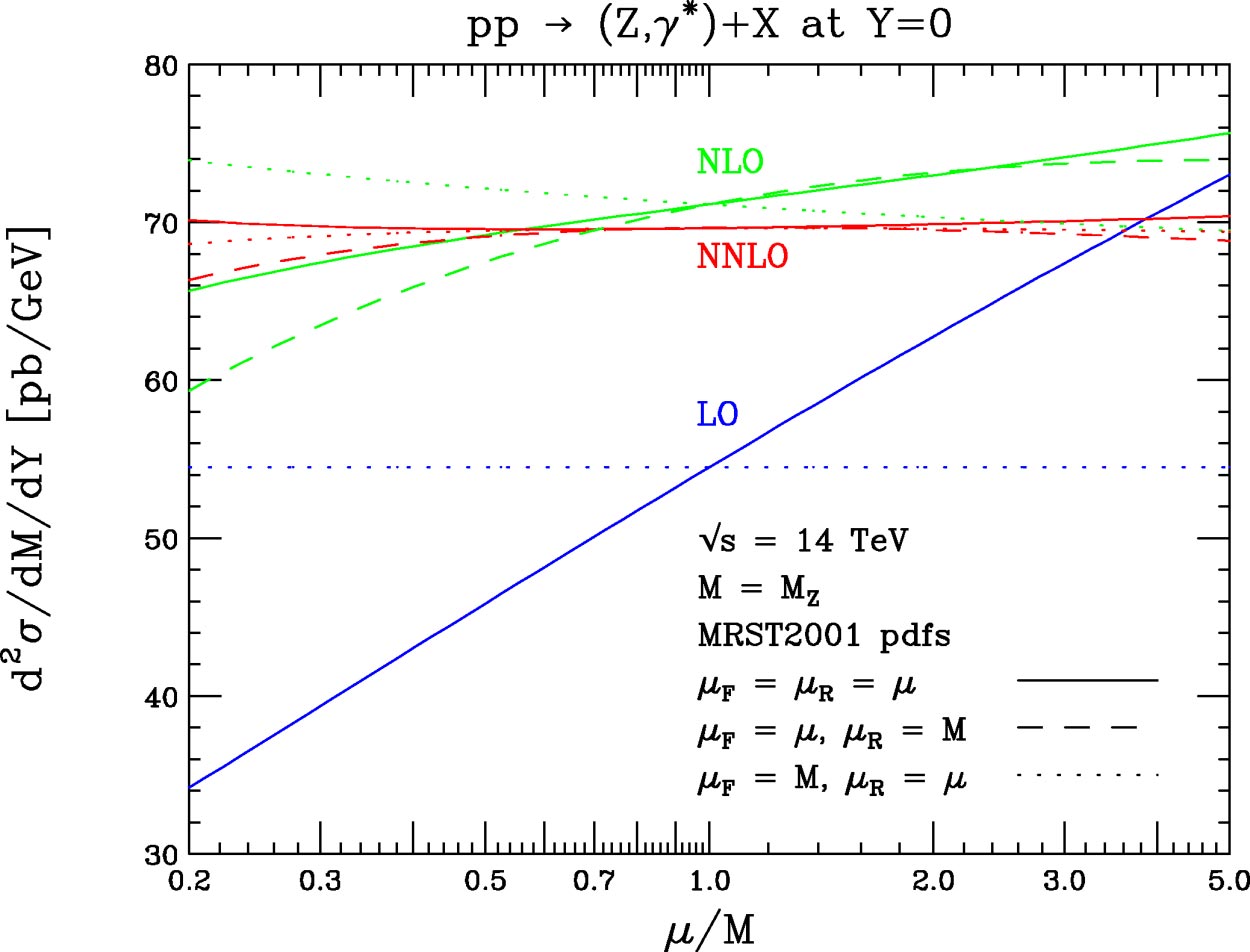}
    \caption{\it Dependence of the production cross section of on-shell $Z$ bosons at
rapidity $y=0$ on the choice of the renormalization and factorization scales. For each order
in perturbation theory (LO, NLO, NNLO), three curves are shown. The solid curve represents
the results obtained under a common variation of $\mu_R = \mu_F = \mu$ over the range
$M/5 < \mu < 5 M$. The dashed (dotted) curves represent the results obtained under
variations of the factorization (renormalization) scale alone, holding the other scale fixed
(from Ref.~\protect\cite{fewz1}).}
    \label{f:WZ_renormalization}
  \end{center}
\end{figure}
Including NNLO contributions slightly
decreases the cross-sections but the result is consistent with the NLO prediction
within the NLO scale uncertainty, indicating that the perturbative expansion converges.
The impact of higher order corrections on the predicted
rapidity\footnote{\noindent The rapidity $y$ of a particle is related to its energy
E and the projection of its momentum on the
beam axis $p_z$ by  \\
$ y = \frac{1}{2} \ln [ (E + p_z) / (E - p_z) ]$.
The pseudorapidity $\eta$ is defined as $\eta = -  \ln  \tan \frac{\theta}{2}$,
where $\theta$ is the polar angle.}
distributions
of the $W$ and $Z$ bosons is shown in Fig.~\ref{fig:TheoryPredictionWZRapidity}
for proton-proton collisions with a centre-of-mass energy of $\sqrt{s}=14$~\TeV.
This figure also illustrates that larger cross sections for $W^+$ production than
for $W^-$ production are expected at the LHC. This asymmetry results from the dominance of
$u$ over $d$ valence quarks in the incoming protons (see also Fig.~\ref{fig:MSTW2008_PDFs}).

\begin{figure}[t]
  \begin{center}
    \includegraphics[width=0.45\textwidth]{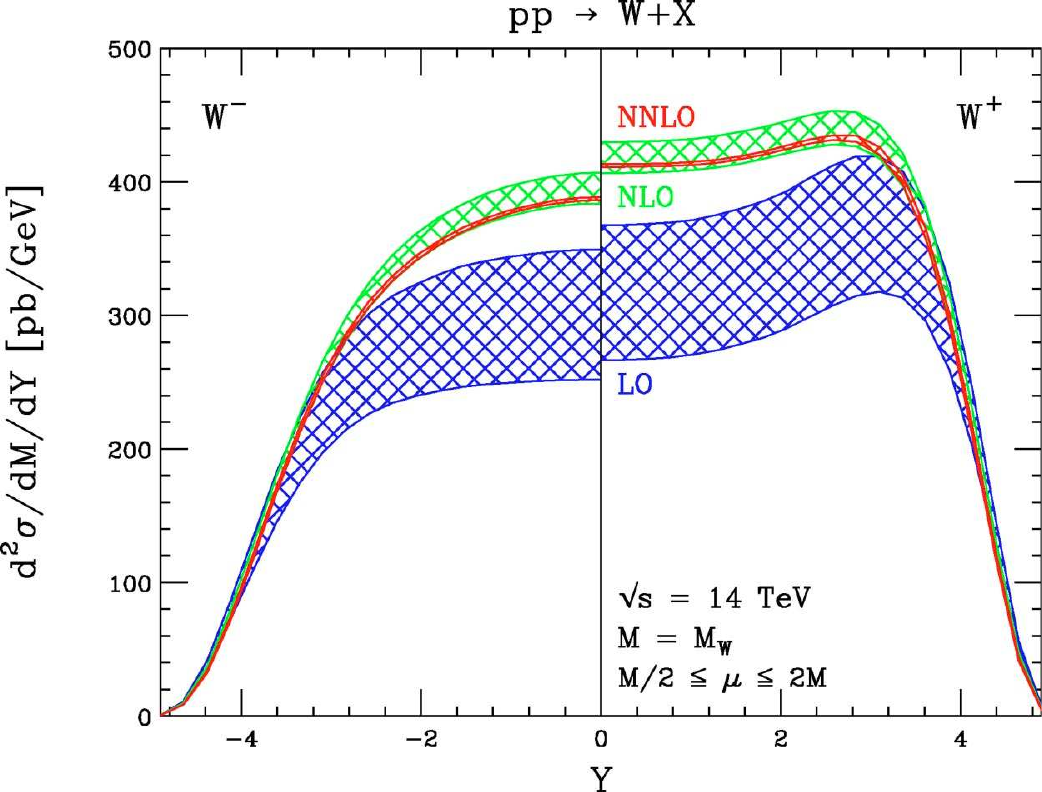}
    \includegraphics[width=0.45\textwidth]{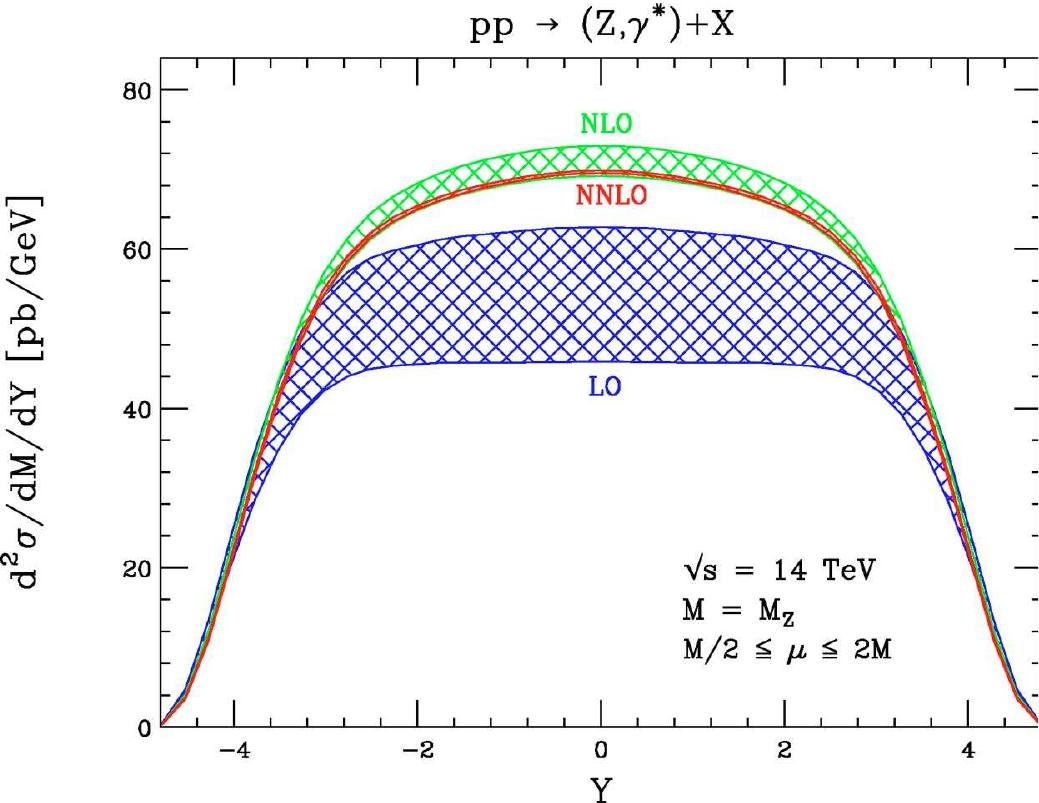}
    \caption{\it Theory predictions at LO, NLO and NNLO of the rapidity distributions
for $W$ (left)
      and $Z$ (right) boson production in proton-proton collisions at $\sqrt{s}=14$~\TeV.
      The bands indicate the factorization and renormalization scale
      uncertainties, obtained by scale variations in the range $M_{W/Z}/2 \leq \mu \leq 2M_{W/Z}$
      (from Ref.~\protect\cite{fewz1}).}
    \label{fig:TheoryPredictionWZRapidity}
  \end{center}
\end{figure}

Electroweak radiative corrections for the $W$ and $Z$ boson production
have been computed up to next-to-leading order~\cite{Dittmaier:WElectroweak,Dittmaier:ZElectroweak}.
These corrections change the production cross sections
and affect kinematic properties like the lepton transverse momenta,
lepton rapidities and the transverse and invariant masses of the lepton pairs.
In particular for precision measurements like that of the $W$ boson mass,
it is therefore important to take electroweak radiative corrections into account.

\subsection{Cross Sections at the LHC}
\label{section:CrossSectionPredictions}

An overview of cross sections of some benchmark processes at proton-proton and
proton-antiproton colliders as a function of the centre-of-mass energy
is shown in Fig~\ref{fig:SM_CrossSections}. The total inelastic proton-proton
cross section is dominant and reaches a huge value of about 70 mb at the LHC.
Processes which can proceed via the strong interaction have a much
larger cross section than electroweak processes. The dominant electroweak process, the
production of $W$ and $Z$ bosons is found to be about six and seven orders of magnitude
smaller than the total inelastic proton-proton cross section. However, this process
constitutes the most copious source of prompt high-\pT\ leptons, which are important
for many physics measurements and searches for new physics at the LHC. The production
processes of the Standard Model Higgs boson and of other non-coloured heavy new
particles have small cross sections and therefore require a correspondingly high
integrated luminosity for their detection. The Higgs boson production cross section
is found to be ten to eleven orders of magnitude smaller than the inelastic pp
cross section. The exact values depend strongly on the mass of the Higgs boson,
as further discussed in Section~\ref{s:higgs_prod}.

\begin{figure}
  \begin{center}
    \includegraphics[width=0.55\textwidth]{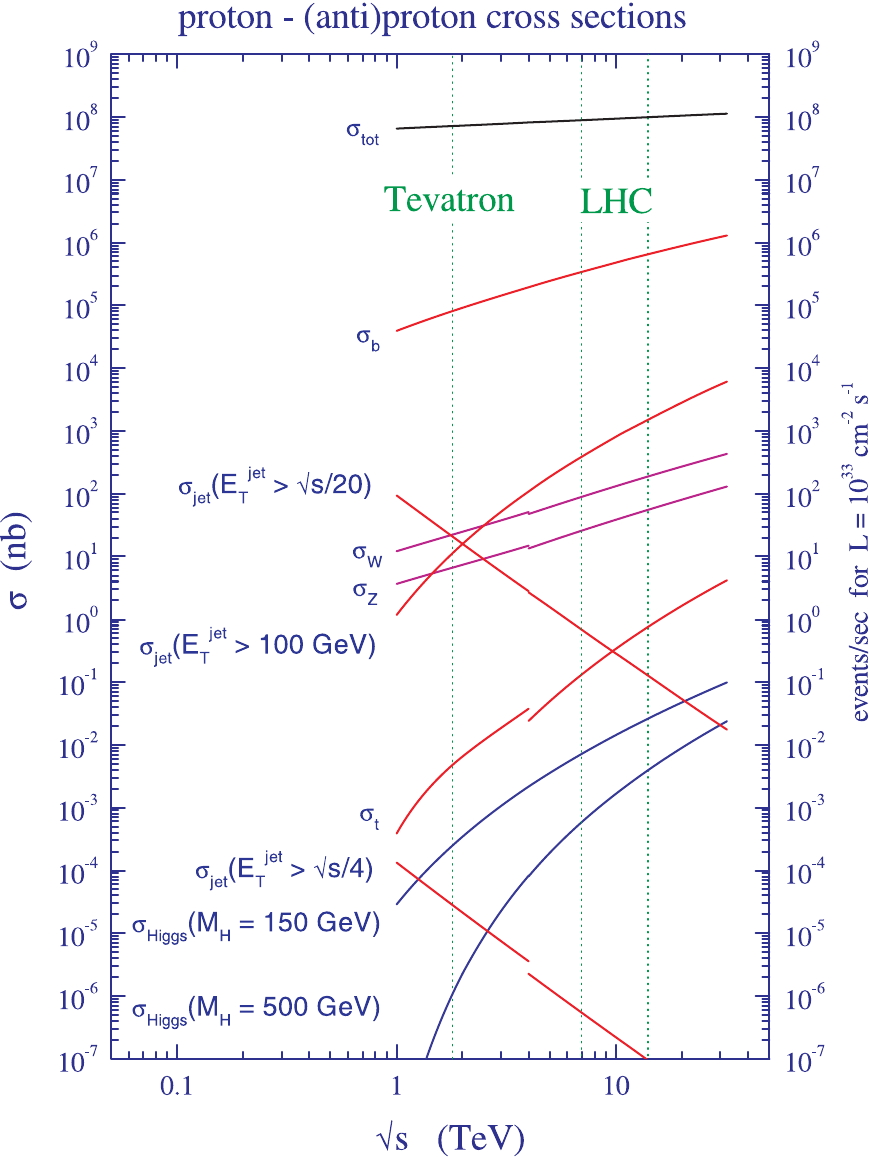}
    \caption{\it Standard Model cross sections at the Tevatron and LHC colliders
      as function of the centre-of-mass energy $\sqrt{s}$. In case of the LHC both
      the energy during
      data taking in 2010/2011 ($\sqrt{s}=7$~\TeV) and the nominal energy ($\sqrt{s}=14$~\TeV)
      are marked (adapted from Ref.~\protect\cite{stirling_QCD}).
}
    \label{fig:SM_CrossSections}
  \end{center}
\end{figure}

\section{Measurement of Standard Model processes}

\subsection{The production of high-\pt\ jets \label{s:QCD_jets}}

A measurement of the production of high-$\pt$ jets constitutes an important test of
QCD in the new energy regime of the LHC. Events with two high transverse momentum jets (dijets) arise
from parton-parton scattering where the outgoing scattered partons manifest
themselves as hadronic jets. The measurements of the inclusive jet production cross section
or the dijet production cross section are therefore also sensitive to the structure
of the proton and may lead to further constrains on the PDFs. In addition, the
precise measurement of jet production is
important for searches of physics beyond the Standard Model. New
physics may lead to significant deviations from the expected
QCD behaviour. For example, a substructure of quarks may manifest itself in deviations
of the measured inclusive jet-production cross section from the expected behaviour at high
transverse momenta. The measurement of the dijet cross section as a function of the dijet
mass $m_{jj}$ allows for a sensitive search for physics beyond the Standard Model,
such as dijet resonances or contact interactions of composite quarks.

The production of multijets provides as well an important background in the search
for physics beyond the Standard Model. In many cases, leptons and missing transverse
energy are used as final states signatures. Although they do not appear at first place in
QCD jet production, they might originate from decays of heavy quarks, from
mis-measurements of jet energies or from mis-identification of jets as leptons.
Although the probability for this to happen is small,
the contributions to the background can be sizeable, give the huge jet production cross
sections.

\subsubsection {Jet reconstruction and calibration}

For the reconstruction of jets
both the ATLAS and CMS experiments use the infrared- and collinear-safe {\em anti-$k_T$} jet clustering
algorithm \cite{anti-kt} with distance parameters $0.4 \le R \le 0.7$.
The inputs are either topological calorimeter cluster energies
\cite{topocluster1,topocluster2} in the ATLAS
experiment or particle flow objects \cite{particle-flow1,particle-flow2} in the
CMS experiment. For the theoretical comparison the input can also be four-vectors from
stable particles in generator-level simulations. In all cases residual jet-level
corrections are needed to account for energy losses not detectable on cluster or
particle flow level. These jet-level calibrations are Monte Carlo based correction
functions in pseudorapidity
$\abseta$ and $\pt$. The jet energy scale and the attached uncertainties
are validated with in-situ methods using the balance of transverse momenta
in dijet and $\gamma$-jet events.
The systematic jet energy scale uncertainties are found to be typically in the range of
$\pm (3-6)\%$ over a large range of $\eta$ and $\pt$. The larger
values are reached at large $\abseta$ as well as at very low and very high $\pt$.

\subsubsection {Jet cross section measurements}

The inclusive jet production cross section has been
measured by both the ATLAS \cite{ATLAS-incl-jets} and CMS \cite{CMS-incl-jets}
experiments as a function
of the jet transverse momentum ($\pt$) and jet rapidity ($y$). In addition,
double differential
cross sections in the maximum jet rapidity $y_{\rm{max}}$ and dijet mass $m_{\rm{jj}}$
for dijet events are measured \cite{ATLAS-incl-jets, CMS-dijets}.
The data are corrected
for migration and resolution effects due to the steeply falling spectra in $\pt$ and mass.
The NLO perturbative parton-level QCD predictions are corrected for
hadronisation and the underlying event activity.
Figure~\ref{f:jets_incl} (left) shows the inclusive jet cross-section measurement
for jets with size
$R = 0.4$ as a function of jet transverse momentum from the ATLAS
collaboration \cite{ATLAS-incl-jets}, based on the total data set collected in 2010 corresponding to
an integrated luminosity of 37~\ipb.
The experimental systematic uncertainties are dominated by the jet energy scale uncertainty.
There is an additional overall uncertainty of $\pm$3.4\% due to the luminosity measurement.
The theoretical uncertainties result mainly from the choice of the renormalization and
factorization scales, parton distribution functions, $\alpha_s (m_Z)$ and the modelling
of non-perturbative effects.

The cross section measurement as a function of the dijet invariant mass from the CMS
collaboration \cite{CMS-dijets} is shown in Fig.~\ref{f:jets_incl} (right).
Like for the inclusive jet cross section measurements, the experimental uncertainties
are in the range 10-20\% and are dominated by uncertainties
on the jet energy scale and resolution.

\begin{figure}[hbtn]
\begin{center}
\includegraphics[width=0.48\textwidth,angle=0]{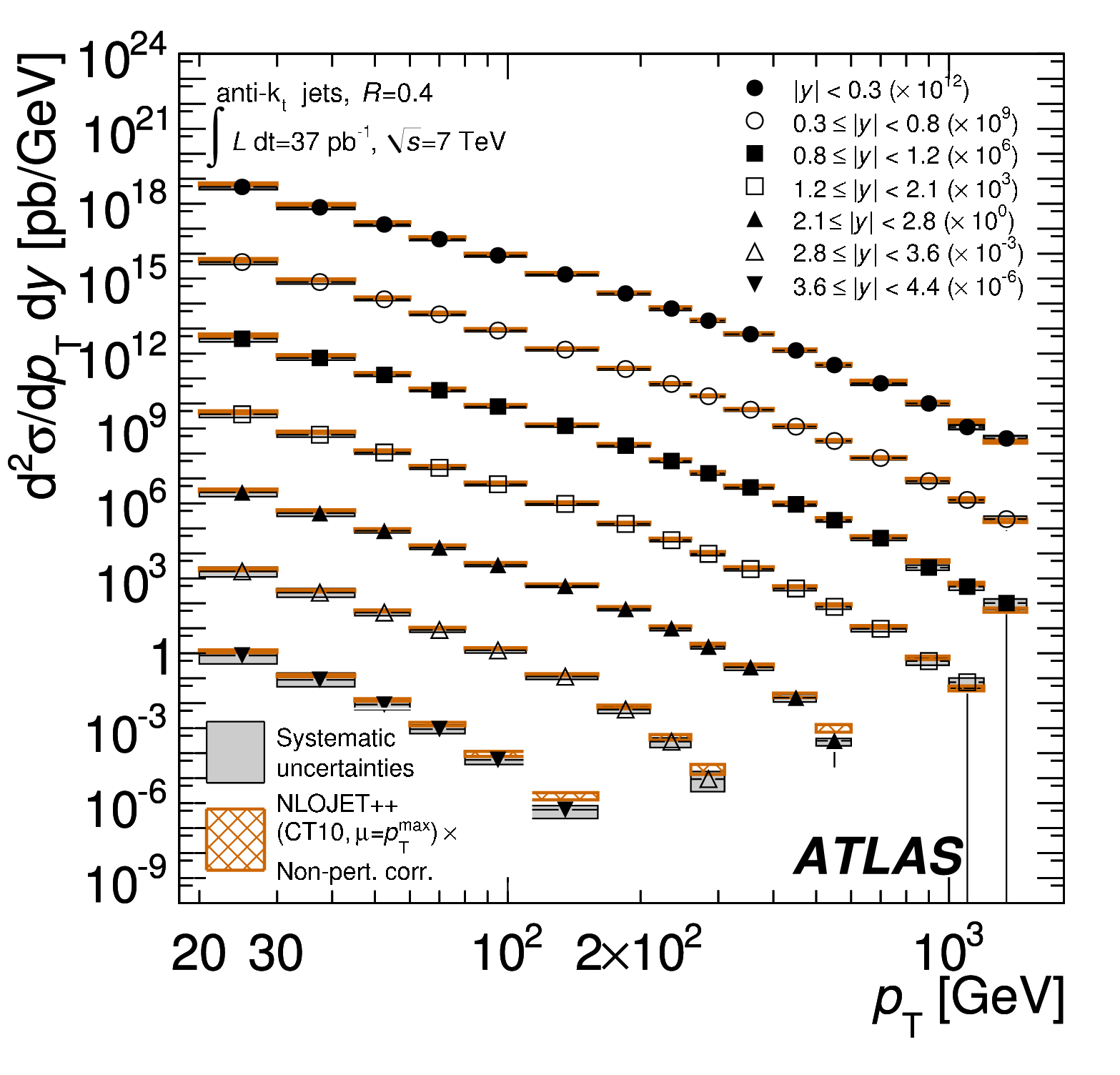}
\includegraphics[width=0.48\textwidth,angle=0]{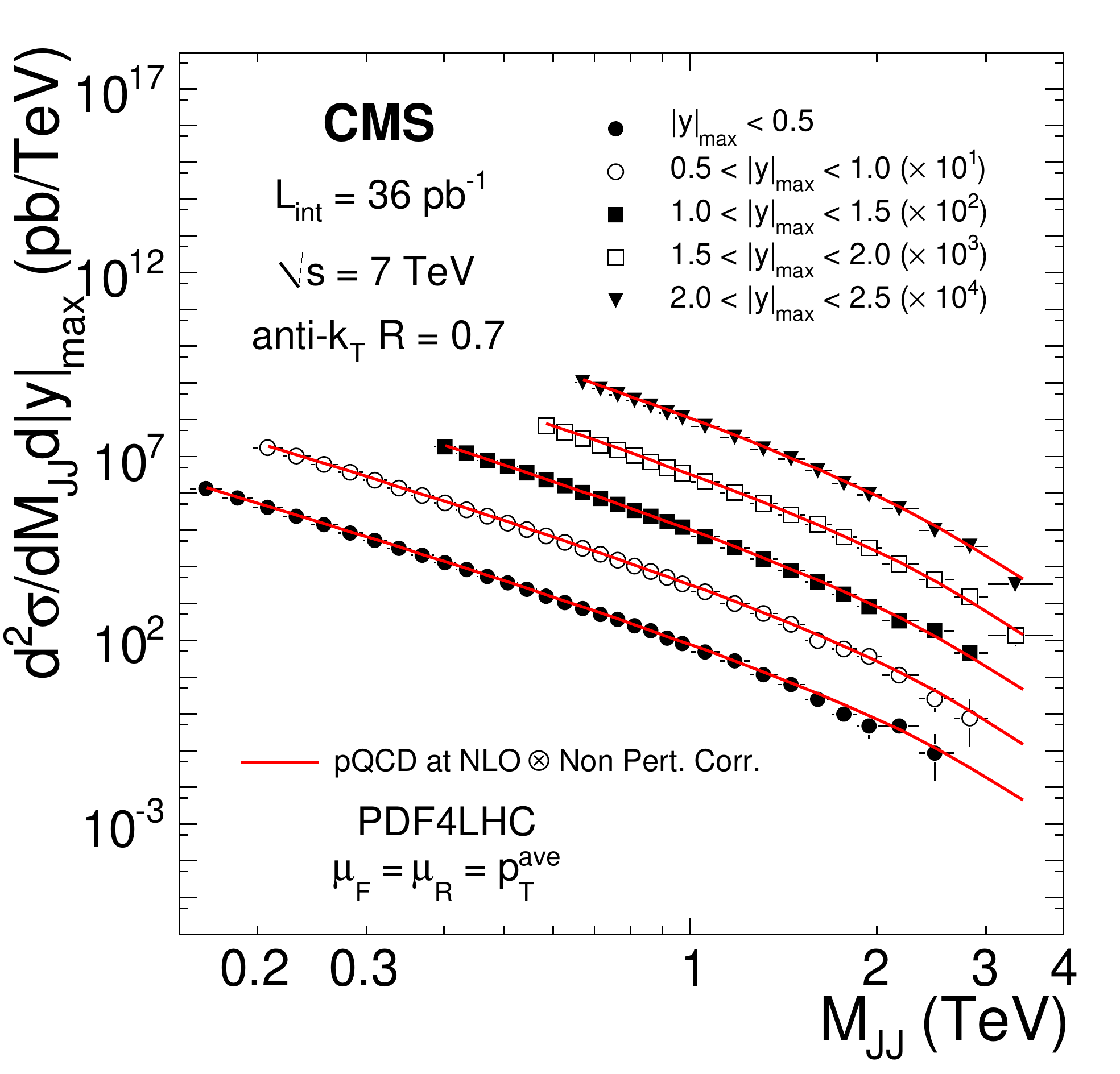}
\end{center}
\caption{\small \it
(Left): Inclusive jet double-differential cross section as a function of jet \pT\ in different regions
of $|y|$ from the ATLAS collaboration.
(Right): Measured double-differential dijet cross sections (points) as a function of the dijet invariant mass $m_{jj}$
in bins of the variable $ y_{\rm{max}}$ from the CMS collaboration.
The data are compared in both cases to NLO pQCD calculations to which non-perturbative corrections have been applied.
The error bars indicate the statistical uncertainty on the measurement. The dark-shaded band indicates the quadratic sum
of the experimental systematic uncertainties, excluding the uncertainties from the luminosity.
The theory uncertainty is shown as the light, hatched band (from Refs.~\protect\cite{ATLAS-incl-jets, CMS-dijets}).
}
\label{f:jets_incl}
\end{figure}

Different NLO pQCD predictions, using different PDF sets, are compared
to the data and the corresponding ratios of data to the NLO predictions.
Figure~\ref{f:jets_incl_ratio} shows an example from the ATLAS
collaboration \cite{ATLAS-incl-jets}. Within
the experimental and theoretical uncertainties the data are well described by the predictions,
although they are found to be systematically higher than the data.
The deviations become larger at large $| y | $ and $\pt$.
However, it is impressive to see that the QCD calculations are able to describe
the data over many orders of magnitude and up to the highest values of \pt\ and mass ever observed.

\begin{figure}[hbtn]
\begin{center}
\includegraphics[width=0.49\textwidth,angle=0]{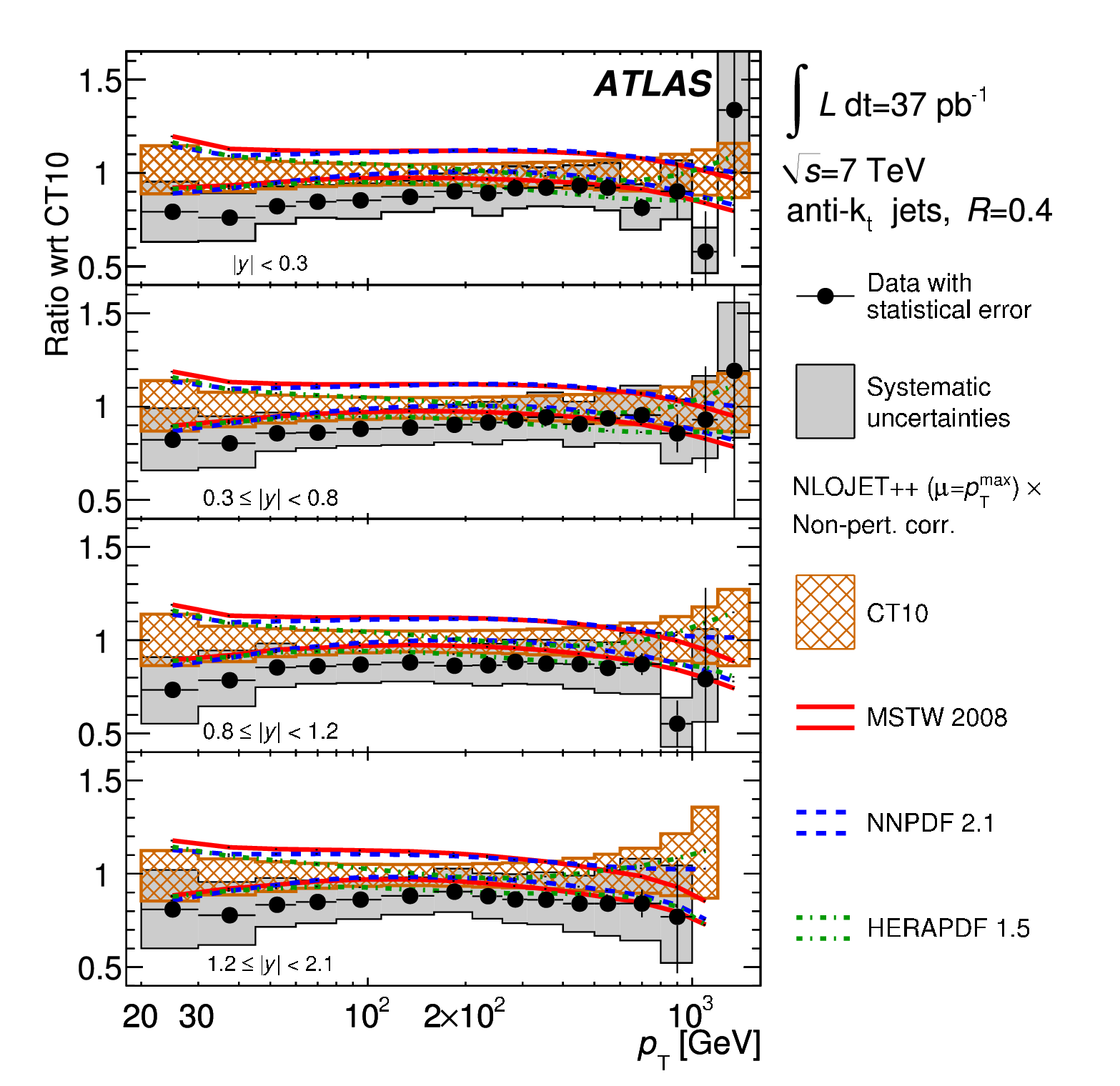}
\includegraphics[width=0.49\textwidth,angle=0]{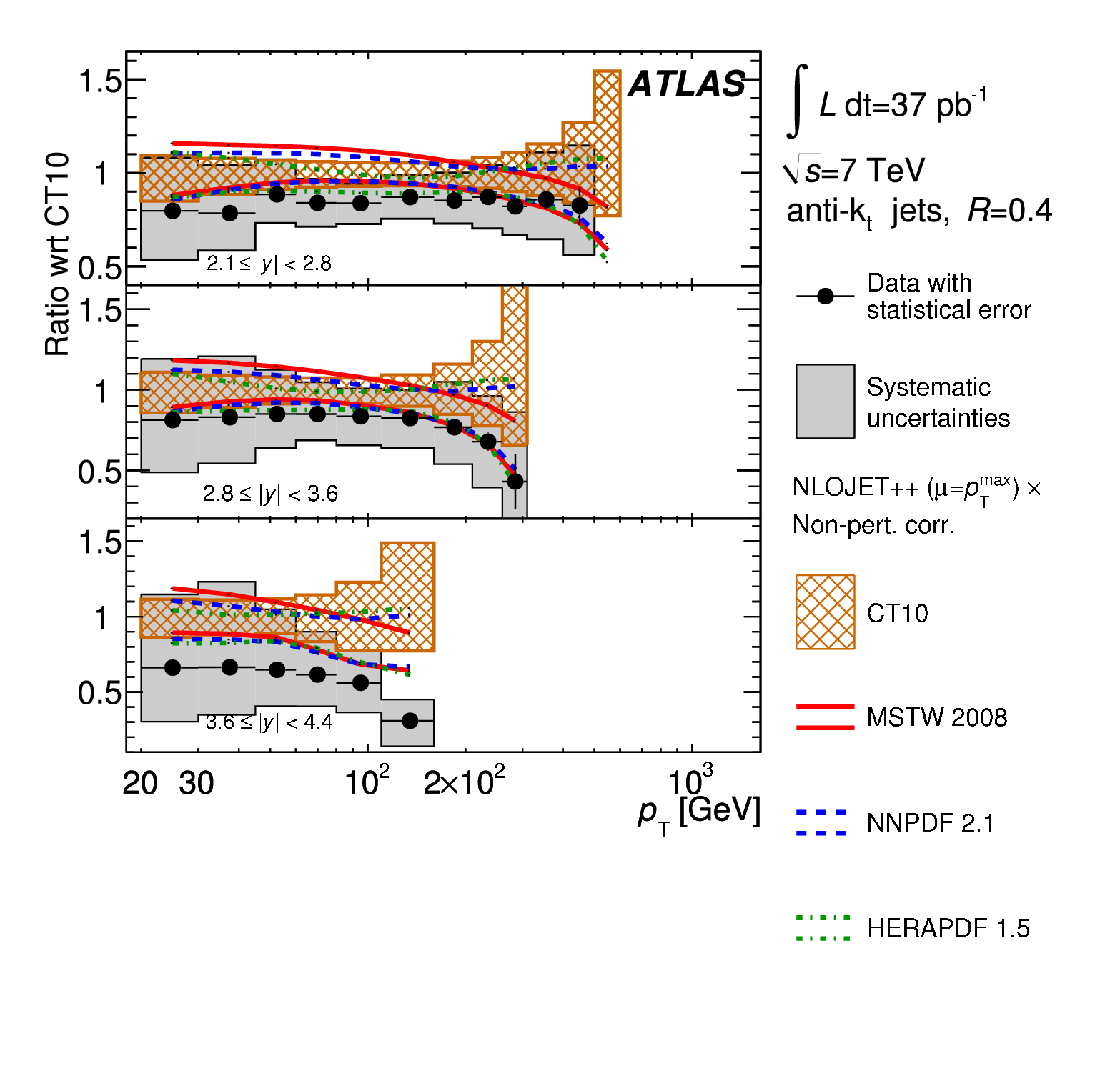}
\end{center}
\caption{\small \it
Ratios of inclusive jet double-differential cross sections to the theoretical predictions. The ratios are shown as a function of jet \pT\ in different regions of $|y|$. The theoretical error bands obtained by using NLOJET++ with different PDF sets (CT10, MSTW 2008, NNPDF 2.1, HERAPDF 1.5) are shown (from Ref.~\protect\cite{ATLAS-incl-jets}).
}
\label{f:jets_incl_ratio}
\end{figure}

The ATLAS and CMS collaborations have performed many further studies on jet production,
including the measurement of dijet angular distributions \cite{CMS-dijet-angles} and
dijet angular decorrelations \cite{ATLAS-incl-jets, CMS-dijet-corr}.
At Born level, dijets are produced with equal transverse momenta \pt\ and back-to-back in the
azimuthal angle ($\Delta \phi_{\rm{dijet}} = | \phi_{\rm{jet1}} - \phi_{\rm{jet2}}|$).
Gluon emission will decorrelate the two highest \pt\ jets and cause smaller angular
separations. The measurement of the angular distribution between the highest $\pt$ jets is therefore
also a sensitive test of perturbative QCD with the
advantage that the measurement is not strongly affected by the dominant systematic
uncertainty on the jet energy scale. The predictions from NLO pQCD are found to be
in reasonable agreement
with the measured distributions \cite{ATLAS-incl-jets, CMS-dijet-corr}.

\begin{figure}[hbtn]
\begin{center}
\includegraphics[width=0.65\textwidth,angle=0]{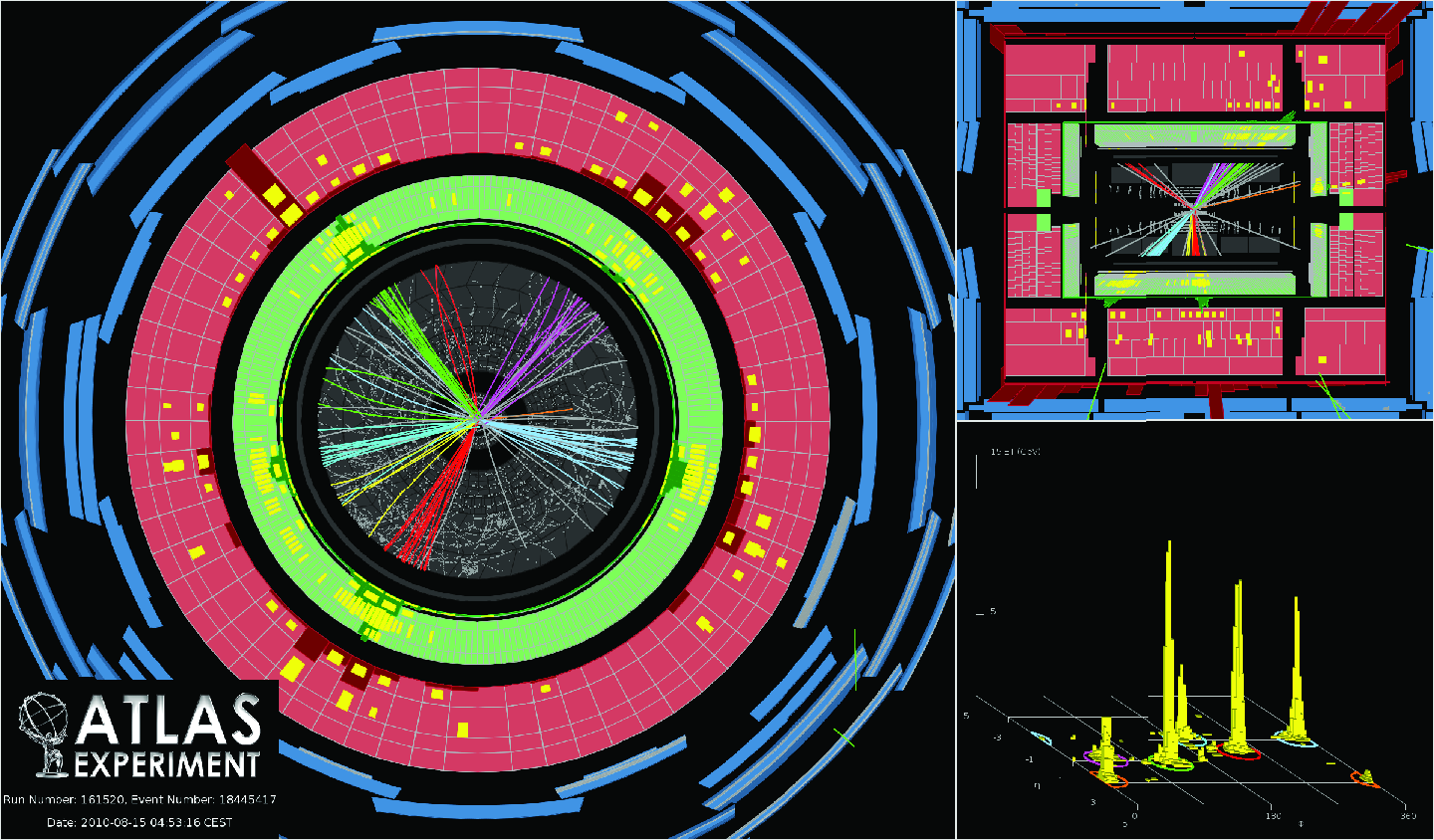}
\end{center}
\caption{\small \it
Event display of a six-jet event passing the ATLAS multijet selection requirements. The towers in the bottom right figure represent transverse energy deposited in the calorimeter projected on a grid of $\eta$ and $\phi$. Jets with transverse momenta ranging from 84 to
203~\GeV\ are measured
in this event (from Ref.~\protect\cite{ATLAS-multijet}).
}
\label{f:ATLAS-six-jets}
\end{figure}

In addition, the production of  multijets was studied \cite{ATLAS-multijet, CMS-multijet}.
In a data sample corresponding to an integrated luminosity of 2.4~\ipb\ the ATLAS collaboration
has identified 115 events with more than six jets. One such event
is shown in Fig.~\ref{f:ATLAS-six-jets}. The transverse energy deposition in the calorimeter
is shown as a function or $\eta$ and $\phi$. For this event the six jets are well separated
spatially.
Leading-order Monte Carlo simulations have been compared to the measured multi-jet inclusive and
differential cross sections.
For events containing two or more jets with $\pT >$
60~\GeV, of which at least one has $\pT >$ 80~\GeV, a reasonable
agreement is found between data and leading-order
Monte Carlo simulations with parton-shower tunes that
describe adequately the ATLAS $\sqrt{s} = 7$~\TeV\ underlying event
data. The agreement is found after the predictions
of the Monte Carlo simulations are normalized to the measured
inclusive two-jet cross section.

\newpage

\subsection{The production of $W$ and $Z$ bosons}

\Wboson~and \Zboson~bosons are expected to be produced abundantly at the LHC. The large dataset and the high LHC energy allow for detailed measurements of their production properties in a previously unexplored kinematic domain. These conditions, together with the proton-proton nature of the collisions, provide new constraints on the parton  distribution functions and allow for precise tests of perturbative QCD. Besides the measurements of the $W$ and $Z$ boson production cross sections, the measurement
of their ratio $R$ and of the asymmetry between the $W^+$ and $W^-$ cross sections (see Section~\ref{s:WZ_prod})
constitute important tests of the Standard Model. This ratio $R$ can be measured with a higher
relative precision because both experimental and theoretical uncertainties partially cancel.
With larger data sets this ratio can be used to provide constraints on the
$W$-boson width $\Gamma_W$ \cite{CDFW}.

\subsubsection{Inclusive cross-section measurements}
Measurement of the $W^+, W^-$ and $\Zg$ boson inclusive production cross sections
are performed using
the leptonic decay modes $W \to \ell \nu$ and $Z \to \ell \ell$.
Already in 2010, the two collaborations published first measurements in the
electron and muon decay modes~\cite{ATLAS:1010.2130, CMS:1012.2466}.
They were updated with the full data sample taken in 2010 corresponding to an integrated
luminosity of 36~$\ipb$~\cite{CMS:1107.4789,ATLAS:1109.5141}. In this data sample
the ATLAS experiment has observed a total of about 270.000 $W \to \ell \nu$ decays
and a total of about 24.000 $\Zg \to \ell \ell$ decays.
The measurements in the electron and muon channels were found to give consistent results
and were combined to obtain a
single joint measurement taking into account the statistical and systematic uncertainties and
their correlations. The results are displayed in Fig.~\ref{f:WZ_xs} together with previous
measurements of the total $W$ and $Z$ production cross sections
by the UA1~\cite{UA1W2} and UA2~\cite{UA2W} experiments at $\sqrt{s} = 0.63$~\TeV\ at the
CERN Sp$\overline{\rm{p}}$S and
by the CDF~\cite{CDFW} and $\dzero$~\cite{D0W} experiments at $\sqrt{s} = 1.8$~\TeV\ and
$\sqrt{s} = 1.96$~\TeV\ at the Fermilab Tevatron collider and by the PHENIX \cite{Adare:2010xa}
experiment in proton-proton collisions
at $\sqrt{s} = 0.5$~\TeV\ at the RHIC collider.
These measurements are compared
to the NNLO theoretical predictions for proton-proton and proton-antiproton collisions.
All measurements are in good agreement with the theoretical predictions and the
energy dependence of the total $W$ and $Z$ production cross sections is well described.
\begin{figure}[htp]
  \begin{center}
{\includegraphics[width=0.49\textwidth]{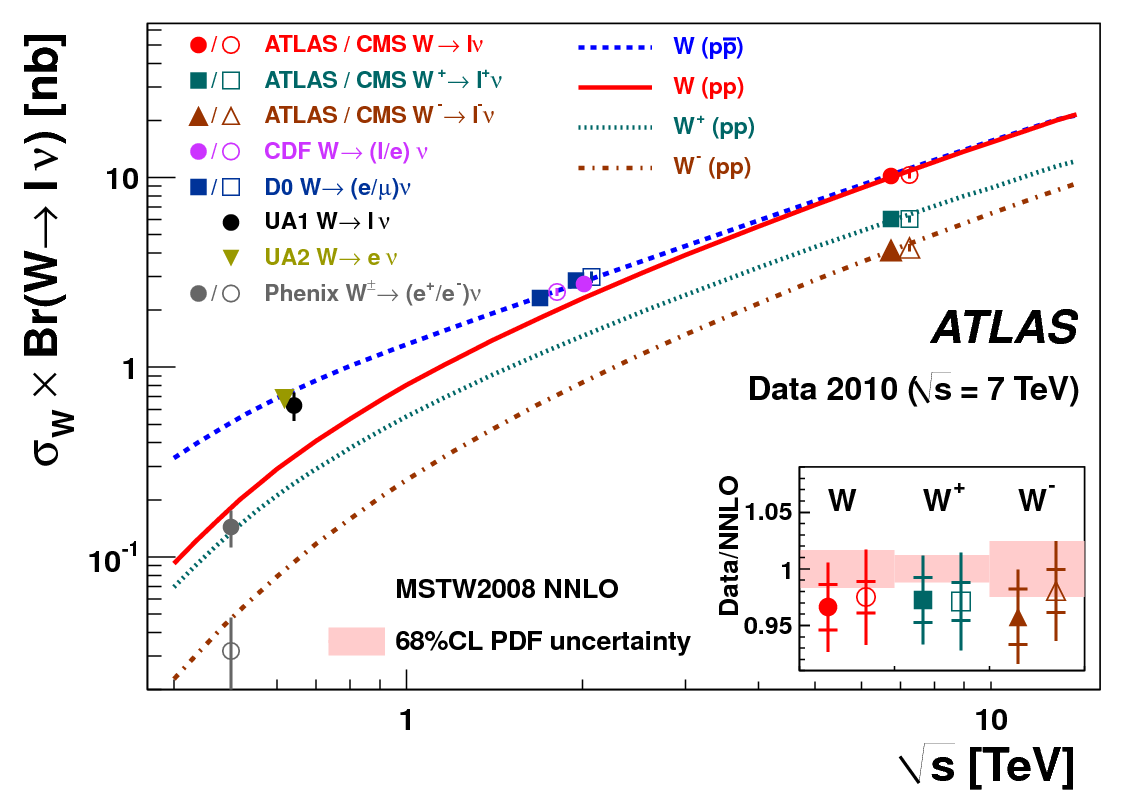}}
{\includegraphics[width=0.49\textwidth]{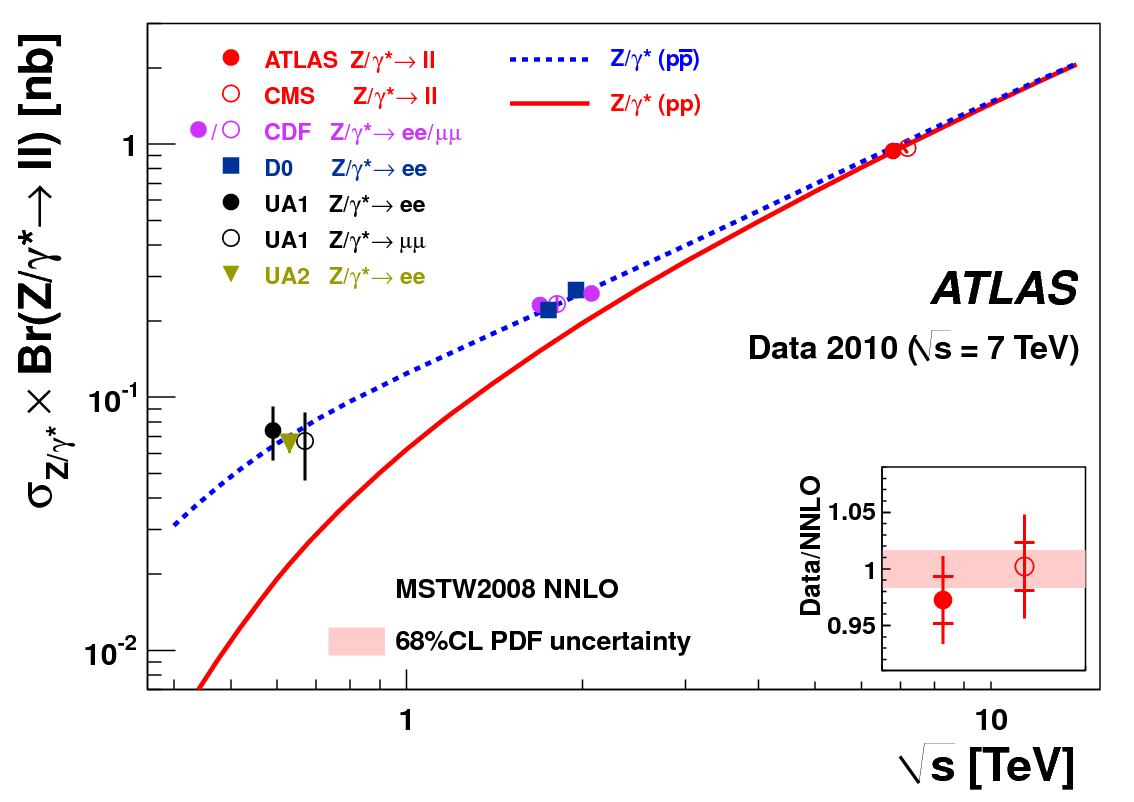}}
\end{center}
\caption{\it\small  The measured values of
$\sigma_W \cdot \mathrm{BR}$ ($\ensuremath{W}\rightarrow \ell \nu)$ for \Wplus, \Wminus~and for their sum (left) and of
$\sigma_{\ensuremath{Z}/\gamma^*} \times \mathrm{BR}$ ($\ensuremath{Z}/\gamma^*\rightarrow \ell \ell$)
(right) compared to the theoretical predictions based on NNLO QCD calculations.
Results are shown for the combined measurements of electron and muon final states. The predictions are shown for both proton-proton (\Wplus, \Wminus and their sum) and proton-antiproton colliders (\Wboson) as a function of $\sqrt{s}$. In addition, previous measurements at proton-antiproton and proton-proton colliders are shown. The data points at the various energies are staggered to improved readability. The data points are shown with their total uncertainty. The theoretical uncertainties are not shown in this figure (from Ref.~\protect\cite{ATLAS:1109.5141}).
}
\label{f:WZ_xs}
\end{figure}
The precision of the integrated $W$ and $\Zg$ cross sections in the fiducial regions
is $\sim \pm 1.2\%$ with an additional uncertainty of $\pm$3.4\% resulting from the
knowledge of the luminosity. It should be noted that the experimental uncertainties
are already dominated by systematic uncertainties.  The total integrated
cross sections are obtained from an extrapolation of the measurement in the
fiducial regions to the full acceptance. Due to uncertainties on the
acceptance corrections, the uncertainties on the total cross sections
are about twice as large.

A summary of the ratios of the measured total $W^+, W^-, W$ and $\Zg$ cross sections by the CMS
collaboration to the theoretical NNLO calculations is shown in Fig.~\ref{f:WZ-R-CMS} (left).
Within the experimental and theoretical uncertainties there is excellent
agreement. This figure also includes a comparison of the measured ratios
$R_{W/Z} = \sigma_W \cdot BR (W \rightarrow \ell \nu) / \sigma_Z \cdot BR (Z\rightarrow \ell\ell)$ and
$R_{+/-} = \sigma_{W^{+}} \cdot BR (W^{+} \rightarrow \ell^{+}\nu) / \sigma_{W^{-}} \cdot BR (W^{-} \rightarrow \ell^{-} \nu) $.
Due to the cancellation of uncertainties, most notably the luminosity uncertainty, the precision of these
ratio measurements is more precise.  Also the measured ratios are well
described by the theoretical NNLO calculations.
\begin{figure}[hbtn]
\begin{center}
\includegraphics[width=0.50\textwidth,angle=0]{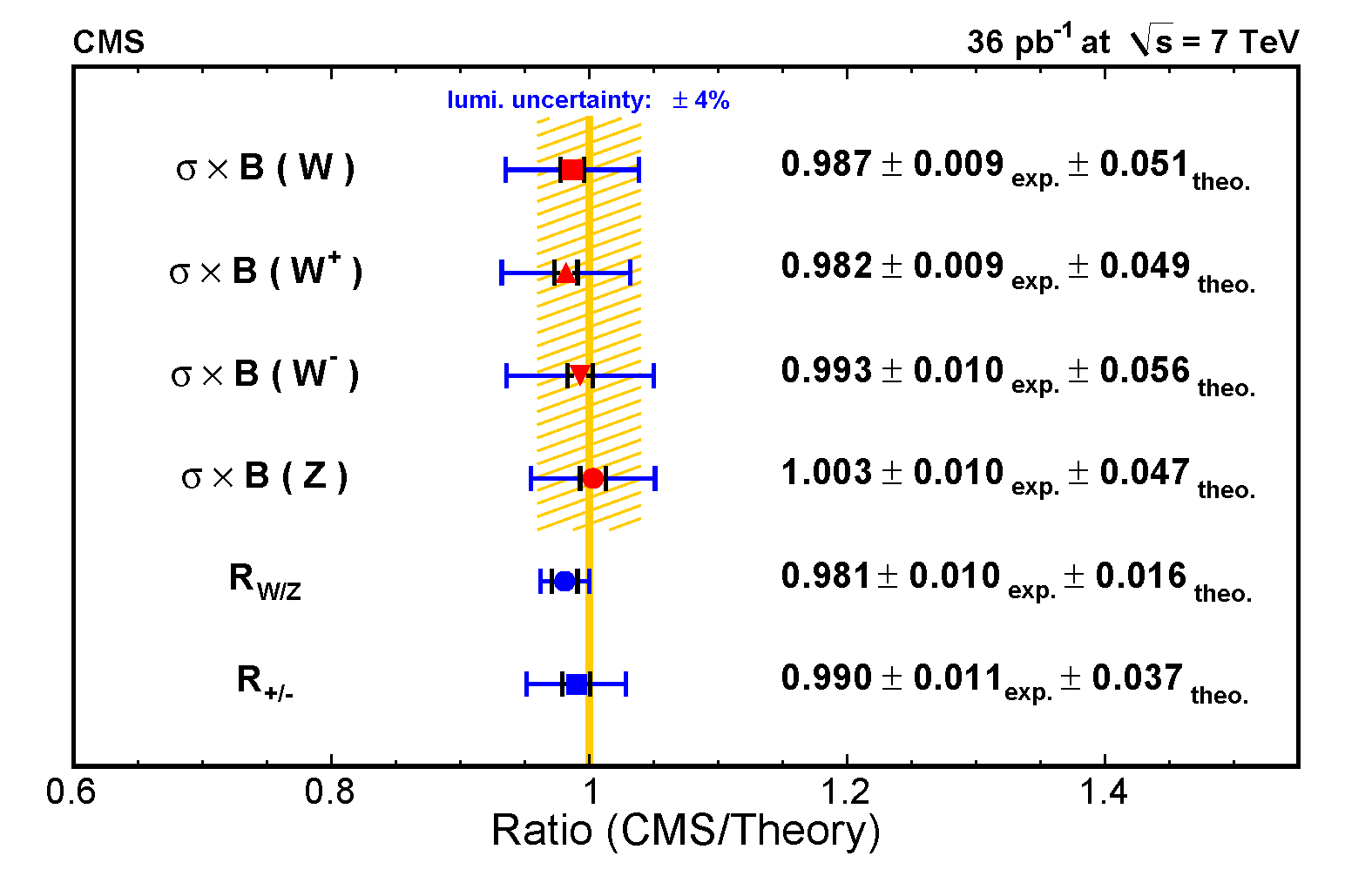}
\includegraphics[width=0.48\textwidth,angle=0]{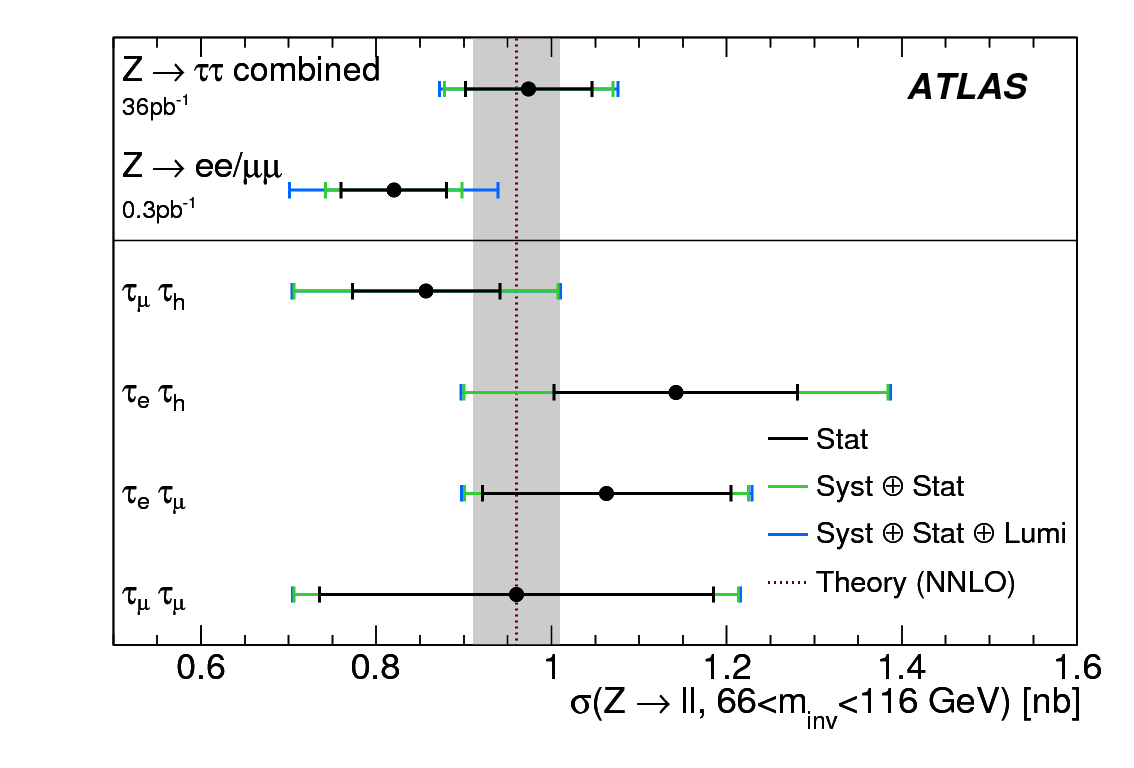}
\end{center}
\caption{\small \it
(Left): Ratio of CMS measurement of $W$ and $Z$ cross sections to theory expectations. The experimental uncertainty is the sum in quadrature of the statistical and the systematic uncertainties not including the uncertainty on the extrapolation to the full acceptance due to parton density functions
(taken from Ref.~\protect\cite{CMS:1107.4789}).
(Right): Measurements of the $Z \to \tau \tau$ cross sections from the ATLAS
experiment in various $\tau$ decay modes and comparison to theoretical predictions and
to the measurements in the electron and muon channels
(taken from Ref.~\protect\cite{ATLAS:1108.2016}).
}
\label{f:WZ-R-CMS}
\end{figure}

Meanwhile the cross sections have also been measured
in the $W \to \tau \nu$~\cite{ATLAS:1108.4101}
and $Z \to \tau \tau$~\cite{CMS:1104.1617, ATLAS:1108.2016} decay modes, where
hadronically decaying $\tau$ leptons are identified and
reconstructed. The results obtained by the ATLAS collaboration in various $\tau$ decay
modes are displayed in Fig.~\ref{f:WZ-R-CMS} (right). They are found to be in good agreement
with theoretical predictions and with the results obtained in the $Z \to \ee$ and
$Z \to \mu \mu$ final states.

\subsubsection{Differential cross section measurements}
With the complete data set collected in 2010 more
differential cross-section measurements became possible.
Both the ATLAS and CMS collaborations have performed measurements as a function of lepton
pseudorapidity $\eta_\ell$, for the \Wpm\ cross sections, and of the
boson rapidity, $y_Z$, for the
\Zg\ cross section~\cite{CMS:1107.4789,ATLAS:1109.5141}.
For the \Zg\ case, all values refer to dilepton mass windows from 66 - 116~\GeV\ and
60~-~110~\GeV\ for the ATLAS and CMS analyses, respectively.
The cross sections are measured in well-defined kinematic regions within the detector
acceptance, defined by the pseudorapidity of the charged lepton and the transverse momentum of the neutrino.
\begin{figure}
\begin{center}
\includegraphics[width=0.45\textwidth,angle=0]{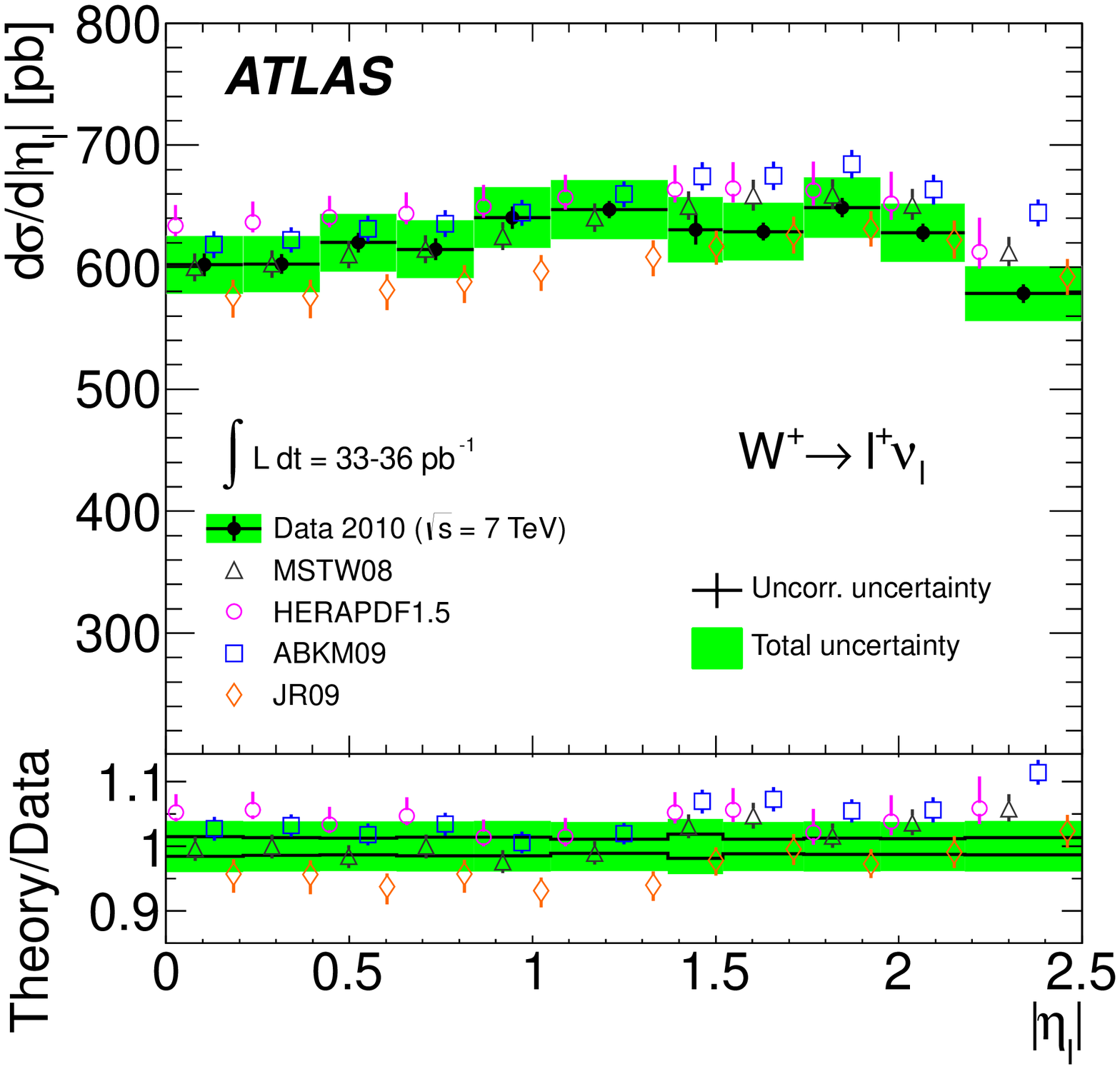}
\includegraphics[width=0.45\textwidth,angle=0]{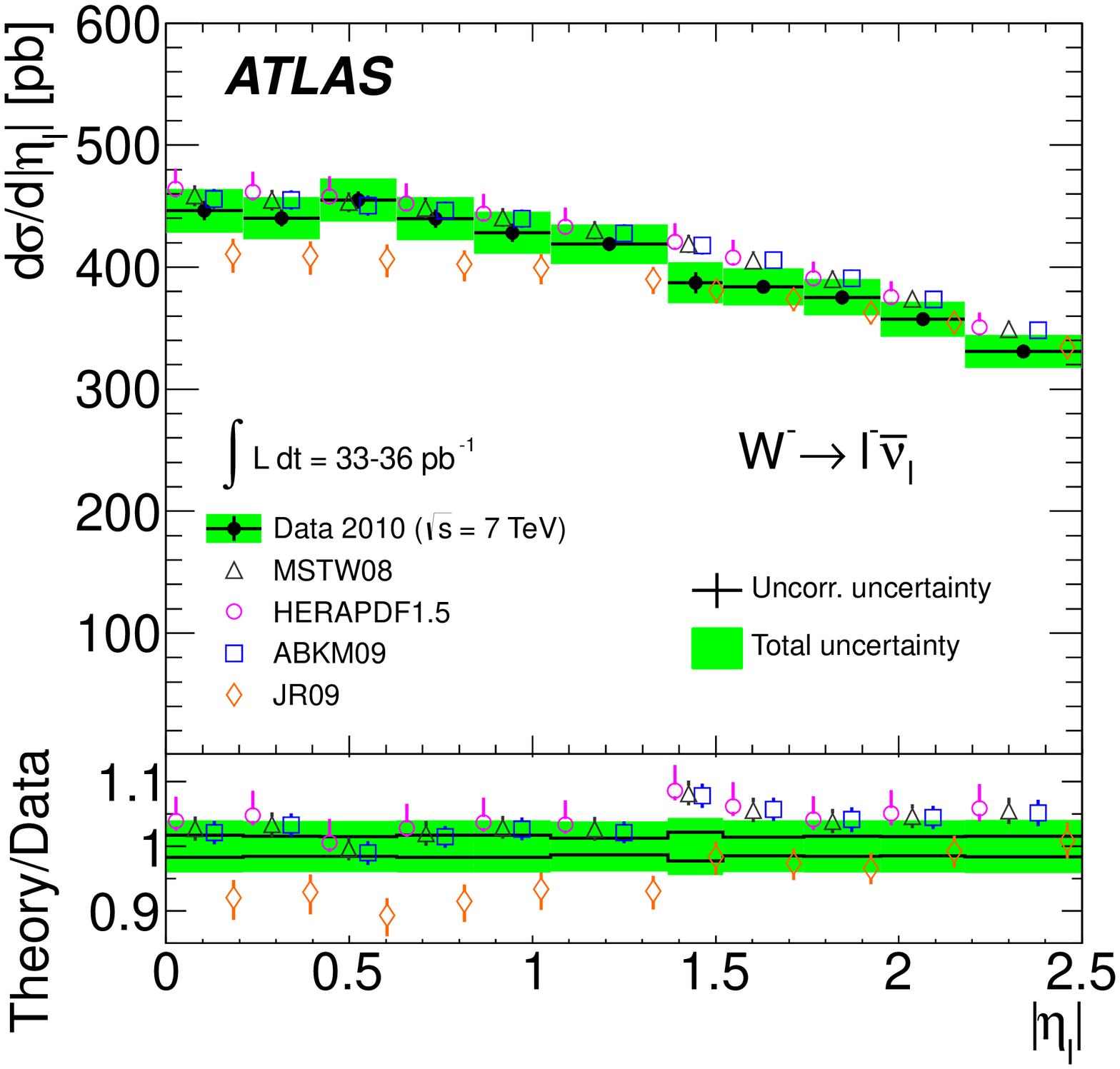}
\end{center}
\caption{\small \it
Differential cross-section measurement $d\sigma/d|\eta_{\ell}|$ for $W^+$ (left) and $W^{-}$ (right) for $W \to \ell \nu$ from the ATLAS
collaboration compared to the NNLO theory predictions using various PDF sets. The kinematic requirements are $\pT(\ell)$ > 20~\GeV, $\pT(\nu)$ > 25~\GeV\ and $m_{T}$ > 40~\GeV. The ratio of theoretical predictions to data is also shown. Theoretical points are displaced for clarity within
each bin (from Ref.~\protect\cite{ATLAS:1109.5141}).
}
\label{f:W-diff}
\end{figure}
The differential $\Wplus$ and $\Wminus$ cross sections as measured by the ATLAS collaboration are shown in Fig.~\ref{f:W-diff}.
The measurements for the electron and muon final states were found to be in good agreement with each other and were combined.
These data are compared with the theoretical NNLO predictions using various NNLO PDF sets
(JR09, ABKM09, HERAPDF1.5 and MSTW08).
The differential $\Zg$ cross section as a function of the boson rapidity as measured by the ATLAS and CMS collaborations are shown in Fig.~\ref{f:Z-diff}.
Although the gross features of these differential $W$ and $Z$ cross-section measurements are well described by the
theoretical calculations, the (pseudo)rapidity dependence shows some disagreement which carries
important information on the underlying parton density functions. It is expected that these differential
measurements will reduce the uncertainties on the parton density functions. Very recently,
these data have been used together with the $ep$ scattering data from HERA to
extract the ratio or the strange-to-down see quark density at Bjorken $x$
values around  0.01. The ratio is found to be consistent with 1 and supports the
hypothesis
that the density of the light sea quarks is flavour independent~\cite{ATLAS-strange}.
The general agreement between theory and experiment is remarkable and provides evidence for the
universality of the PDFs and the reliability of perturbative QCD calculations in the
kinematic regime of the LHC.

\begin{figure}[hbtn]
\begin{center}
\includegraphics[width=0.45\textwidth,angle=0]{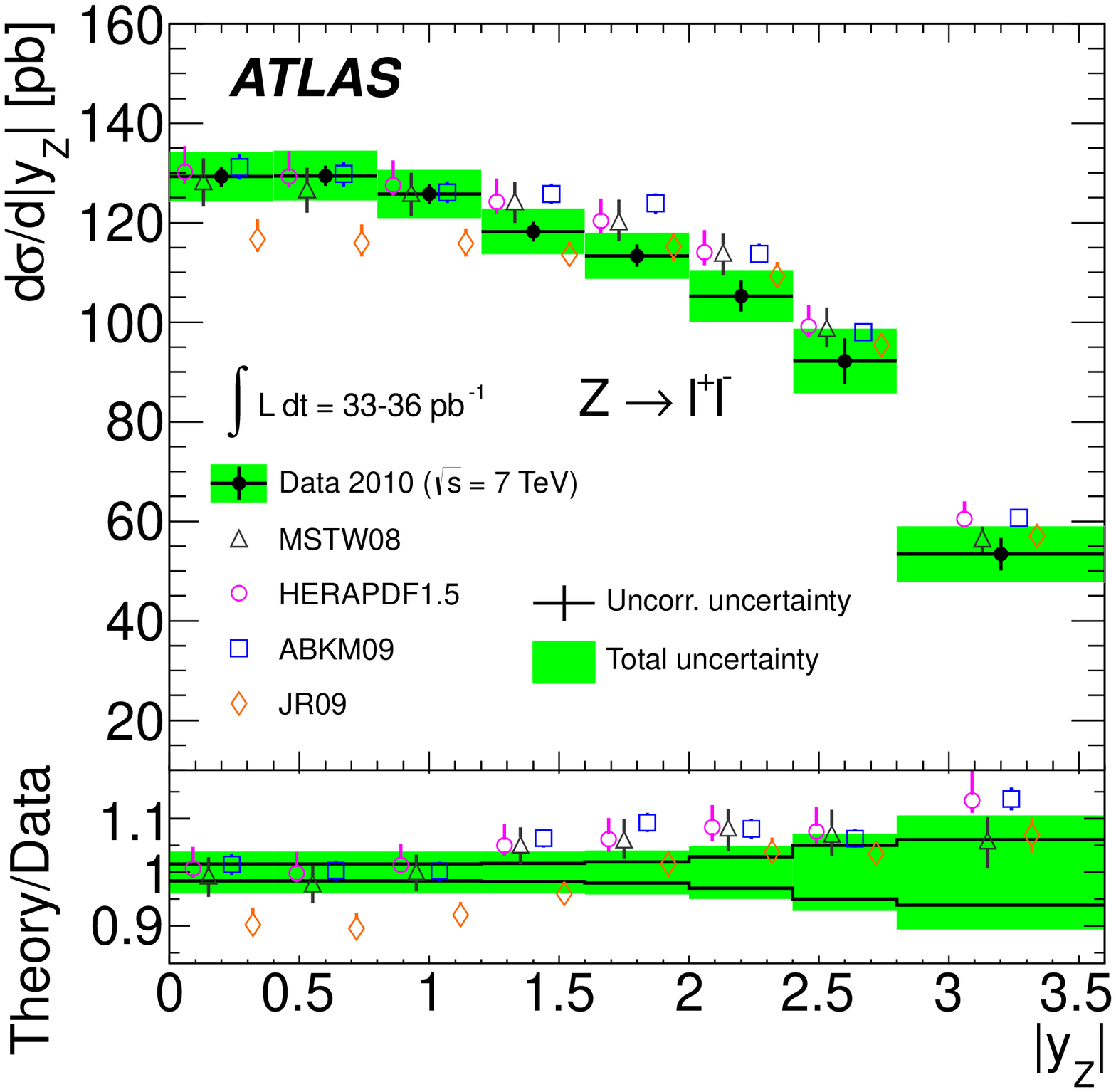}
\includegraphics[width=0.45\textwidth,angle=0]{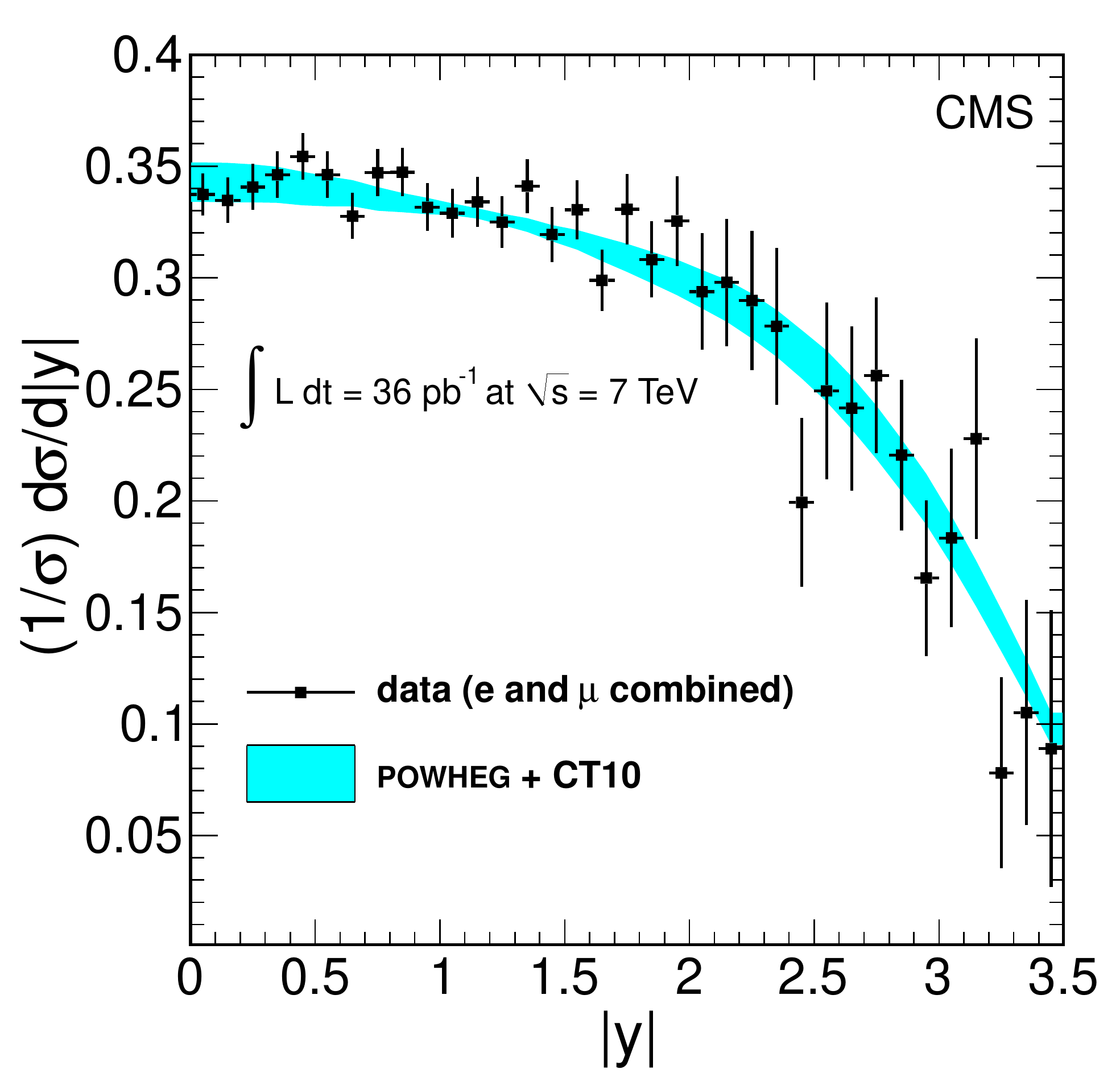}
\end{center}
\caption{\small \it
Differential cross-section measurements ${\rm d}\sigma/{\rm d}|y_Z|$ for $Z \to \ell\ell$ from the ATLAS (left) and CMS (right) collaborations compared to NNLO theory predictions using various PDF sets. The kinematic requirements are $66 < m(\ell\ell) < 116$~\GeV\ and $\pT(\ell) > 20$~\GeV. The ratio of theoretical predictions to data is also shown for the ATLAS measurements. Theoretical points are displaced for clarity within each bin
 (from Refs.~\protect\cite{ATLAS:1109.5141,CMS:1107.4789}).
}
\label{f:Z-diff}
\end{figure}

\subsubsection{Measurements of the associated $W$ and $Z$ + jet production}

The study of the associated production of vector bosons with high-\pT\
jets constitutes another
important test of the perturbative QCD. In addition, these final states are a significant
background to studies of other Standard Model processes, such as \ttbar\ or diboson production,
as well as for searches for the Higgs boson and for physics beyond the Standard Model.

The ATLAS and CMS collaborations have presented detailed measurements of these processes
based on the complete dataset from 2010, corresponding to an integrated luminosity of
36~$\ipb$~\cite{CMS:1110.3226, CMS:1110.4973, ATLAS:1111.2690, ATLAS:1201.1276}.
Cross sections have been determined for the associated $W$ and $Z$+jet production
as a function of inclusive jet multiplicity, $N_{\rm{jet}}$, for up to five jets. At each
multiplicity, the cross sections have also been presented as a function of jet
transverse momenta of all jets. The results,
corrected for all detector effects and for all backgrounds such as
diboson and top quark pair production, are compared with
particle-level predictions from perturbative QCD.
As an example, the $W$+jets cross-section measurements as a function of
jet multiplicity are shown in Fig.~\ref{f:wjets} (left) and as a function
of the \pT\ of the first jet in the event in Fig.~\ref{f:wjets} (right).
Leading-order multiparton event generators like {\em ALPGEN}~\cite{Mangano:2002ea}~or {\em SHERPA}~\cite{Gleisberg:2008ta},
normalized to the NNLO total cross
section for inclusive $W$-boson production, describe the data
reasonably well for all measured inclusive jet multiplicities.
Next-to-leading-order calculations from {\em MCFM}~\cite{MCFM}, studied for
$N_{\rm{jet}} \le 2$, and {\em BlackHat-Sherpa}~\cite{BlackHat-Sherpa}, studied for
$N_{\rm{jet}} \le 4$, are found to be mostly in good agreement with
the data. This also holds for the measurement of the transverse momentum
distributions of the $W$ and $Z$ boson, which are correlated to the jet activity
in the $W$ and $Z$ events.

\begin{figure}[hbtn]
\begin{center}
\includegraphics[width=0.4\textwidth,angle=0]{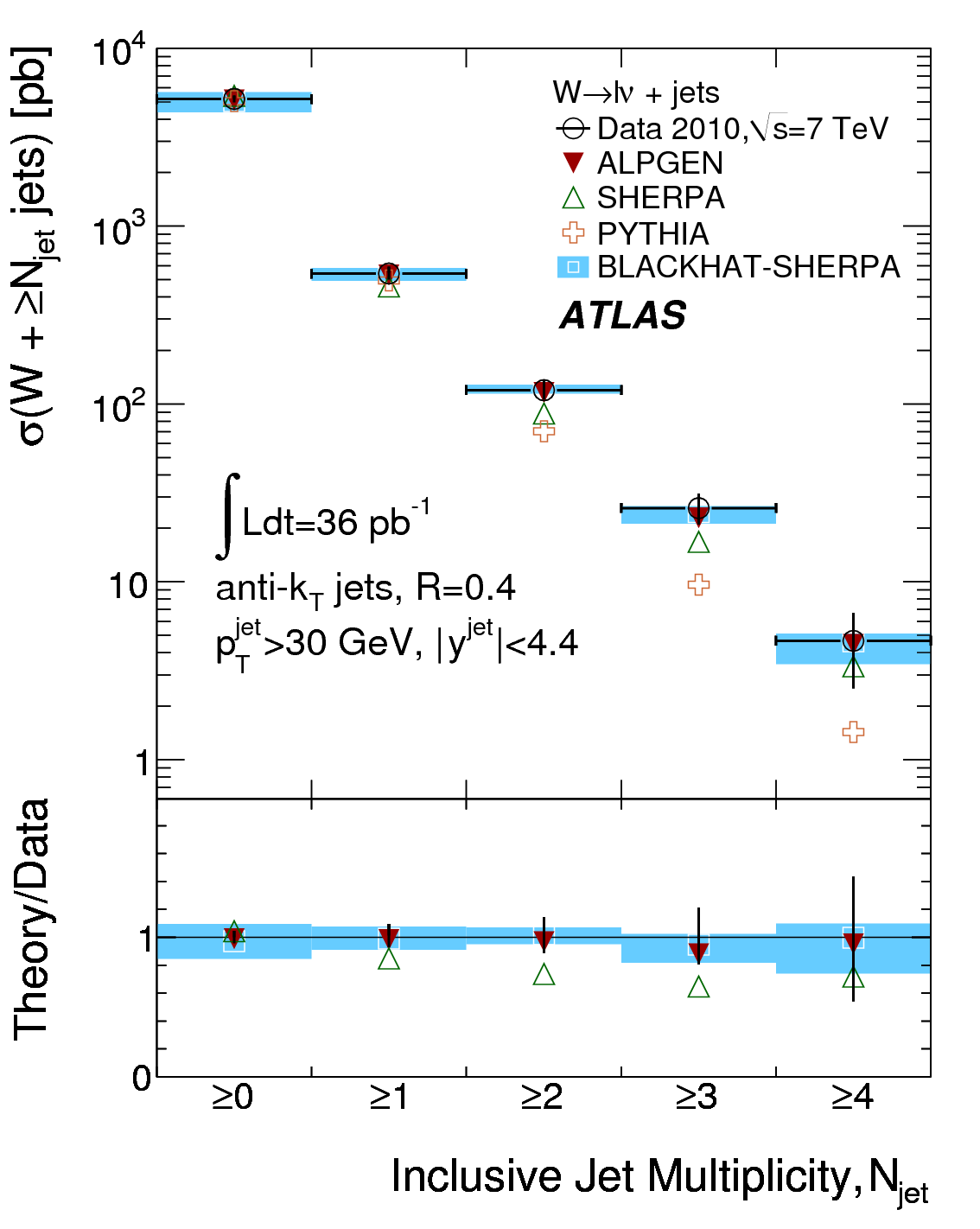}
\includegraphics[width=0.4\textwidth,angle=0]{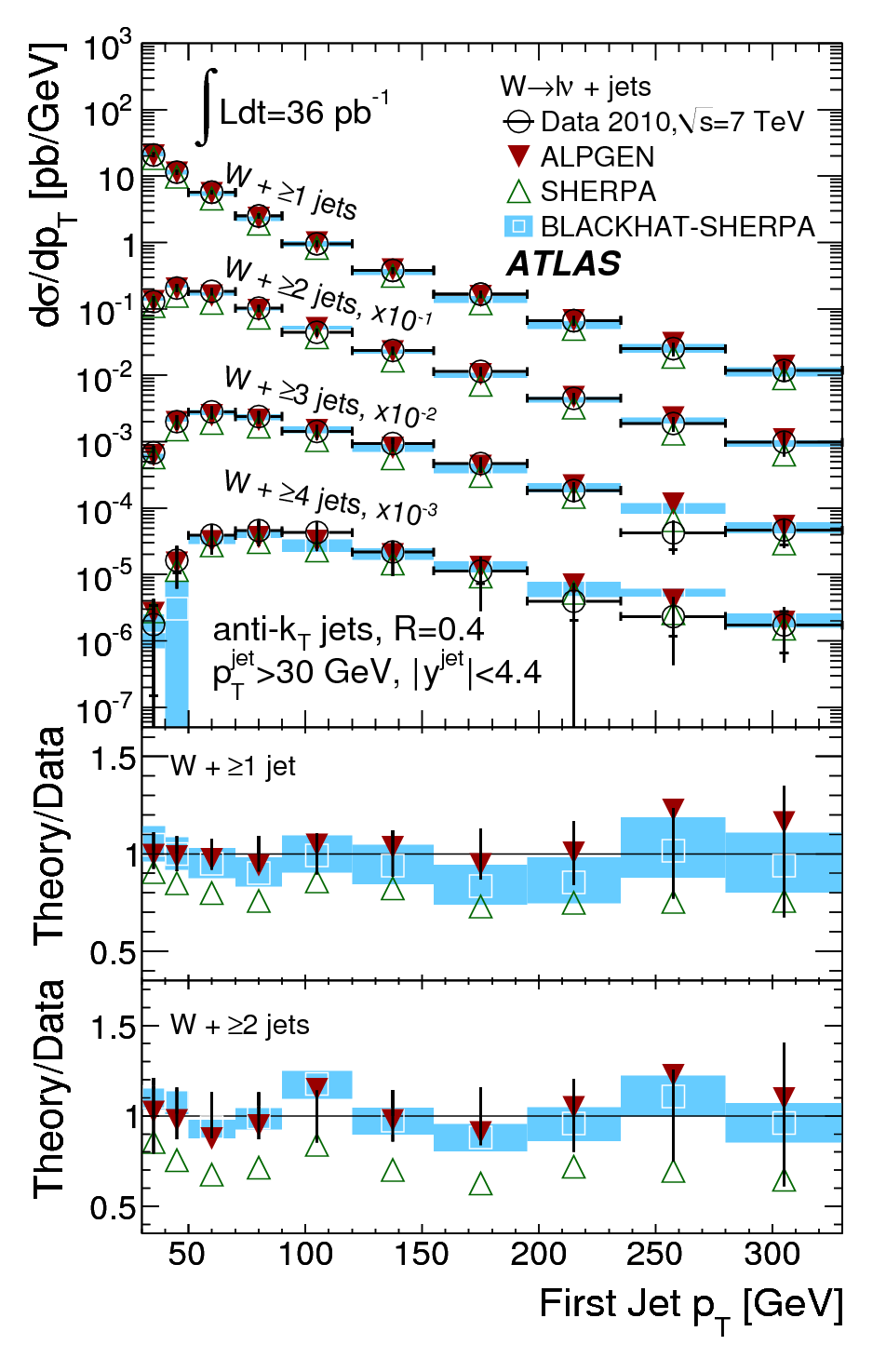}
\end{center}
\caption{\small \it
The measured $W$+jets cross sections as a function of jet multiplicity (left) and as a function
of the \pT\ of the first jet in the event (right). The \pT\ of the first jet is shown
separately for events with $\ge 1$ jet to $\ge 4$ jets. Shown are predictions from
ALPGEN, SHERPA and BlackHat-SHERPA, and the ratio of theoretical predictions
to data (from Ref.~\protect\cite{ATLAS:1201.1276}).
}
\label{f:wjets}
\end{figure}

\subsection{Production of top quarks}

\subsubsection{Measurement of production cross sections}

The top quark is the heaviest known elementary particle with a mass of about 173~\GeV.
Due to its high mass it is believed to play a special role in the electroweak symmetry
breaking and
possibly in models of new physics beyond the Standard Model. In that context it
is remarkable that its Yukawa coupling $\lambda_t$ is close to one. We still know little
about the properties of the top quark, like spin, charge, lifetime, decay properties
(rare decays) or the gauge and Yukawa couplings. Another important parameter is the mass
of the top quark, which has, however, been relatively precisely measured
at the Tevatron collider to be $m_{\rm top} = 173.2 \pm 0.90$~\GeV, i.e. with a precision
of 0.5\%. A further improvement here is important for a precise test of electroweak
radiative corrections.

Due to the high mass, the top quarks decays mainly via $t \to W b$ before it hadronizes.
The production of \ttbar\ pairs at the LHC proceeds mainly via
gluon initiated processes and is expected to be a factor of 20 larger at the LHC with
$\sqrt{s}$ = 7~\TeV\ than at the Tevatron.
The decays studied are characterized by the number of
charged leptons in the final state. A large fraction of the branching ratio is devoted to
lepton-hadron decays, where one of the $W$s decays leptonically and the other one
into a pair of jets, i.e. $\ttbar \to Wb \ Wb \to \ell \nu b \ qq b$.
The final state in this case consists
of a lepton, missing transverse momentum (carried away by the neutrino) and four jets out of which
two are originating from b-quarks. The complementary dilepton decay mode is also important for
top-quark physics at hadron colliders. The fully hadronic channel is more difficult to trigger
on and shows a worse signal-to-background ratio, but despite this has also
been measured at the LHC.

Both the ATLAS and CMS collaborations measured the production cross section
for the pair production of top quarks in all above-mentioned final
states~\cite{CMS:1105.5661, CMS:1108.3773, ATLAS:1108.3699, ATLAS:1201.1889, ATLAS-CONF-2012-032}.  The results are displayed in Fig.~\ref{f:sigma_tt}.
The most precise measurement comes from the single lepton channel, which shows already a precision
of the order of $\pm$7\%.
In this channel the cross sections are measured with and without the requirement of a
 b-tagged jet. The results obtained in the dilepton channel are consistent with these results.
The measurements are found to be
in good agreement with the approximate NNLO calculations~\cite{Moch:0804.1476, Moch:0907.2527},
although the experimental values are found to be about 1$\sigma$ higher in each experiment.
The experimental measurement is already limited by the
experimental systematic uncertainties (jet energy scale, b-tagging, ...) and by the
uncertainty on the luminosity.
\begin{figure}[t]
\begin{center}
\includegraphics[width=0.49\textwidth,angle=0]{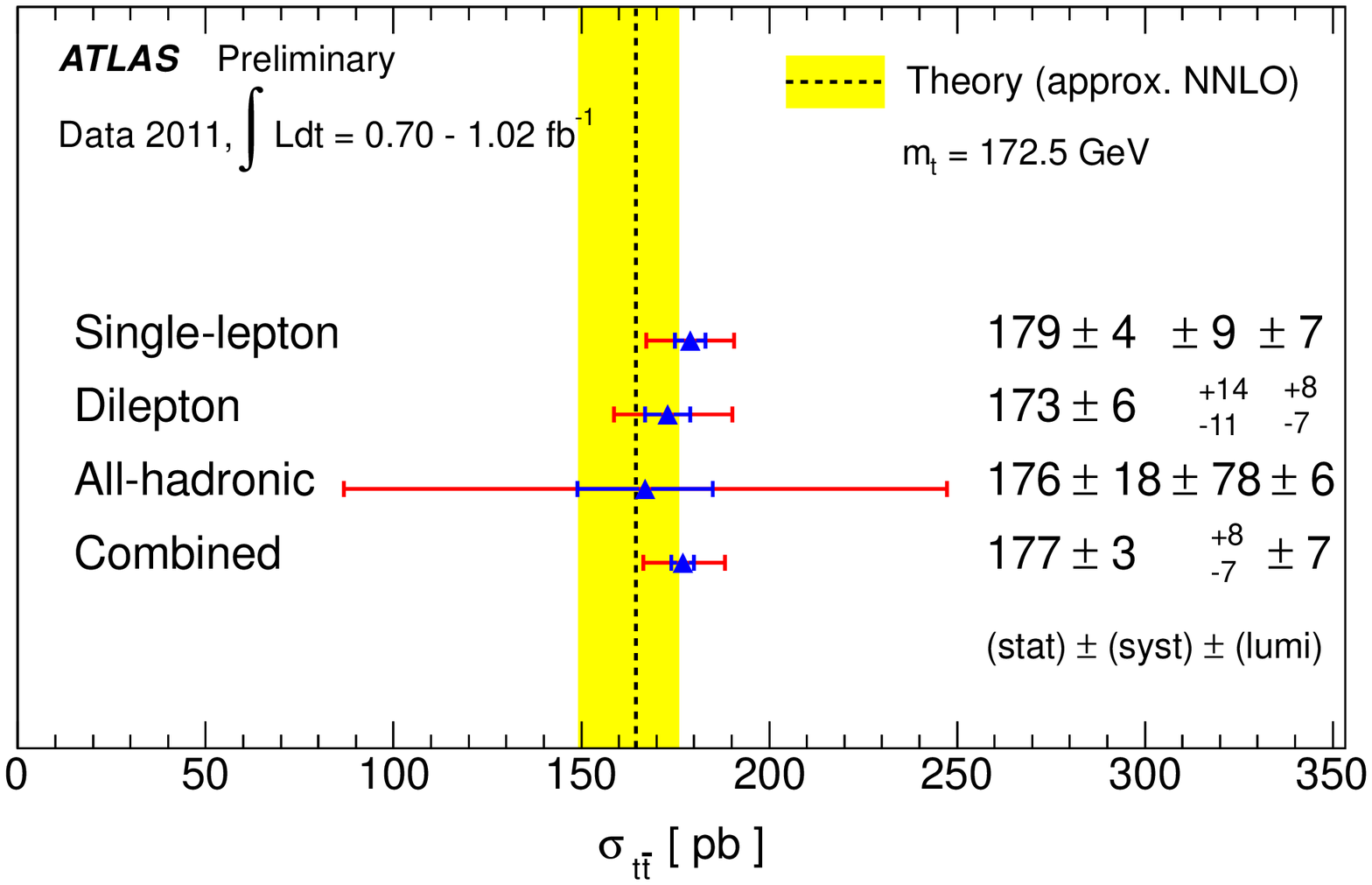}
\includegraphics[width=0.49\textwidth,angle=0]{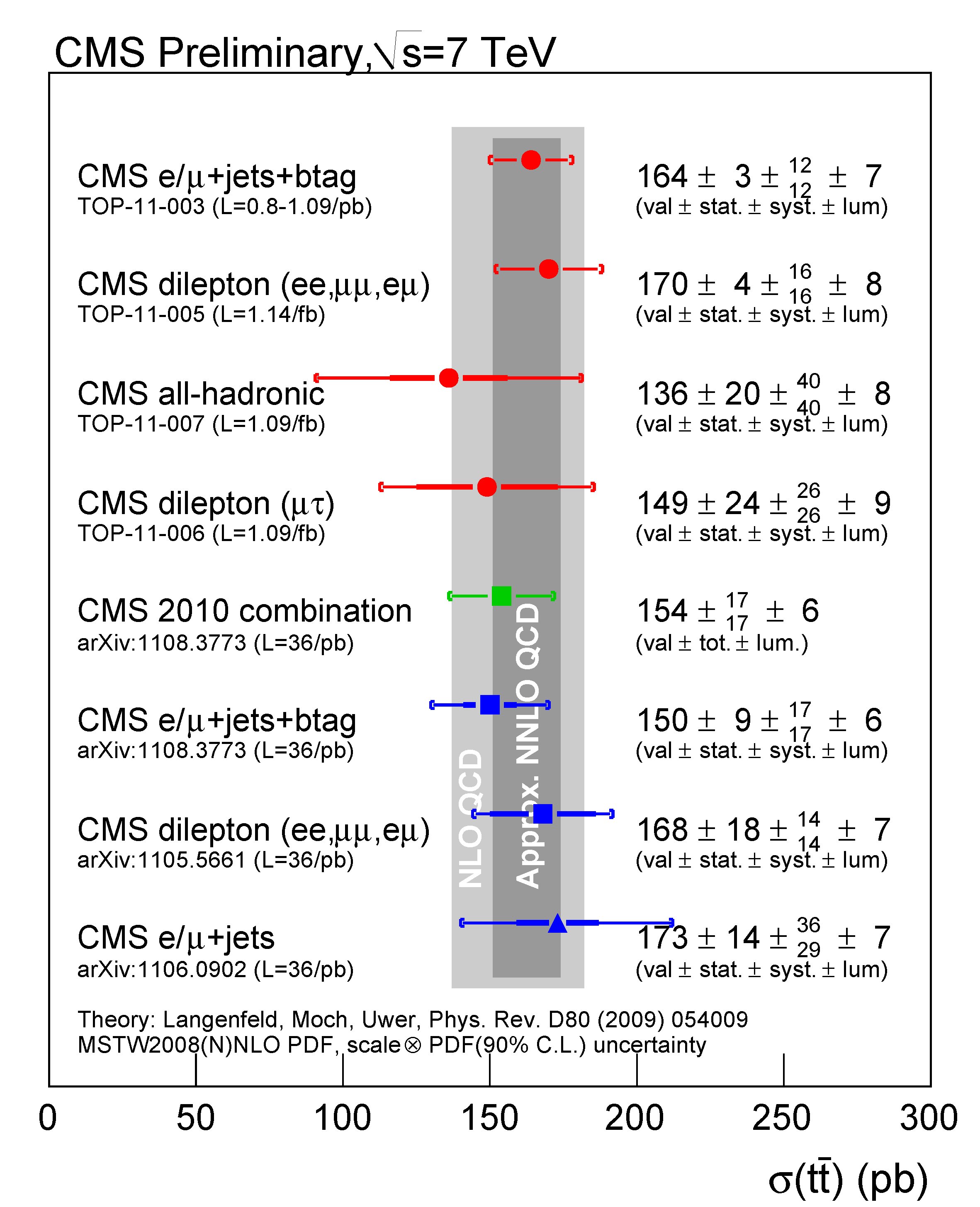}
\end{center}
\caption{\small \it
The measured value of $\sigma_{\ttbar}$ in the various decay modes and the combination of these measurements from the ATLAS (left) and CMS (right) experiments. The approximate
NNLO prediction with its uncertainty is also shown
(from Refs.~\protect\cite{ATLAS-CONF-2012-032,CMS:1108.3773}).
}
\label{f:sigma_tt}
\end{figure}

Single top quarks can be produced at the LHC via electroweak processes. The $t$-channel production
of single top-quarks has been measured by both the ATLAS~\cite{ATLAS-CONF-2011-101}
(L~=~0.7~\ifb)
and CMS~\cite{CMS:1106.3052} (L~=~1.5~\ifb) collaborations. The results are found to be consistent
with the Standard Model expectations.  The measured cross section value from the
CMS collaboration is shown in Fig.~\ref{f:top-mass}
in comparison to the theoretical expectation and the measurements at the Tevatron.

\subsubsection{Measurement of the top-quark mass}
Among the various top-quark properties, the ATLAS and CMS collaborations have already
presented first measurements on the top-quark mass $m_t$ in several final
states~\cite{ATLAS-CONF-2011-120, ATLAS-CONF-2012-030, CMS:PAS-TOP-11-015, CMS:PAS-TOP-11-016}.
The most precise result was presented recently as a preliminary result
by the CMS collaboration~\cite{CMS:PAS-TOP-11-015}. The top-quark mass has been measured
using a sample of \ttbar\ candidate events with one muon and at least four jets in the
final state. The full 2011 data set corresponding to an integrated luminosity of
4.7~\ifb\ was used. In this sample 2391 candidate events were selected and using
a likelihood method the top-quark mass was measured from fits to kinematic
distributions, simultaneously with the jet energy scale (JES). The result of
$m_t = 172.6 \pm 0.6 \ \rm{(stat + JES)} \ \pm 1.2 \ \rm{(syst)}$~\GeV\ is consistent
with the Tevatron result (see Fig.~\ref{f:top-mass} (right)).
It is impressive that such a precision, in particular on the systematic uncertainty,
can be claimed after only two years of operation of the LHC. The dominant contribution
to this systematic uncertainty results from uncertainties on the $b$-jet energy scale
and from modelling uncertainties estimated via changes of the factorization
scale~\cite{CMS:PAS-TOP-11-015}. However, it remains to be seen whether the small overall uncertainty can be
confirmed by the ATLAS experiment. The results of the present measurements at the
LHC are summarized in Fig.~\ref{f:top-mass} together with the Tevatron results.

\begin{figure}[hbtn]
\begin{center}
\includegraphics[width=0.49\textwidth,angle=0]{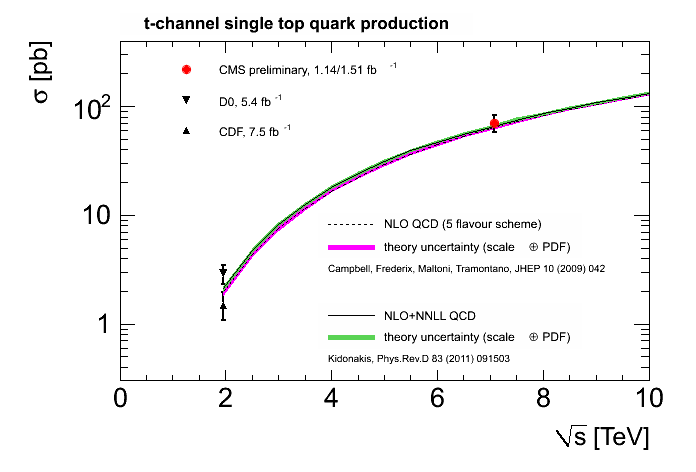}
\includegraphics[width=0.49\textwidth,angle=0]{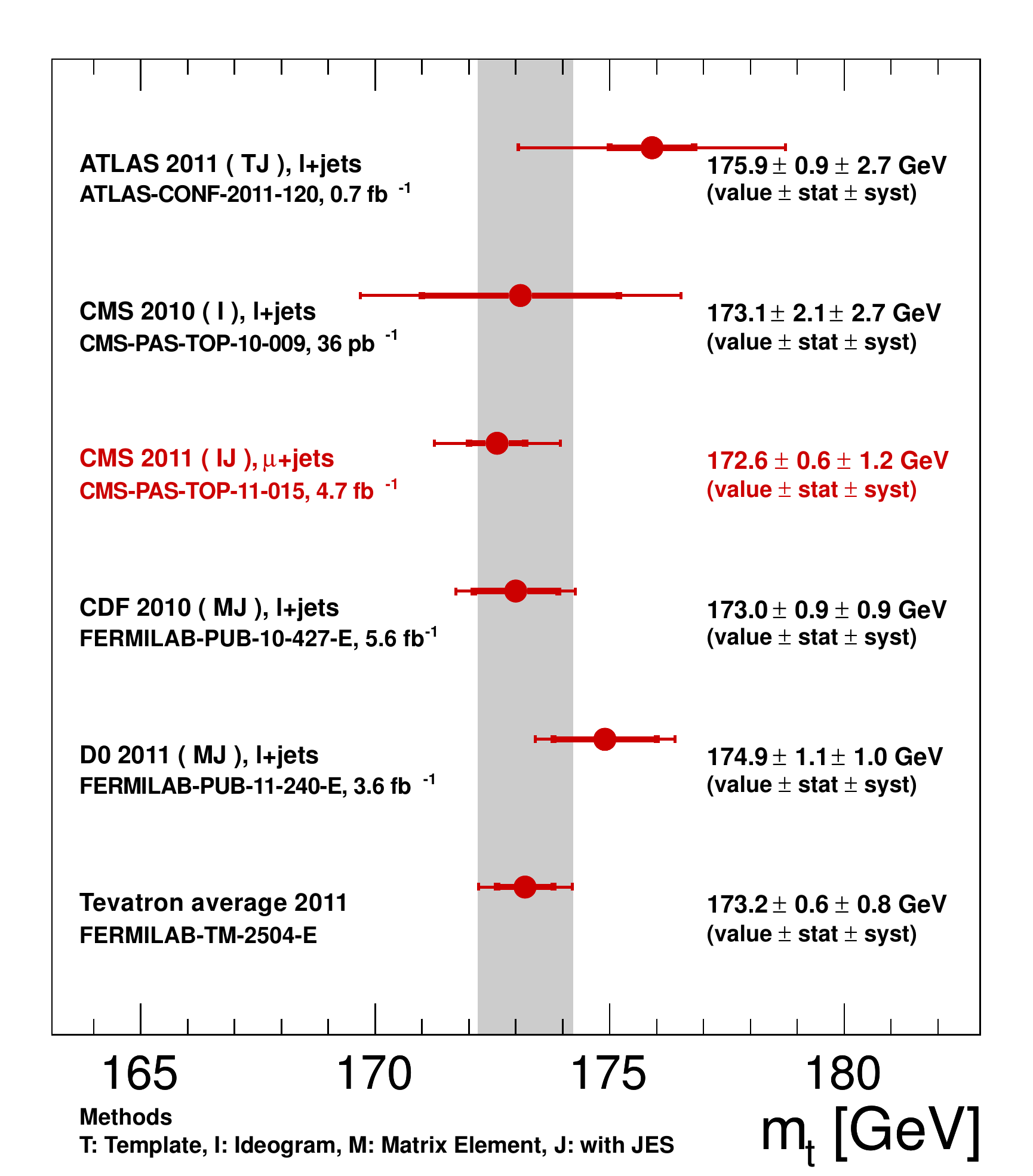}
\end{center}
\caption{\small \it
(Left): Measured cross section for single-top quark production via the $t$-channel process in the
CMS experiment in comparison to the theoretical expectation and the measurements at the Tevatron
(from Ref.~\protect\cite{CMS:1106.3052}).
(Right): Compilation of the top-quark mass measurements from the ATLAS, CDF, CMS and \dzero\
experiments (from Ref.~\protect\cite{CMS:PAS-TOP-11-015}).
}
\label{f:top-mass}
\end{figure}

\subsection{The production of diboson pairs}

The production of pairs of bosons ($W\gamma$, $WW$, $WZ$, $ZZ$) at the LHC is of great interest since it provides
an excellent opportunity to test the predictions on the structure of the gauge couplings of the electroweak sector
of the Standard Model at the \TeV\ energy scale. In addition,
$WW$ and $ZZ$ pairs constitute the irreducible background in important Higgs boson search
channels like $H \to WW$ and $H \to ZZ$.

The dominant Standard Model $W^+W^-$
production mechanisms are $s$-channel and $t$-channel quark-antiquark annihilation. The $s$-channel production
occurs only through the triple gauge coupling vertex and accounts for $\sim$10\% of the full $W^+W^-$
production cross section. The leading-order Feynman diagrams for the dominant $q \bar q' \to W^+W^-$ production
mechanisms at the LHC are shown in the left and middle diagrams of Fig.~\ref{f:FD_WW}.
The total cross section $ \sigma (q \bar q, q \bar q')  \to W^+W^-$ are known at next-to-leading order. The gluon
fusion through quark loops, shown in the right diagram of Fig.~\ref{f:FD_WW}, contributes an additional 2.9\%.

\begin{figure}[h]
\begin{center}
\includegraphics[width=0.6\textwidth,angle=0]{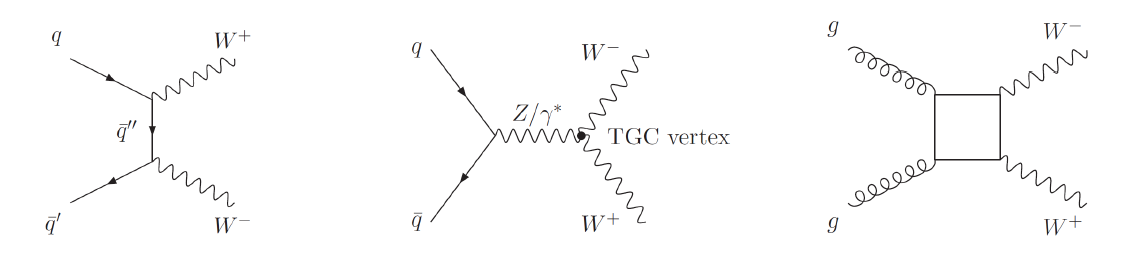}
\end{center}
\caption{\small \it
(Left):  The Standard Model tree-level Feynman diagram for $W^+W^-$ production through the $q \bar q$ initial state
in the $t$-channel. (Middle): The corresponding Standard Model tree-level diagram in the $s$-channel, which contains the
$WWZ$ and $WW\gamma$ triple gauge boson coupling (TGC) vertices. (Right): The gluon fusion process,
mediated by quark loops.
}
\label{f:FD_WW}
\end{figure}

The $ZZ$ production proceeds at leading order via $t$-channel
quark-antiquark interactions. The $ZZZ$ and $ZZ\gamma$ triple gauge boson couplings (nTGCs)
are absent. Hence there is no contribution from $s$-channel $q\bar q$ annihilation at
tree level.  Many models of physics beyond the Standard Model predict values of nTGCs at the level of
10$^{-4}$ to 10$^{-3}$ \cite{TGC_ZZ}.
The signature of nonzero nTGCs is an increase of the $ZZ$ cross section at high $ZZ$ invariant mass
and high transverse momentum of the $Z$ bosons \cite{TGC_Baur}.

The ATLAS and CMS collaborations have measured the cross sections for all diboson production processes,
$W\gamma$~\cite{CMS:1105.2758, ATLAS:1106.1592},
$WW$~\cite{ATLAS:1104.5225, CMS-PAS-EWK-11-010, ATLAS-CONF-2012-025},
$WZ$~\cite{ATLAS:1111.5570, CMS-PAS-EWK-11-010},
$ZZ$~\cite{ATLAS:1110.5016, CMS-PAS-EWK-11-010}.
Several analyses are already based on the full data set from 2011.
The results are found to be in good agreement with the predictions from the Standard Model and
first constraints on anomalous triple gauge boson couplings have been set.
The agreement between the measurements and the Standard Model predictions is shown
for the $WW$ and $ZZ$ production in Fig.~\ref{f:diboson}. The constraints on the triple
gauge boson couplings are still limited by the number of observed diboson events.

\begin{figure}[hbtn]
\begin{center}
\includegraphics[width=0.45\textwidth,angle=0]{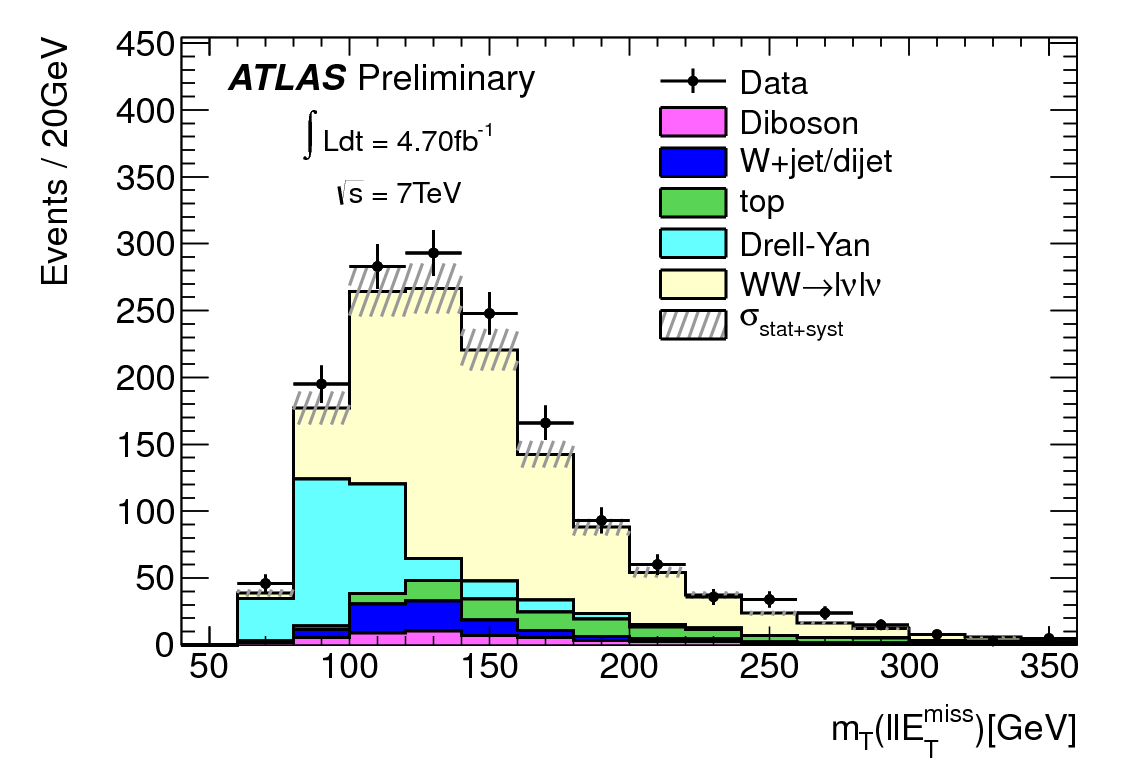}
\includegraphics[width=0.35\textwidth,angle=0]{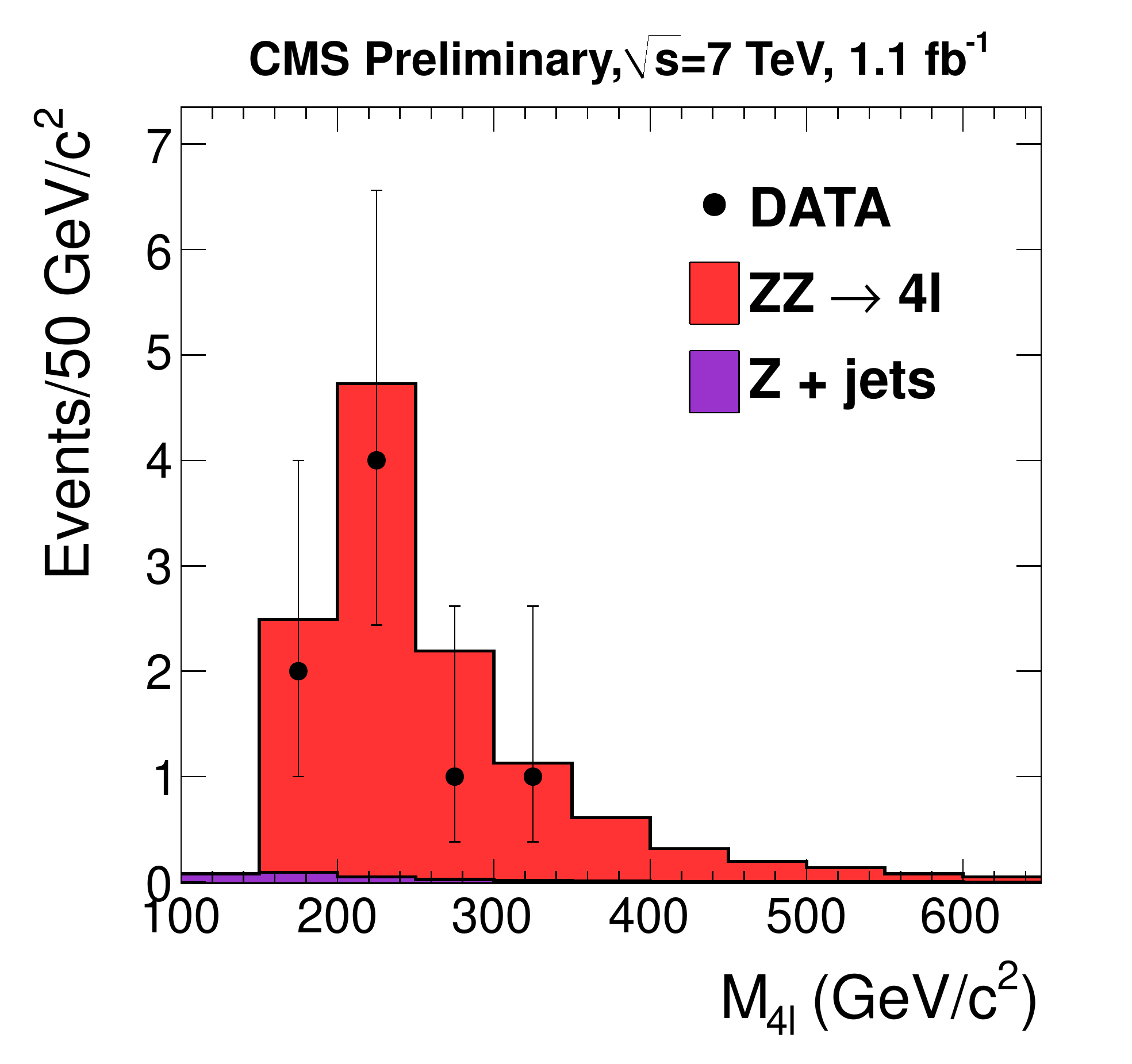}
\end{center}
\caption{\small \it
(Left):  The distributions of $m_T$ of the dilepton+$\ETmiss$ system for the $W^+W^-$ candidates in the ATLAS experiment
(from Ref.~\protect\cite{ATLAS-CONF-2012-025}).
(Right): The distribution of the four-lepton invariant mass for the $ZZ$ candidate events in the CMS experiment
(from Ref.~\protect\cite{CMS-PAS-EWK-11-010}).
}
\label{f:diboson}
\end{figure}

\subsection{Summary}
As discussed in the previous subchapters, the first two years have seen a very successful operation of the LHC
collider and of the experiments. The data collected have been used to extract precise measurements of many
Standard Model processes. They range from the measurement of $W$ and $Z$ production with cross sections in
the order of picobarns via the measurement of \ttbar\ production to the measurement of important diboson processes.
The summary of all measured cross sections by the ATLAS collaboration is shown in Fig.~\ref{f:SM_summary} together
with the theoretical predictions. Within the uncertainties, excellent agreement is found for all
the processes considered. This is a remarkable achievement of the Standard Model and the underlying theoretical
concepts, including QCD and factorization.
The smallest cross sections measured, the diboson production cross sections, are extremely relevant for the
Higgs boson search, as discussed in the next section.

\clearpage

\begin{figure}[hbtn]
\begin{center}
\includegraphics[width=0.80\textwidth,angle=0]{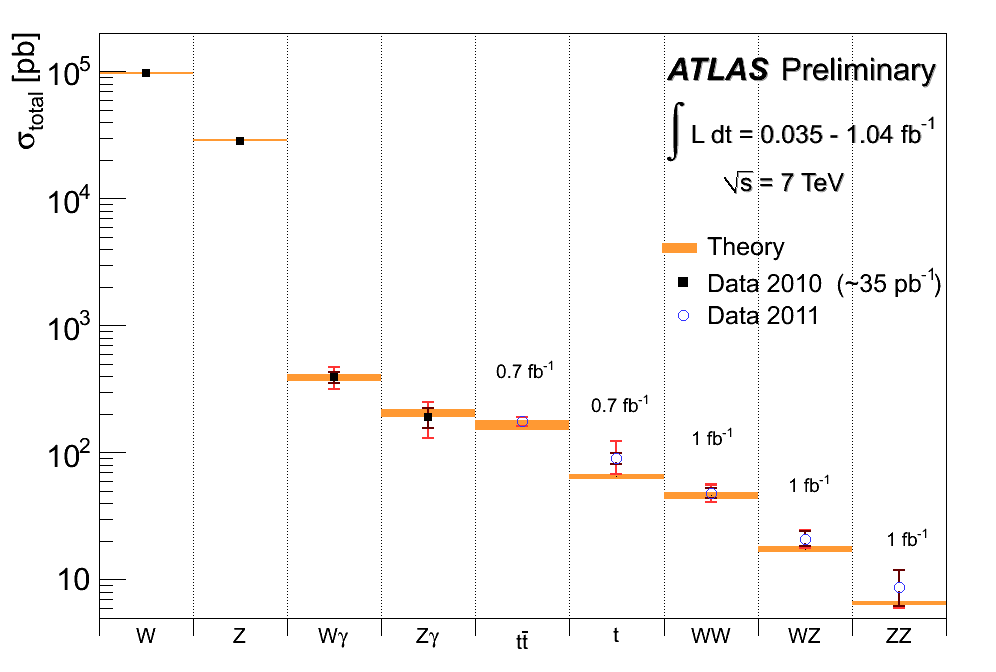}
\end{center}
\caption{\small \it
Summary of several Standard Model total production cross-section measurements from the ATLAS collaboration compared to the
corresponding theoretical expectations. The integrated luminosities used for the measurements are indicated on the figure.
The dark error bars represent the statistical uncertainties.
The red error bars represent the
full uncertainties, including systematics and luminosity uncertainties. All
theoretical expectations were calculated at NLO or higher.
}
\label{f:SM_summary}
\end{figure}

\section{Search for the Higgs boson}

The Higgs boson is the only Standard Model particle that has not been discovered so far.
Indirectly, high precision electroweak data constrain its mass via their
sensitivity to radiative corrections. Assuming the overall validity of the Standard Model,
a global fit \cite{LEP-EWWG} to all electroweak data leads to
$m_H = 94 ^{+29}_{-24}$~\GeV. On the basis of the present theoretical knowledge, the Higgs sector
in the Standard Model remains largely unconstrained. While there is no direct prediction for the
mass of the Higgs boson, an upper limit of $\sim$1~\TeV\ can be inferred from unitarity
arguments \cite{higgs-unitarity,*higgs-unitarity1,*higgs-unitarity2}.

Direct searches at the \epem\ collider LEP has led to a lower bound on its mass of
114.4~\GeV~\cite{LEP-higgs-limit}. Before the LHC started its operation, the Fermilab
Tevatron \ppbar\ collider with a centre-of-mass energy of 1.96~\TeV\ was the
leading accelerator exploring the energy frontier. Until the end of data
taking in September 2011, the two
experiments CDF and \dzero\ have collected data corresponding to an integrated luminosity
of 11.9~\ifb.
During the past years, they were able to exclude Higgs boson masses
around 160~\GeV, mainly based on the search for
the \hwwsll\ decay mode. In Summer 2011, the combination
of the results from the two experiments, based on data corresponding to
an integrated luminosity of 8.6~\ifb, excluded a mass range from 156 - 177~\GeV\
\cite{Tevatron:1107.5518}.
At the same time, the first exclusions from the ATLAS and CMS experiments, based on a data
corresponding to an integrated luminosity of up to 2.3~\ifb, were presented. The ATLAS
experiment excluded mass ranges from 146 - 230~\GeV, 256-282~\GeV\ and 296 - 459~\GeV\
\cite{ATLAS-CONF-2011-135}. The CMS analysis was based on data
corresponding to an integrated luminosity of up to 1.7~\ifb\ and the Higgs boson was
excluded in the mass ranges from 145 - 216~\GeV, 226 - 288~\GeV\ and 310 - 400~\GeV
\cite{CMS-PAS-HIG-11-022}.

Since then the full data set of the LHC taken until the end of 2011 has been analyzed.
Preliminary results were presented in a special seminar at CERN in December 2011 and
updates were presented at the Moriond conference 2012.
They created a lot of attention and excitement
in the community since the data show tantalizing hints of a possible Higgs boson
signal in the low mass region around 125~\GeV. However, it must be stressed that the
background-only probability still shows acceptable values, in particular if the
look-elsewhere effect \cite{LEE} is taken into account.

In the following, these results are summarized since they
supersede those presented at the CERN school in September 2011.
Before entering the discussion, the Higgs boson production at hadron colliders and
the Higgs boson decay properties as well as a few statistical issues are briefly summarized.

\subsection{Higgs boson production at the LHC \label{s:higgs_prod}}
At hadron colliders, Higgs bosons can be produced via four different production
mechanisms:

\begin{itemize}
\item gluon fusion, $gg \to H$, which is mediated at lowest order by a heavy
quark loop;
\item vector boson fusion (VBF), $ qq \to qq H $;
\item associated production of a Higgs boson with weak gauge bosons, \\
$qq \to W/Z \ H$
(Higgs strahlung, Drell-Yan like production);
\item associated Higgs boson production with heavy quarks, \\
$gg, qq \to \ttbar H$, $gg, qq \to bb H$  (and $gb \to b H$).
\end{itemize}

The dominant production mode is the gluon-fusion process, followed by the vector boson fusion.
In the low mass region it amounts at leading order to about 20\% of the
gluon-fusion cross section, whereas it reaches the same level for masses around 800~\GeV.
The associated $WH$, $ZH$ and $\ttbar H$ production
processes are relevant only for the search of
a light Standard Model Higgs boson with a mass close to the LEP limit.

The Higgs boson production cross sections are computed up to next-to-next-to-leading order
(NNLO) \cite{a1a2GGF} in
QCD for the gluon-fusion process. In addition, QCD soft-gluon resummations up to next-to-next-to-leading log
(NNLL) improve the NNLO calculation \cite{Catani:2003zt}. The next-to-leading order (NLO) electroweak corrections  \cite{Aglietti:2004nj,*AActis:2008ug} are also applied,
assuming factorization between QCD and electroweak corrections.
The cross sections for the VBF process are calculated with full NLO QCD and
electroweak corrections \cite{VBFNLO}, and approximate NNLO
QCD corrections are available \cite{Bolzoni}. The $W/Z\ H$ processes are calculated
at NLO \cite{Han:1991ia} and at NNLO \cite{Brein:2003wg}, and NLO
electroweak radiative corrections \cite{Ciccolini:2003jy} are applied.
Also for the associated $\ttbar H$ production the full NLO QCD corrections are
available \cite{TTHALLNEW}.
For a more detailed review of the theoretical aspects of Higgs boson production
the reader is referred to Ref.~\cite{higgs-xsections}.
The results of the calculations and the estimated theoretical uncertainties are
shown for the different production processes in Fig.~\ref{f:prod_xs} (left)
as a function of the Higgs boson mass \cite{higgs-xsections}.

\begin{figure}
\begin{center}
\includegraphics[width=0.49\textwidth,angle=0]{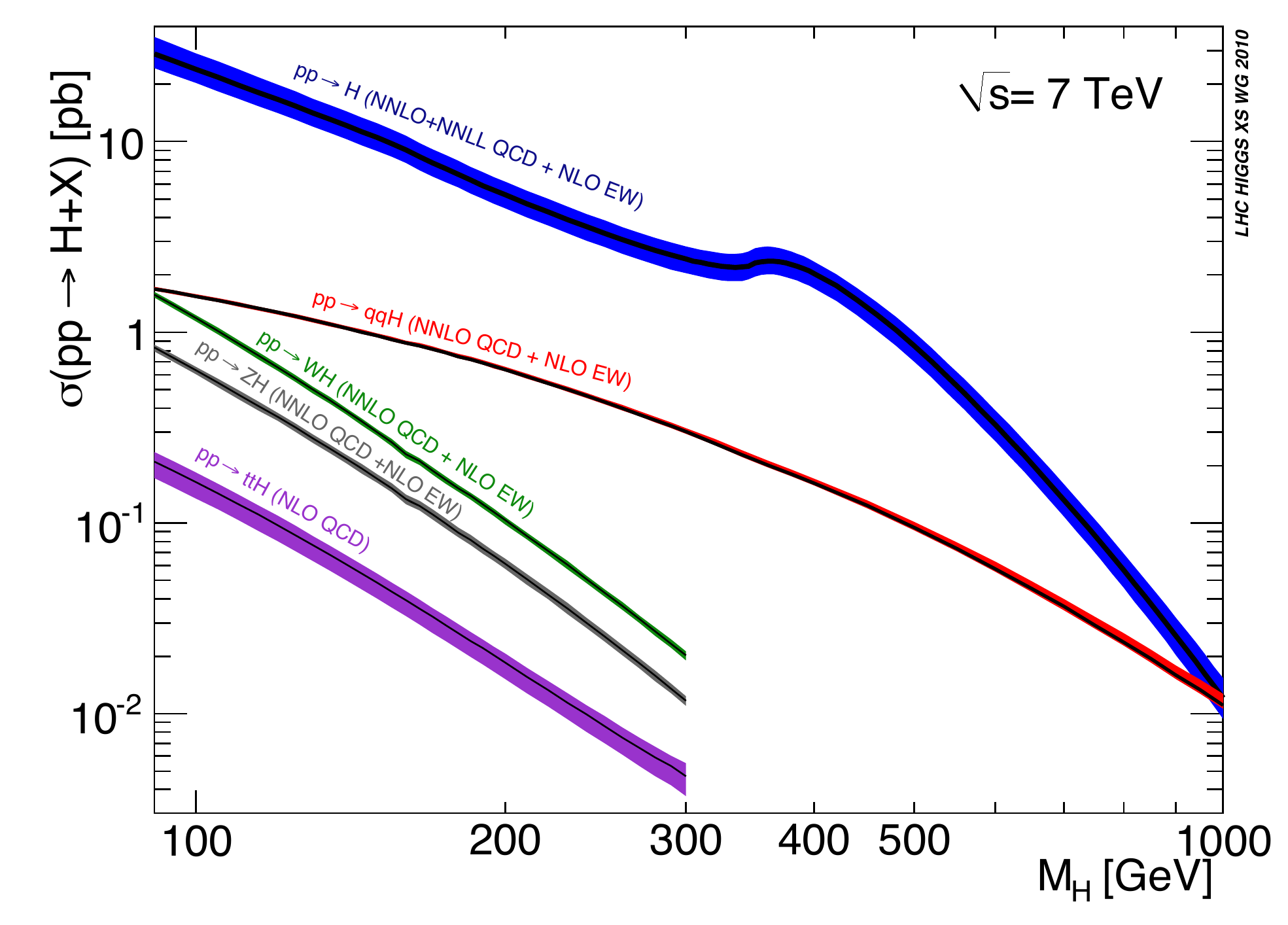}
\includegraphics[width=0.49\textwidth,angle=0]{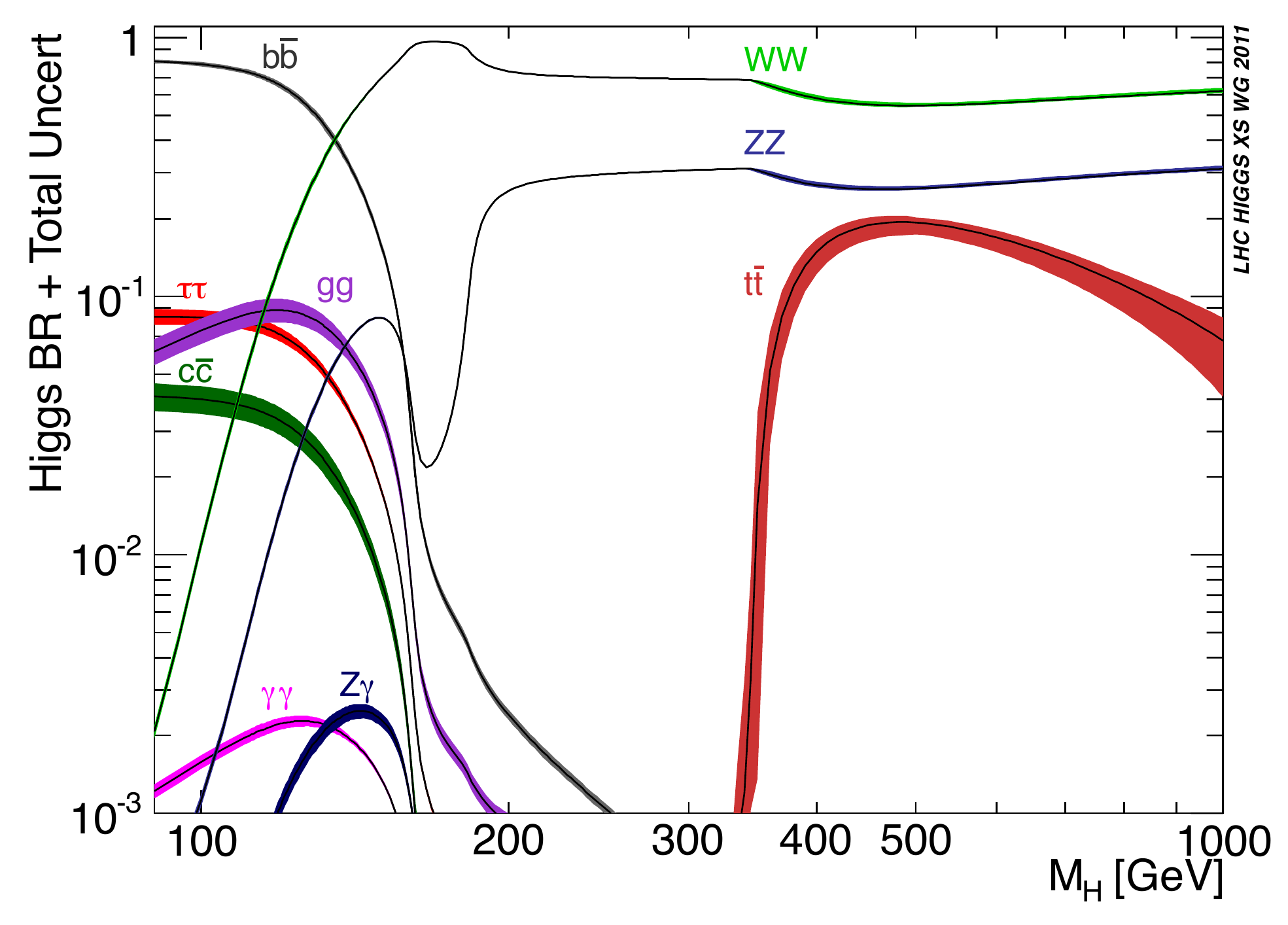}
\end{center}
\caption{\small \it
(Left): Production cross sections for the different production processes
for a Standard Model Higgs boson as a function of
the Higgs boson mass at the LHC.
(Right): Branching ratios of the Standard Model Higgs
boson as a function of Higgs boson mass (from Ref.~\protect\cite{higgs-xsections}).
}
\label{f:prod_xs}
\end{figure}

\subsection{Higgs boson decays}

The branching ratios and the total decay width of the Standard Model Higgs boson
are shown in Fig.~\ref{f:prod_xs} (right) as a function of the Higgs boson mass.
They have been calculated taking into account both electroweak and
QCD corrections \cite{Djouadi:1997yw,spira-decays}.
When kinematically accessible, decays of the Standard Model Higgs boson into
vector boson pairs $WW$ or $ZZ$ dominate over all other decay modes.
Above the kinematic threshold, the branching fraction into \ttbar\ can
reach up to 20\%.
All other fermionic decays are only relevant for
Higgs boson masses below $\sim$140~\GeV, with \Hbb\ dominating.
The branching ratios for both \Htau\ and $H \to  gg$ reach up
to about 8\% at Higgs boson masses between 100 and
120~\GeV. Decays into two photons, which are of interest due to their
relatively clean experimental signature, can proceed via charged fermion and $W$ loops
with a branching ratio of up to 2~$\cdot$~10$^{-3}$ at low Higgs
boson masses.

Compared to the mass resolution of hadron collider experiments, the total decay
width of the Standard Model Higgs boson is small at low masses and
becomes significant only above the threshold for decays into $ZZ$.
For a Higgs boson with a mass of $\sim$1~\TeV\ the resonance is broad
with a width of about 600~\GeV.
In this mass regime, the Higgs field is coupling strongly,
resulting in large corrections \cite{higgs-xsections, binoth-gamma-h}.

\subsection{Search for the Standard Model Higgs boson at the LHC} \label{s:sm-higgs}

The Standard Model Higgs boson is searched for at the LHC in various decay
channels, the choice of which is given by the signal rates and the
signal-to-background ratios in the different mass regions.
The search strategies and background rejection methods have been established
in many studies over the past years \cite{physics-tdr,*physics-tdr-cms,*atlas-csc}.
Among the most important channels are the inclusive \Hgg\ and \Hllll\
decay channels. These channels are characterized by a very good mass resolution.
In the low mass region, the Higgs boson appears as a sharp resonance, the width of which is dominated by the
detector resolution, on top of flat backgrounds which are dominated by \gamgam\
and $ZZ$ continuum production, respectively.
In addition, the \HWW\ decay mode contributes in particular in the
mass region around 160~\GeV, however, due to the neutrinos in the final state, no mass peak can
be reconstructed. Evidence for Higgs boson production is given by a broad peak in
the transverse mass distribution (see below).
From the fermionic decays, only the modes \Htau\ and \Hbb\ have a chance to be detected.
For the \Htau\ decays the selection of the vector boson fusion production mode is important
to improve the signal-to-background ratio by exploiting forward
jet tagging \cite{vbf_zeppenfeld}. The \bbbar\ decay mode is searched for in the associated
production of the Higgs boson with a vector boson or with a \ttbar\ pair
\cite{butterworth_WH_boosted, ATLAS_WH_boosted}.

In the following the present status (March 2012) of the Higgs boson search in the various
final states at the LHC is described. Before the individual channels are discussed some
common issues on the statistical treatment are given. At the end of this section
the combination of the results is presented for both the ATLAS and CMS collaborations.

\subsubsection{Limit setting, statistical issues}

In the following the distributions of reconstructed masses or other distributions
as measured in data are compared to the expectations from Standard Model
background processes. In order to test the compatibility of the data with
the background-only hypothesis a so-called $p_0$ probability value is calculated.
It quantifies the probability that a background-only experiment is more
signal-like than that observed. The  {\em local} $p_0$ probability is
assessed for a fixed \mH\ hypothesis and the equivalent formulation in
terms of number of standard deviations is referred to as the {\em local}
significance.  Since fluctuations of the background could occur at any point
in the mass range the results have to be corrected for the
look-elsewhere effect~\cite{LEE}. The probability for a
background-only experiment to produce a local significance of this size or
larger anywhere in a given mass region is referred to as the {\em global} $p_0$.
The corresponding reduction in the significance is estimated using the prescription
described in Ref.~\cite{HCG}.

The data can also be used to set 95\% confidence level (C.L.) upper limits ($\sigma_{95}$)
on the cross section for Higgs boson production. These cross sections are usually
normalized to the expected Standard Model value ($\sigma_{95}/ \sigma_{\rm{SM}}$).
All exclusion limits quoted in the following have been calculated using the
$CL_s$ method~\cite{Read:2002hq}. In addition to the {\em observed} limits based on the
observed data, also the {\em expected} limits are calculated. They are calculated
as a function of $m_H$ and are based on
the central values of the expected background in case no Higgs signal is present.
The 1$\sigma$ and 2$\sigma$ fluctuations around the expected exclusion limits are
calculated as well.

Excluded mass regions are determined from a comparison of the observed cross-section limit
to the Standard Model cross-section value. If the observed value of $\sigma_{95}/ \sigma_{\rm{SM}}$
is smaller than 1 (Standard Model cross-section expectation) for a particular hypothetical
Higgs boson mass, this mass value can be excluded with a confidence level of 95\%.
Systematic uncertainties are incorporated by introducing nuisance parameters
with constraints. Asymptotic formulae~\cite{asym} are used to derive the
limits and $p_0$ values. This procedure has been validated
using ensemble tests and a Bayesian calculation of the exclusion
limits with a uniform prior on the signal cross section. These
approaches to the limits typically agree with the asymptotic
median results to within a few percent~\cite{combinationLetter}.

\subsubsection{Search for \Hgg\ decays}
The decay $\Hgg$ is a rare decay mode, which is only detectable in a limited
Higgs boson mass region between 100 and 150~\GeV,
where both the production cross section and the decay
branching ratio are relatively large.
Excellent energy and angular resolution are required to observe the
narrow mass peak above the irreducible prompt $\gamgam$ continuum.
In addition, there is a reducible background resulting from direct photon
production or from two-jet production via QCD processes. These processes contribute if one
or both jets are misidentified as a photon.
The background can be determined from a
fit to the data (sidebands) such that uncertainties from Monte Carlo predictions or uncertainties
due to normalizations in control regions are avoided. Due to the rather large amount of
material in the tracking detectors of the LHC experiments there is a high probability for a
photon to undergo conversion and therefore both unconverted and converted photons need
to be reconstructed.

\begin{figure}[hnt]
\begin{center}
\includegraphics[width=.48\textwidth]{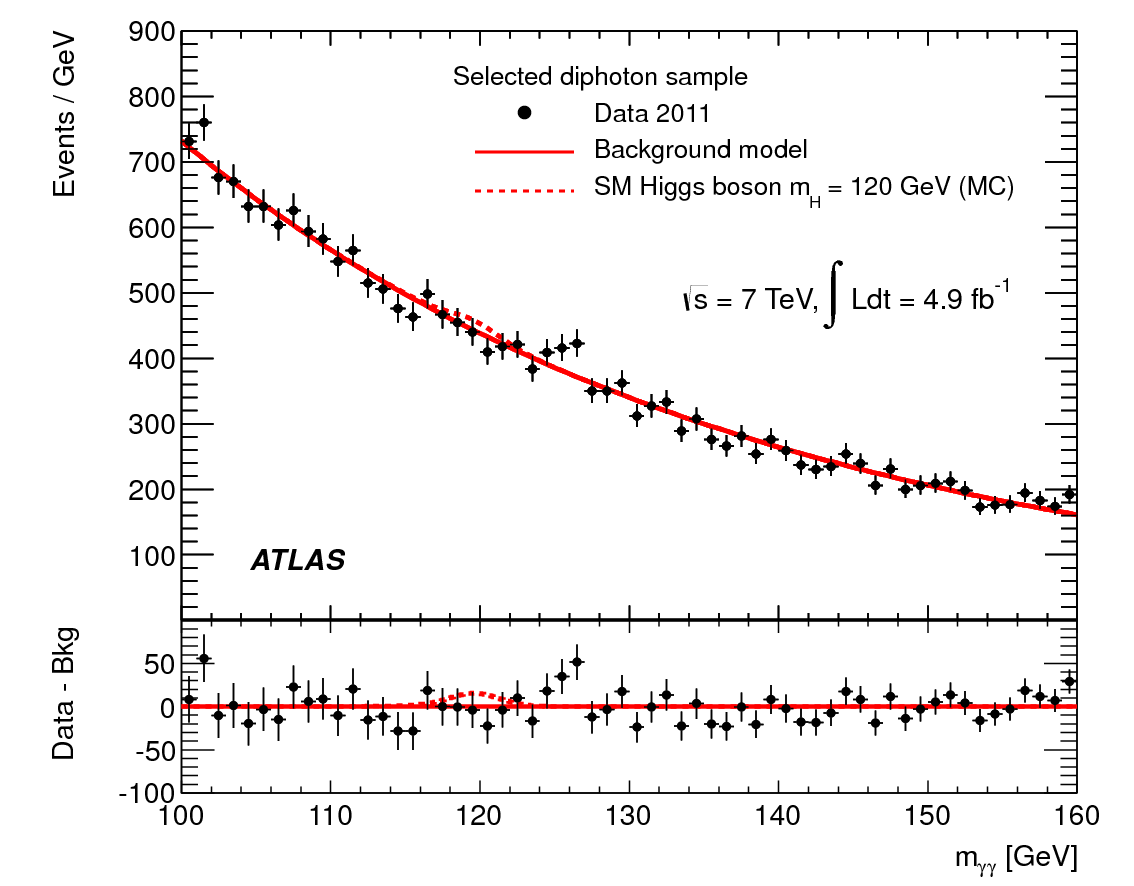}
\includegraphics[width=.50\textwidth]{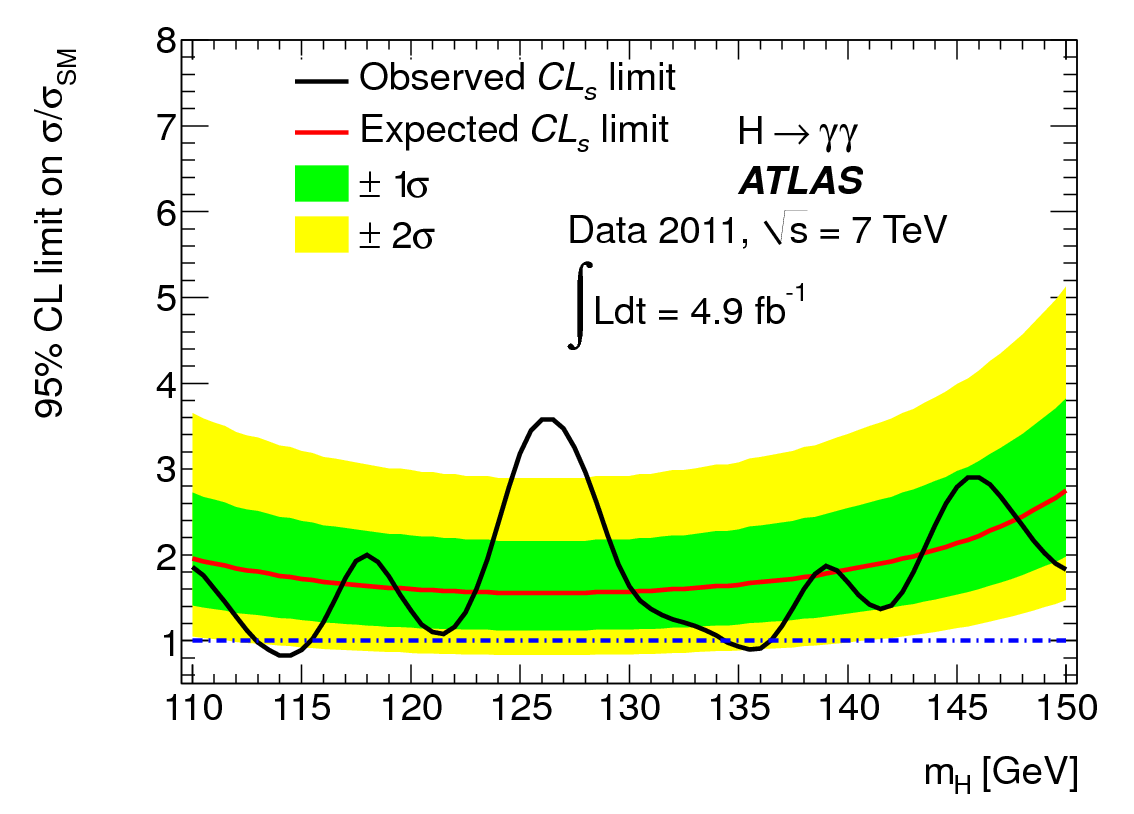}
\end{center}
\caption{\small \it
(Left): Invariant mass distribution for the selected data sample in the ATLAS experiment, overlaid with
the total background (see text). The bottom inset displays the residual of the data with respect to the total background.
The Higgs boson expectation for a mass hypothesis of 120~\GeV\ corresponding to the Standard
Model cross section is also shown.
(Right): Observed and expected 95\% C.L. limits on the Standard Model Higgs boson production cross section
normalized to the predicted one as a function of $m_H$
(from Ref.~\protect\cite{:2012sk}).
}
\label{f:H_gg_mass_ATLAS}
\end{figure}

Both collaborations have presented results on the \Hgg\ search in a mass range between
110 and 150~\GeV\ based on the full data set
collected until the end of
2011 \cite{:2012sk, Chatrchyan:1422388}.
The ATLAS analysis \cite{:2012sk} separates events into nine independent
categories. The categorisation is based on the direction of each
photon and whether it was reconstructed as a converted or unconverted photon, together with
the momentum component of the diphoton system transverse to the thrust axis.
The distribution of the invariant mass of the diphoton
events, $m_{\gamma\gamma}$, summed over all categories, is shown in
Fig.~\ref{f:H_gg_mass_ATLAS} (left).
The fit to the background is performed separately in each category in the mass range
100 - 160~\GeV\
by using an exponential function with free slope and normalization parameters. The result
for the total sample is superimposed in Fig.~\ref{f:H_gg_mass_ATLAS}.
The signal expectation for a Higgs boson with $m_H=120$~\GeV\  is also shown.
The mass resolution depends on the classification of the photon
(calorimeter $\eta$ region, conversion
status) and is found to be 1.4~\GeV\ in the best category and 1.7~\GeV\ on average.
The residuals of the data with respect to the
total background as a function of $m_{\gamma\gamma}$ is also shown in
Fig.~\ref{f:H_gg_mass_ATLAS}.
Around a mass of 126~\GeV\ an excess of events above the background is seen (see discussion
below). The 95\% C.L. upper limits on the cross section for Higgs boson
production normalized to the Standard Model value, $\sigma_{95} / \sigma_{SM}$,
are shown in Fig.~\ref{f:H_gg_mass_ATLAS} (right).
The observed exclusion limits follow well the
expectations over a large mass range, except in the region around 126~\GeV.
The ATLAS data allow for a 95\% C.L. exclusion of a Standard Model Higgs boson in the
mass ranges between
$113{\--}115$~\GeV\ and $134.5{\--}136$~\GeV.
\begin{figure}[hbtn]
\begin{center}
\includegraphics[width=.30\textwidth]{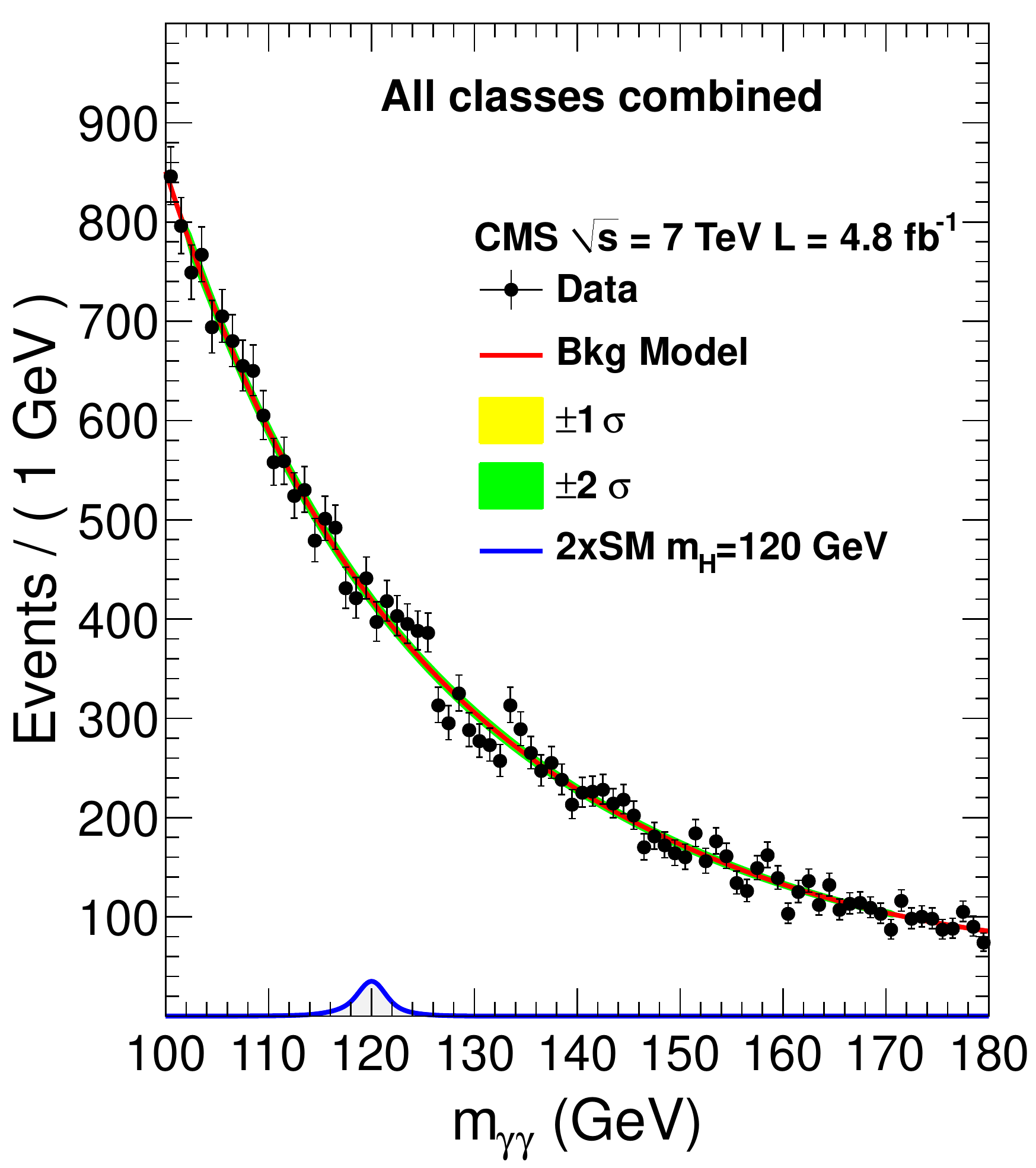}
\includegraphics[width=.30\textwidth]{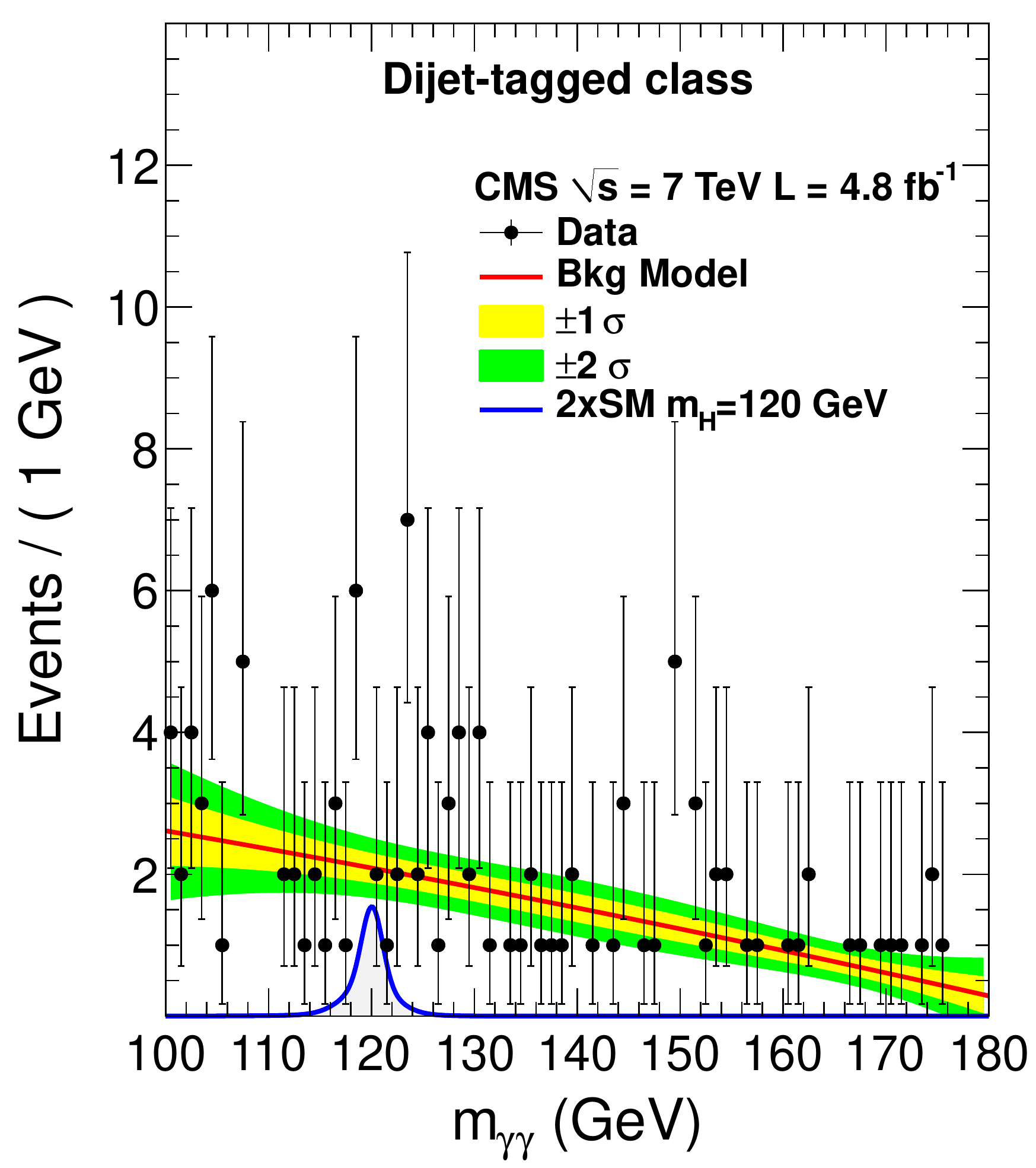}
\includegraphics[width=.38\textwidth]{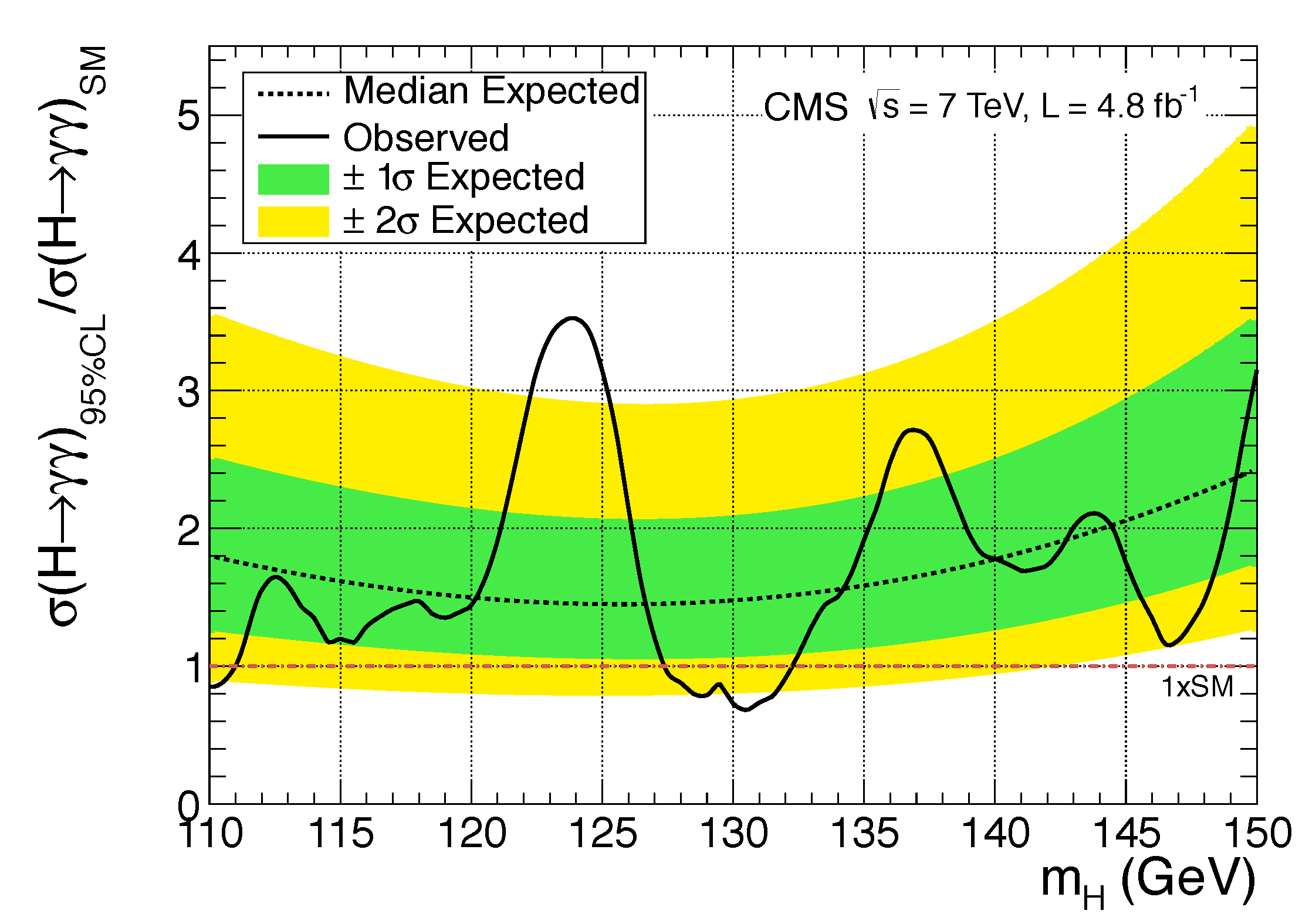}
\end{center}
\caption{\small \it
(Left): Invariant mass distribution for the selected data sample in the CMS experiment,
overlaid with the total background.
(Middle): The invariant mass distribution for diphotons fulfilling the VBF selection (see text).
The Higgs boson expectation for a mass hypothesis of 120~\GeV\ corresponding to the Standard
Model cross section multiplied by a factor of two is also shown.
 (Right): Observed and expected 95\% C.L. limits on the Standard Model Higgs boson
production cross section normalized to the predicted one as a function of $m_H$
(from Ref.~\protect\cite{Chatrchyan:1422388}).
}
\label{f:higgs_gg_mass_CMS}
\end{figure}

The analysis of the CMS collaboration \cite{Chatrchyan:1422388} is done in a similar way.
Diphoton events are split into four categories depending on their $\eta$ value and shower
shape characteristics (to distinguish converted
from unconverted photons). The diphoton mass resolution is best for the class of two central unconverted
photons and reaches a value of 1.2~\GeV\ (full width at half maximum divided by 2.35)
for a Higgs boson mass of 120~\GeV. Including the other classes a weighted average resolution of
$\sim$1.8~\GeV\ is found.
A further class of events is introduced to select the vector boson fusion topology
(VBF topology).
By requiring two jets with a large separation in pseudorapidity,
a class of events is defined for which the expected signal-to-background
ratio is about an order of magnitude larger than for the events in the four
classes defined by photon properties.
The $m_{\gamma\gamma}$  distributions observed in the data for the sum of the five
event classes and for the VBF topology separately are shown in
Fig.~\ref{f:higgs_gg_mass_CMS} (left, middle)
together with the background fits based on polynomial functions.
The uncertainty bands shown are computed from the fit uncertainty on
the background yields.
The limit set on the cross section of a Higgs boson decaying to two photons
normalized to the Standard Model value is shown in Fig.~\ref{f:higgs_gg_mass_CMS} (right).
The CMS analysis excludes at the 95\% C.L.
the Standard Model Higgs boson decaying into two photons in the mass range 128 to 132~\GeV.
However, it should be noted that this exclusion, as well as the ATLAS exclusions in this channel,
are {\em lucky} since the expected sensitivities are
larger than one and the observed values of $\sigma_{95}/ \sigma_{\rm{SM}}$
are at the edges of the 2$\sigma$ bands.
The fluctuations of the observed limit about the expected limit are consistent with statistical
fluctuations to be expected in scanning the mass range. The largest deviation in the CMS
experiment is seen at $m_{\gamma\gamma}$ =124~\GeV.

In order to quantify the fluctuations seen in both experiments, the probabilities for  the
background-only hypothesis have been calculated.
The observed and expected local $p_0$ values obtained are displayed in
Fig.~\ref{f:higgs_p0} for the ATLAS (left) and CMS (right) data.
Before considering the uncertainty on the signal mass position, the largest excess
with respect to the background-only hypothesis in the mass range $110{\--}150$~\GeV\
is observed at 126.5~\GeV\ in the ATLAS data with a local significance of 2.9$\sigma$.
The uncertainty on the mass position ($\pm0.7$~\GeV) due to the imperfect knowledge
of the photon energy scale has a
small effect on the significance. When this uncertainty is taken into account,
the significance is slightly reduced to 2.8$\sigma$.
The local $p_0$ value corresponding to the largest upwards fluctuation in the CMS data
at 124~\GeV\ has a significance of 3.1$\sigma$.
The observed significances reduce to 1.5$\sigma$ for ATLAS and 1.8$\sigma$ for CMS,
when the look-elsewhere effect is taken into account over the mass range $110{\--}150$~\GeV.

\begin{figure}[hbtn]
\begin{center}
\includegraphics[width=.51\textwidth]{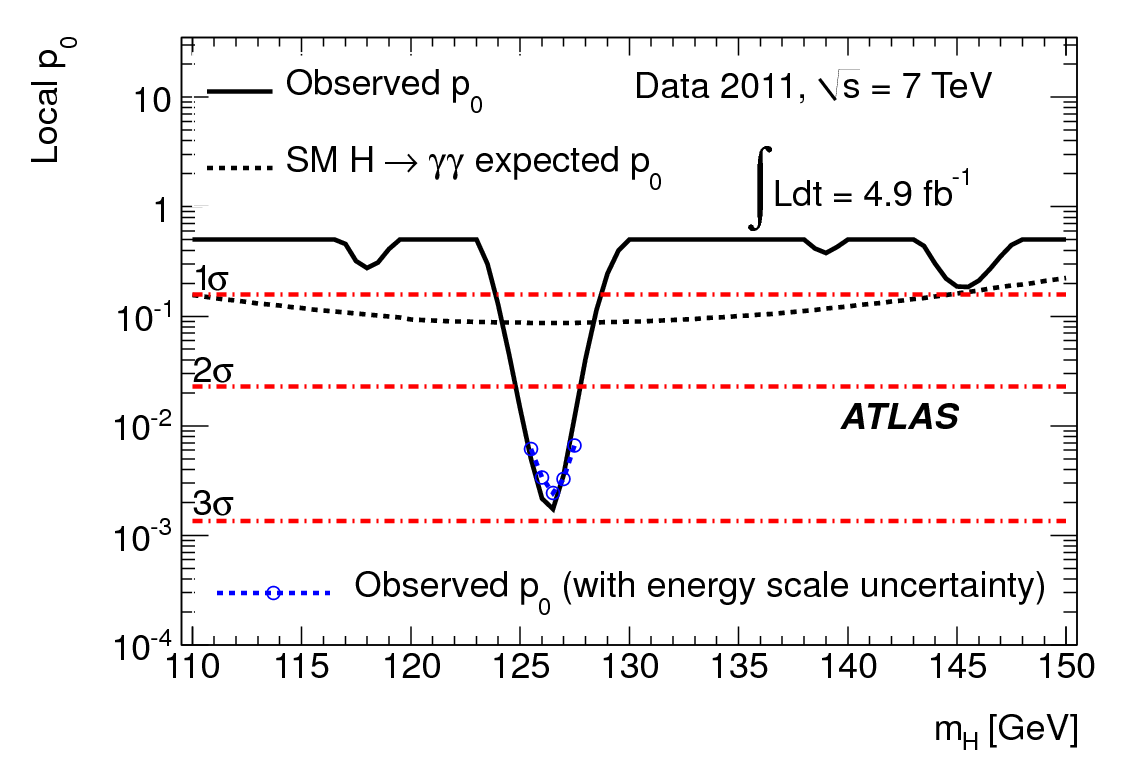}
\includegraphics[width=.47\textwidth]{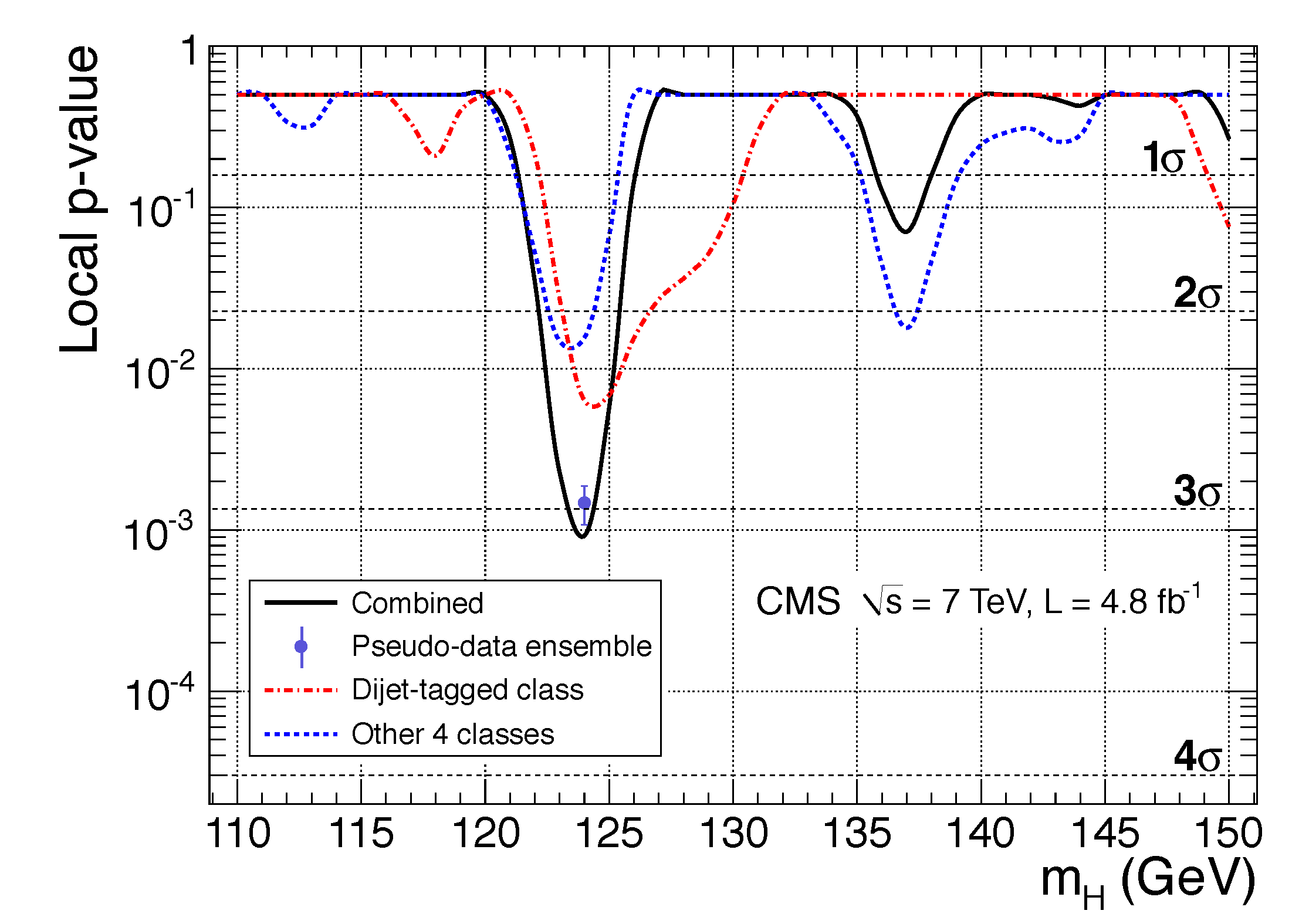}
\end{center}
\caption{\small \it
The observed local $p_0$, the probability that the background fluctuates to the observed
number of events or higher, for the ATLAS (left) and CMS (right) data. In the ATLAS case, the open points indicate
the observed local $p_0$ value when energy scale uncertainties are taken into
account. The dotted line shows the expected median local $p_0 \ $
for the signal hypothesis
when tested at $m_H$. In the CMS case, the $p_0$ values are shown for the VBF-tagged class separately
(from Refs.~\protect\cite{:2012sk, Chatrchyan:1422388}).
}
\label{f:higgs_p0}
\end{figure}

\subsubsection{Search for \Hllll\ decays}

The decay channel $H \to Z Z^{(*)} \to \ell \ell \ \ell \ell$ provides a rather
clean signature in the mass range 115~\GeV\ $< m_H < 2 \ m_Z$.
In addition to the irreducible backgrounds from $ZZ^*$ and $Z \gamma^*$ production,
there are large reducible backgrounds from \ttbar\ and $Z \bbbar$ production.
Due to the large production cross section, the \ttbar\ background
dominates at production level, whereas the $Z \bbbar$ events contain a
genuine $Z$ boson in the final state and are therefore more difficult to reject.
In addition, there is background
from $ZZ$ continuum production,
where one of the $Z$ bosons decays into a
$\tau$ pair, with subsequent leptonic decays of the $\tau$ leptons, and the other
$Z$ decays into an electron or muon pair.

\begin{figure}[hbtn]
\begin{center}
\includegraphics[width=.35\textwidth]{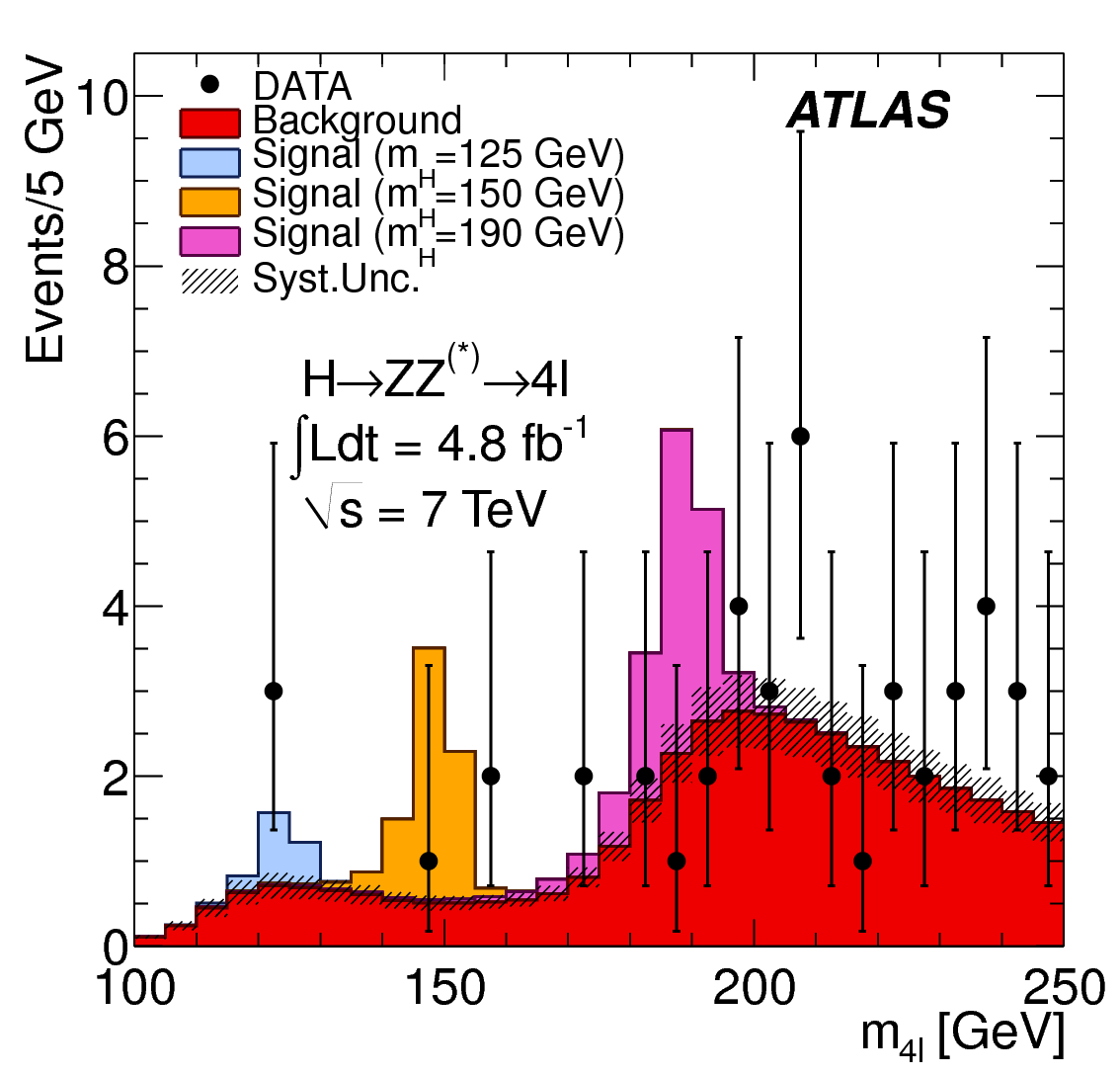}
\includegraphics[width=.35\textwidth]{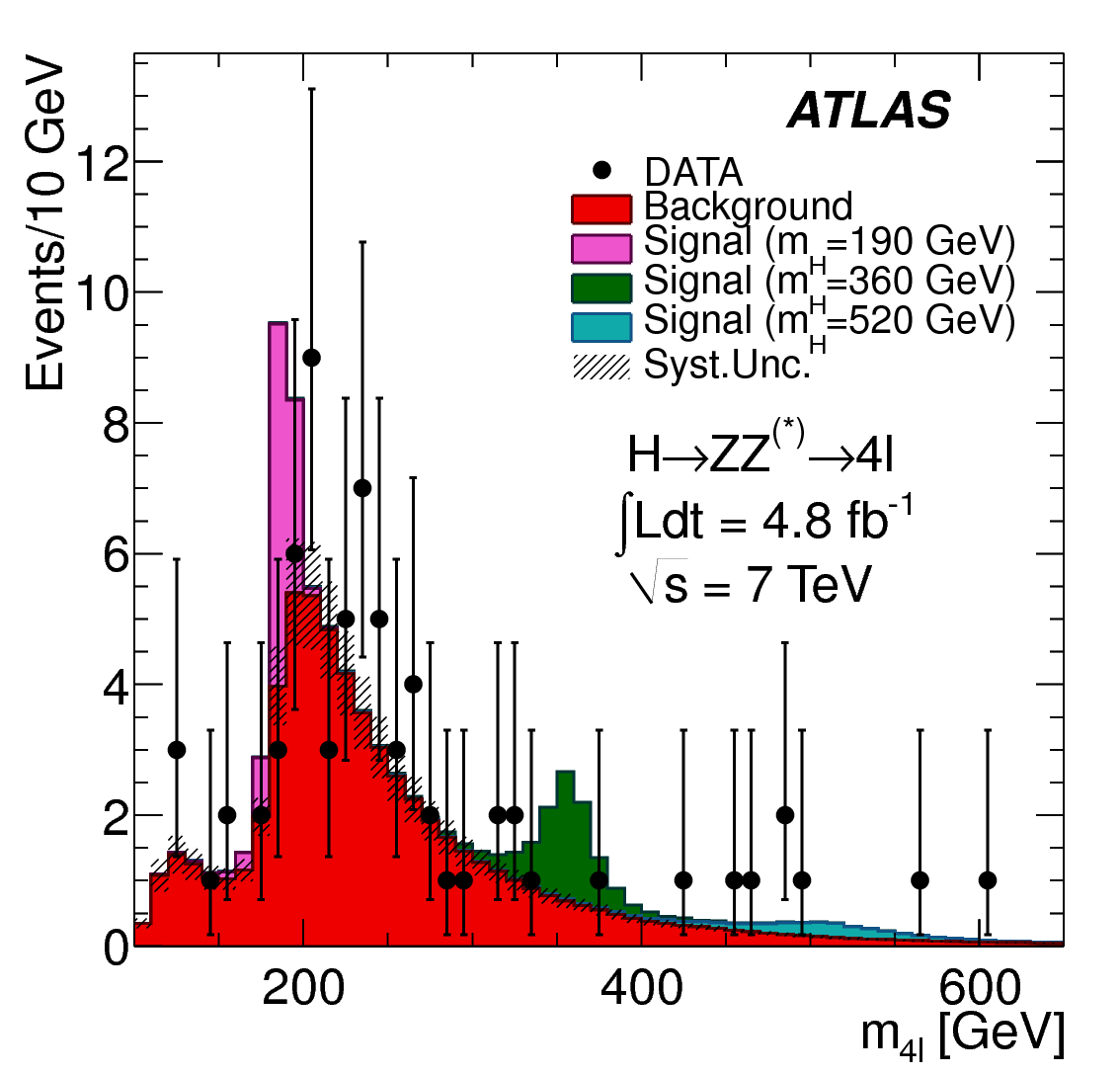}
\end{center}
\caption{\small \it
Distributions of the four-lepton invariant mass, $m_{4\ell}$ of the selected candidates, compared to the background expectation for
the 100 - 250~\GeV\ mass range (left) and the full mass range (right)
in the ATLAS experiment. The signal expectations for several $m_H$ hypotheses are
also shown (from Ref.~\protect\cite{:2012sm}).
}
\label{f:higgs_ZZ4l_mass_ATLAS}
\end{figure}

\begin{figure}[hbtn]
\begin{center}
\includegraphics[width=.35\textwidth]{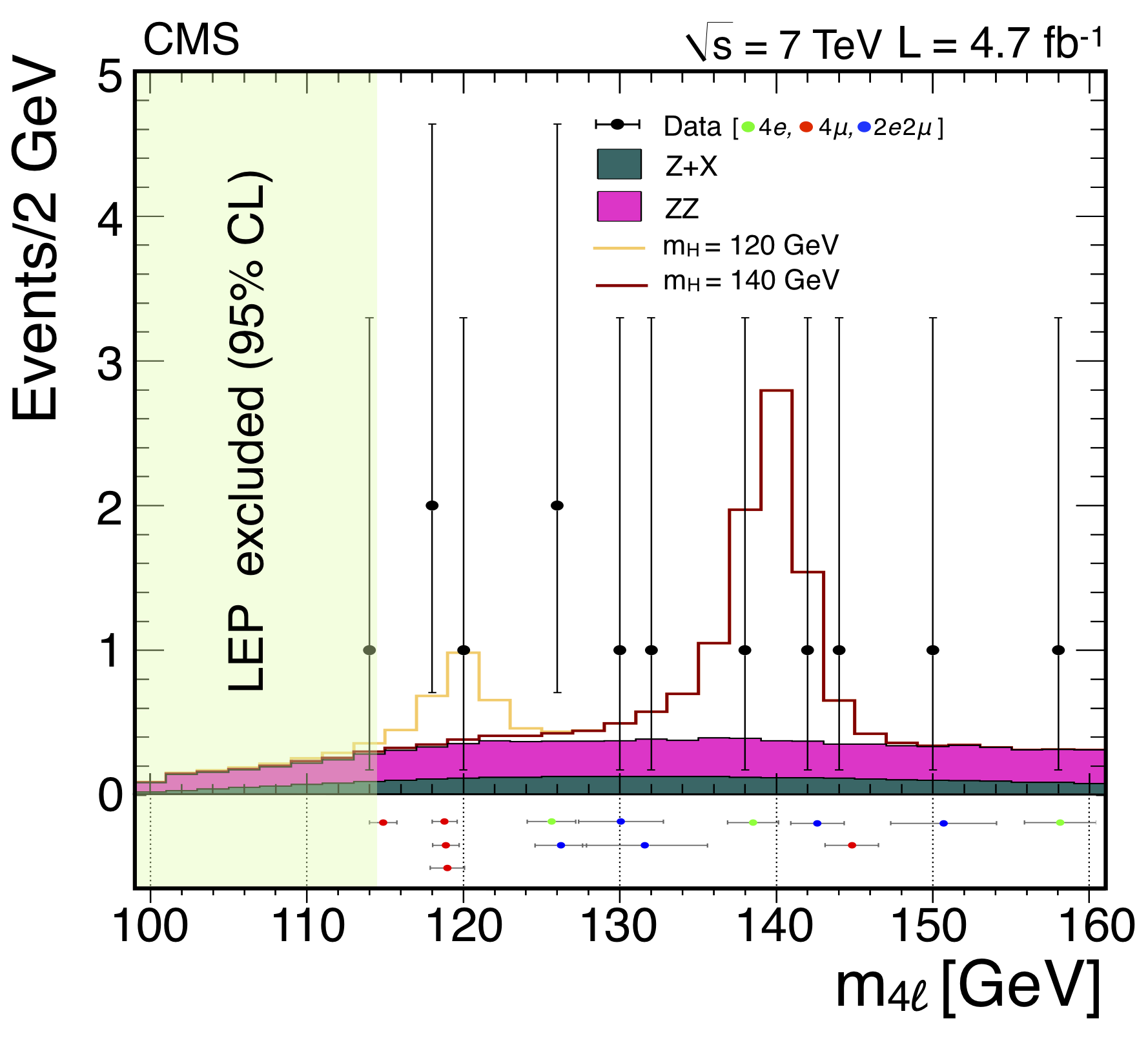}
\includegraphics[width=.35\textwidth]{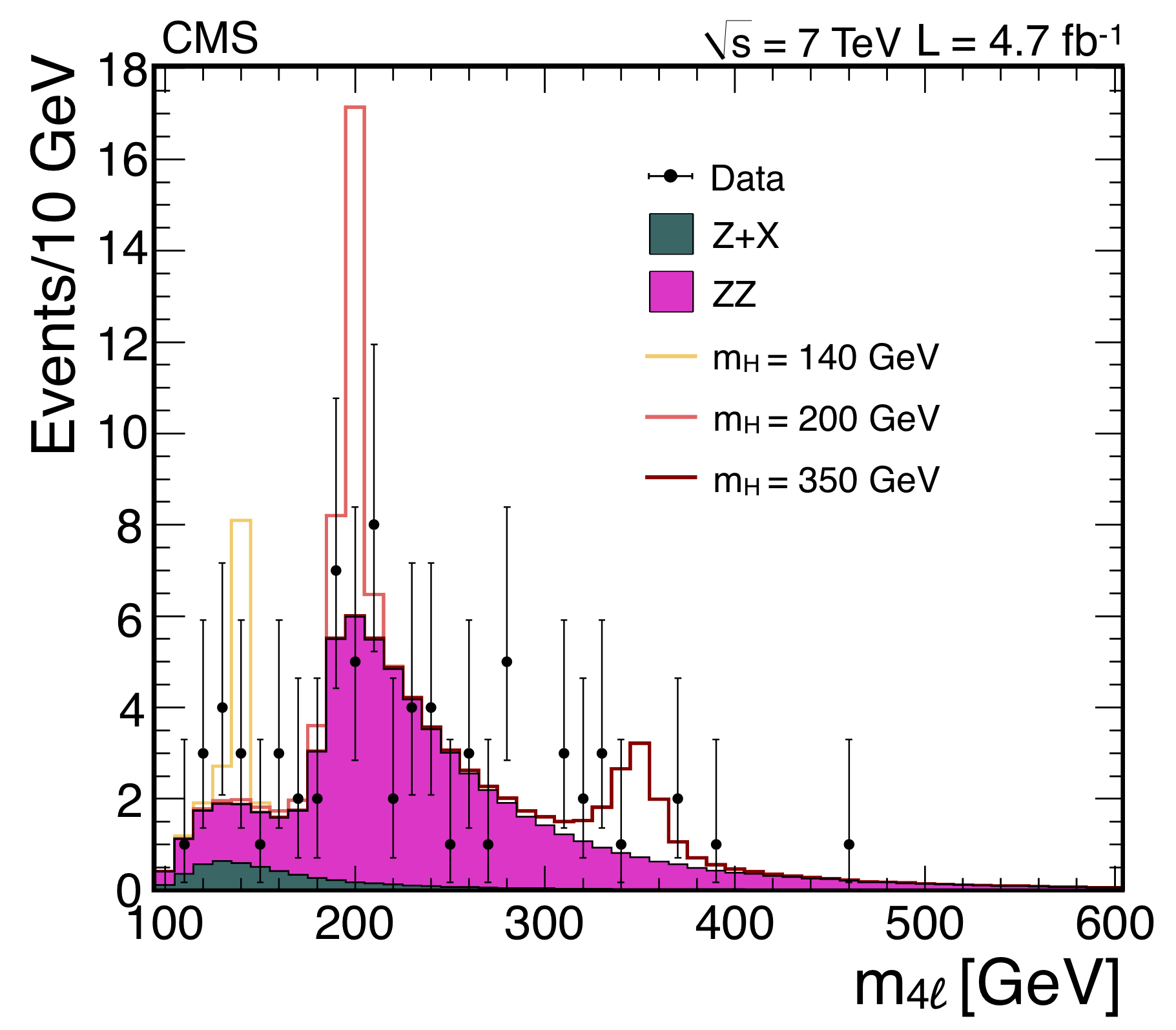}
\end{center}
\caption{\small \it
Distributions of the four-lepton invariant mass, $m_{4\ell}$ of the selected candidates, compared to the background expectation for
the 100 - 160~\GeV\ mass range (left) and the full mass range (right) in the CMS experiment.
The signal expectations for several $m_H$ hypotheses are
also shown (from Ref.~\protect\cite{Chatrchyan:1422706}).
}
\label{f:higgs_ZZ4l_mass_CMS}
\end{figure}

Both collaborations have performed the $\Hllll$ search for \mH\ hypotheses
in the full 110 to 600~\GeV\ mass range using data corresponding
to an integrated luminosity of $\sim$4.8~\ifb~\cite{:2012sm,Chatrchyan:1422706}.
It has been shown that in both experiments calorimeter and track isolation requirements
together with impact parameter requirements can be used to suppress the irreducible
background well below the irreducible $ZZ^*$ continuum background.
The residual irreducible $Z$+jets and \ttbar\ backgrounds, which have an impact
mostly for low invariant four-lepton masses, are estimated from
control regions in the data.
The irreducible $ZZ^*$ background is estimated using Monte Carlo
simulation.  The events
are categorised according to the lepton flavour combinations. Mass
resolutions of approximately 1.5\%\ in the four-muon channel and 2\%\
in the four-electron channel are achieved at \mH$\sim$120~\GeV\ \cite{:2012sm}.
The four-lepton invariant mass is used as a discriminant variable. The observed and
expected mass distributions for events selected after all cuts are
displayed in Figs.~\ref{f:higgs_ZZ4l_mass_ATLAS} and \ref{f:higgs_ZZ4l_mass_CMS}
for the ATLAS and CMS experiments, respectively.

The measured mass distributions are again confronted to the background-only
hypotheses. The corresponding $p_0$ values are shown in Fig.~\ref{f:higgs_ZZ4l_p0}
for the two experiments. In the ATLAS experiment
large upward deviations from the background-only hypothesis are observed
for $m_H$ = 125~\GeV, 244~\GeV\ and
500~\GeV\ with local significances of 2.1$\sigma$, 2.2$\sigma$ and 2.1$\sigma$, respectively.
After accounting for the look-elsewhere effect none of these excesses is significant.
The CMS collaboration observes excesses of events around 119~\GeV, 126~\GeV\ and
320~\GeV. The most significant excess for a mass value near 119~\GeV\ corresponds
to a local (global) significance of about 2.5$\sigma$ (1.0$\sigma$).

\begin{figure}[t]
\begin{center}
\includegraphics[width=.40\textwidth]{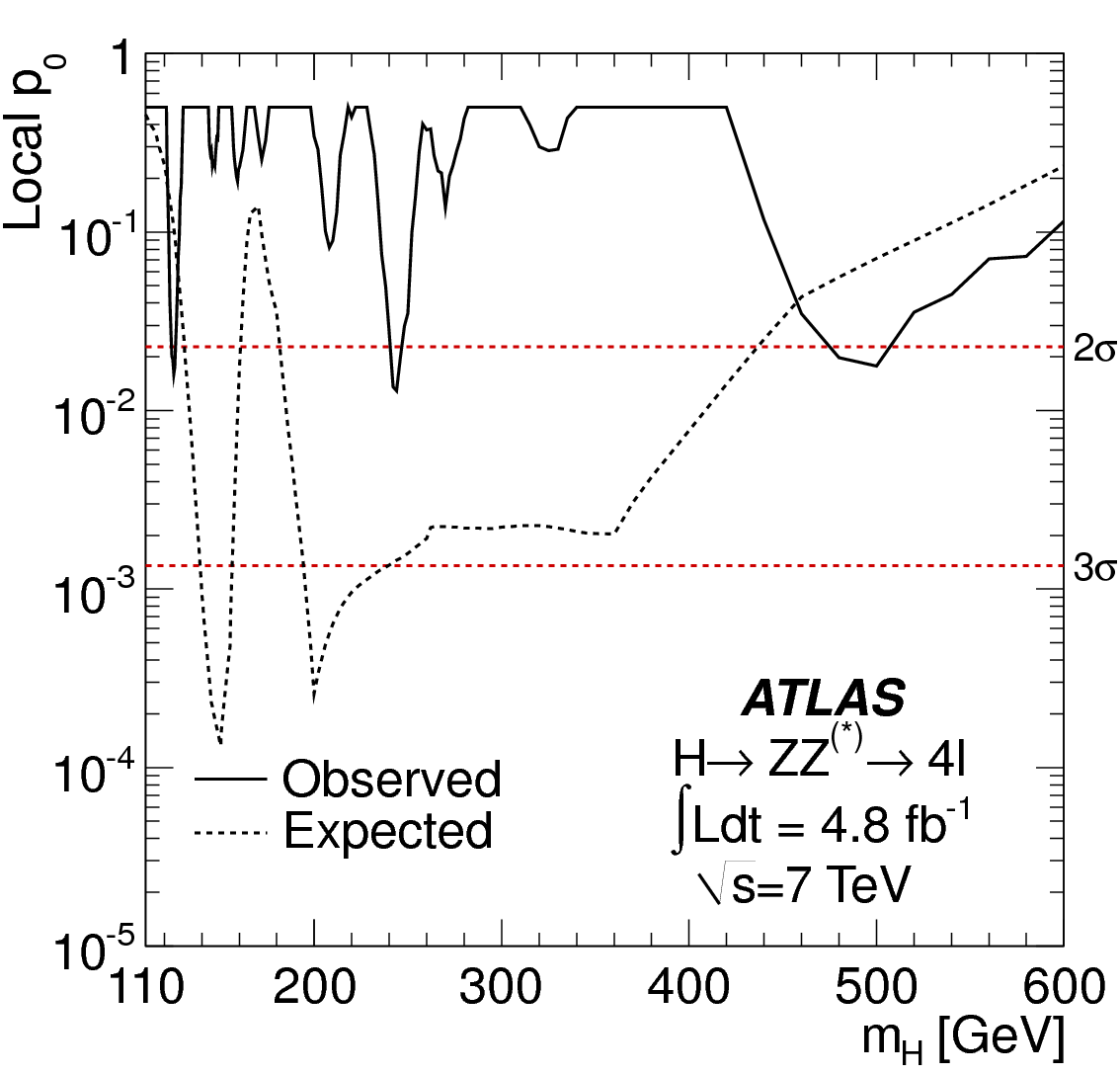}
\includegraphics[width=.45\textwidth]{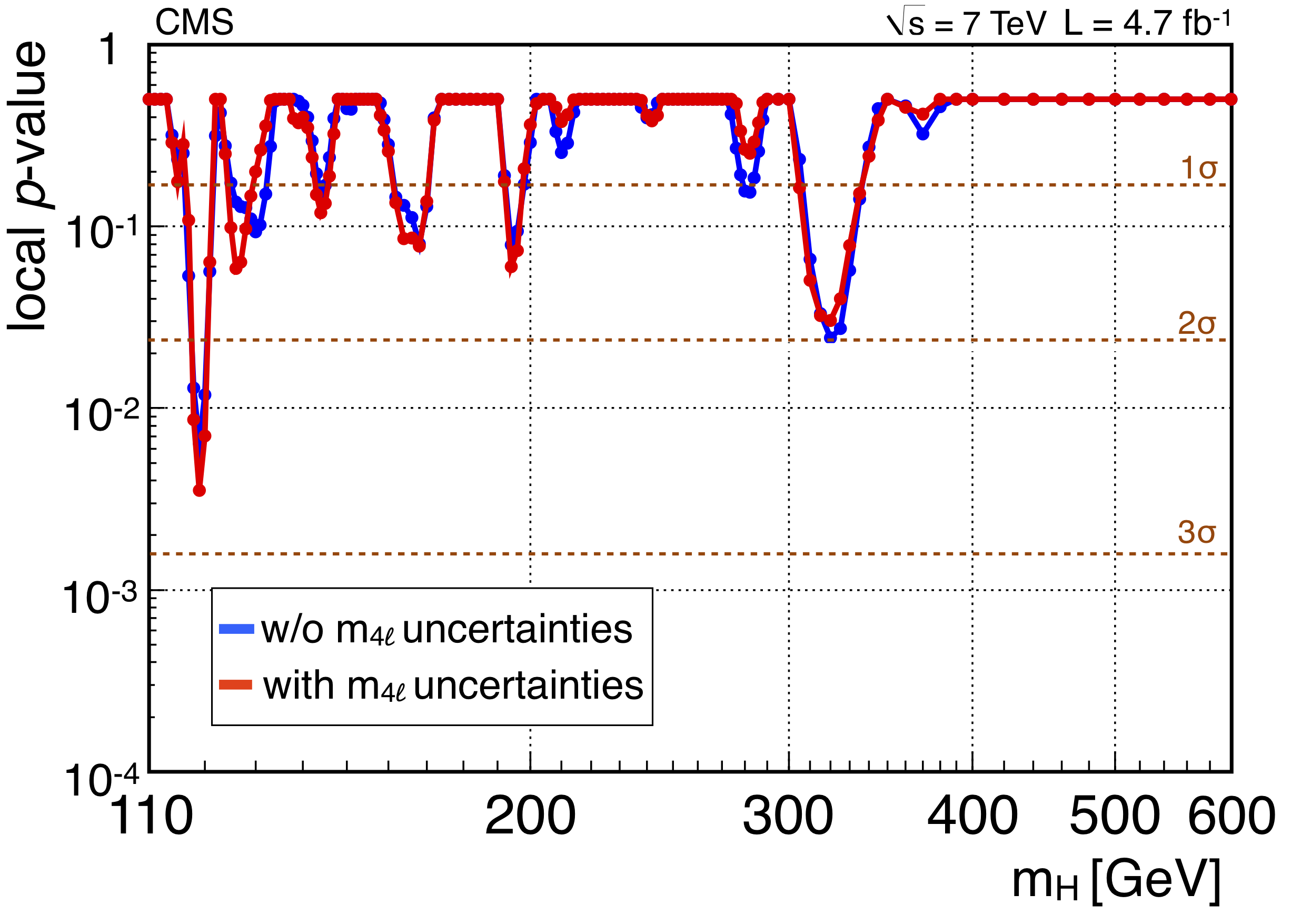}
\end{center}
\caption{\small \it
The observed local $p_0$ values as a function of the Higgs boson mass in the $\Hllll$ channel in the ATLAS (left) and CMS (right) experiments.
The dashed curve shows the expected median local $p_0$. The horizontal lines indicate values of constant significance of 1$\sigma$, 2$\sigma$ and 3$\sigma$ (from Refs.~\protect\cite{:2012sm, Chatrchyan:1422706}).
}
\label{f:higgs_ZZ4l_p0}
\end{figure}

\subsubsection{Search for \hwwsll\ decays}

The decay mode \hwwsll\ has the highest sensitivity for Higgs boson
masses around 170~\GeV. Based on searches in this channel, mass regions could be
excluded by both the Tevatron and the LHC experiments already in Summer 2011
\cite{Tevatron:1107.5518, ATLAS-CONF-2011-135, CMS-PAS-HIG-11-022}.
However, this channel is more challenging in the low mass region around 125~\GeV\ since due
to the reduced \HWW\ branching ratio the expected signal rates are small. Due to
the presence of neutrinos it is not possible to reconstruct a
Higgs boson mass peak and evidence for a signal must be extracted from
an excess of events above the expected backgrounds. Usually,
the $WW$ transverse mass ($m_T$),
computed from the leptons and the missing transverse momentum,	
\begin{displaymath}
	  \label{eq:mT}
	  m_T = \sqrt{(E_{\rm T}^{\ell\ell}+\met)^{2} - |{\bf p}_{\rm T}^{\ell\ell}+{\bf p}_{\rm T}^{\rm miss}|^{2}},
\end{displaymath}
\noindent where $E_{\rm T}^{\ell\ell} = \sqrt{|{\bf p}_{\rm T}^{\ell\ell}|^{2}+m_{\ell\ell}^{2}}$,
$|{\bf p}_{\rm T}^{\rm miss}|=\met$ and
$|{\bf p}_{\rm T}^{\ell\ell}| = p_{\rm T}^{\ell\ell}$,
is used to discriminate between signal and background.
The $WW$, \ttbar\ and single-top production processes constitute
severe backgrounds and the signal significance depends
critically on their absolute knowledge.

The analyses of the ATLAS and CMS collaborations are based on the full data set ($\sim$4.7~$\ifb$)
\cite{ATLAS-CONF-2012-012,CMS:1202.1489}.
In order to optimize the sensitivity, the analyses are split into different lepton
final states ($ee$, $e \mu$ and $\mu \mu$) and different jet multiplicities. In addition, they
have been optimized for different mass regions (low and high mass). Typical selection cuts
require the presence of two isolated high \pT\ leptons with a significant missing transverse
energy and a small azimuthal angular separation. The latter requirement is motivated by the
decay characteristics of a spin-0 Higgs boson decaying into two $W$ bosons with their subsequent
$W\to \ell \nu$ decay \cite{dittmar}.  The various jet categories are sensitive to different
Higgs boson
production mechanisms and have very different background compositions. The 0-jet category is
mainly sensitive to the gluon-fusion process and has the non-resonant $WW$ production as major
background. The 2-jet category is more sensitive to the vector-boson fusion process, with
\ttbar\ as dominant background.
As a final discriminant the $WW$ transverse mass distribution is used.
This distribution is shown in Fig.~\ref{f:higgs_WW_ATLAS} (left) for events passing the
0-jet selection in ATLAS. The observed data are well described by the expected
background contributions which are dominated by the $WW$ production.
\begin{figure}[t]
\begin{center}
\includegraphics[width=.48\textwidth]{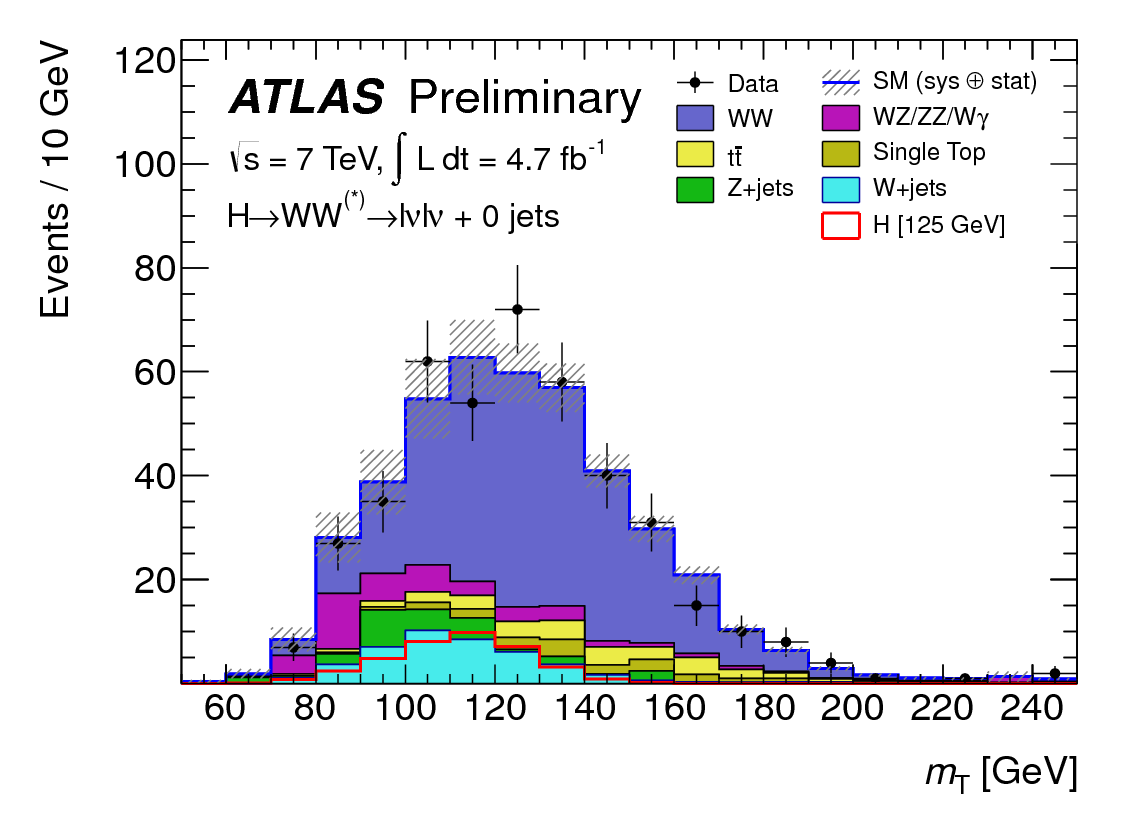}
\includegraphics[width=.48\textwidth]{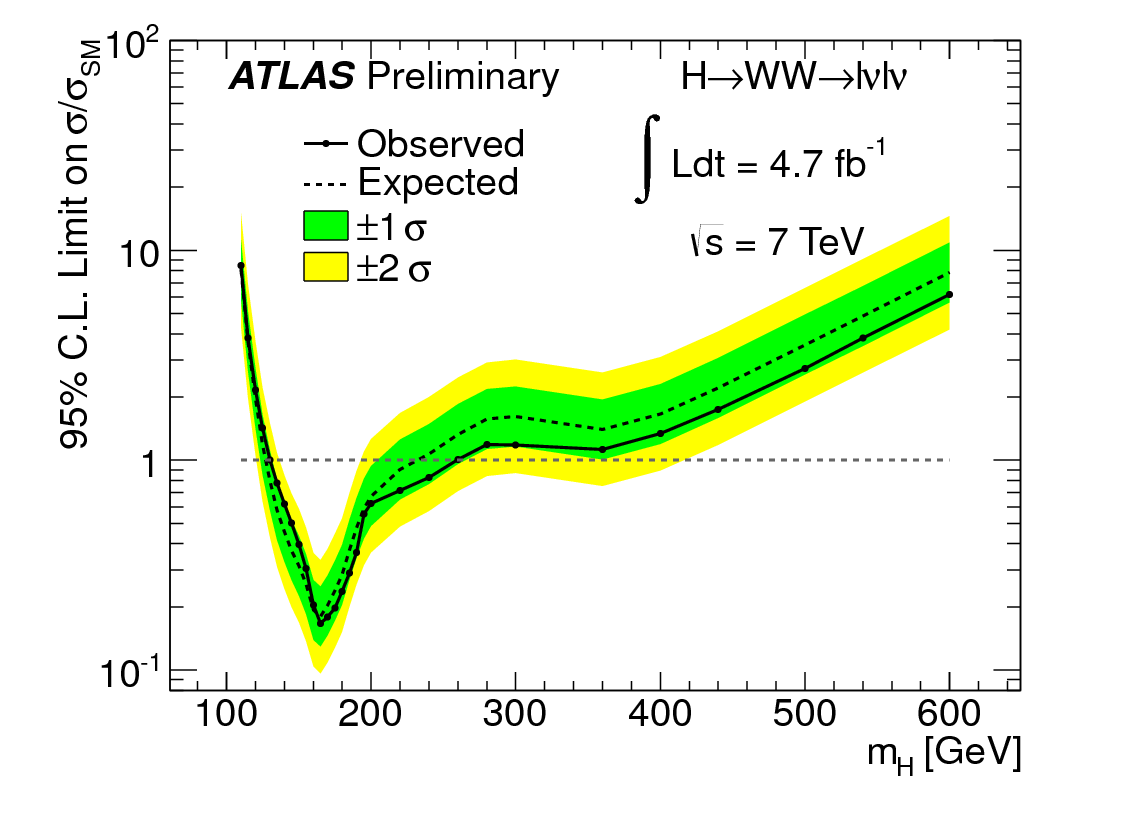}
\end{center}
\caption{\small \it
(Left):
The distribution of the transverse mass $m_T$ in the H+0 jet channel of the ATLAS analysis.
The expected signal for a Standard Model Higgs boson with $m_H$ = 125~\GeV\ is superimposed.
(Right): Expected (dashed) and observed (solid) 95\%~C.L. upper limits on the cross section, normalized
to the Standard Model cross section, as a function of $m_H$. The results at neighbouring mass points are highly correlated due to the limited
mass resolution in this final state (from Ref.~\protect\cite{ATLAS-CONF-2012-012}).
}
\label{f:higgs_WW_ATLAS}
\end{figure}
\begin{figure}[hbtn]
\begin{center}
\includegraphics[width=.40\textwidth]{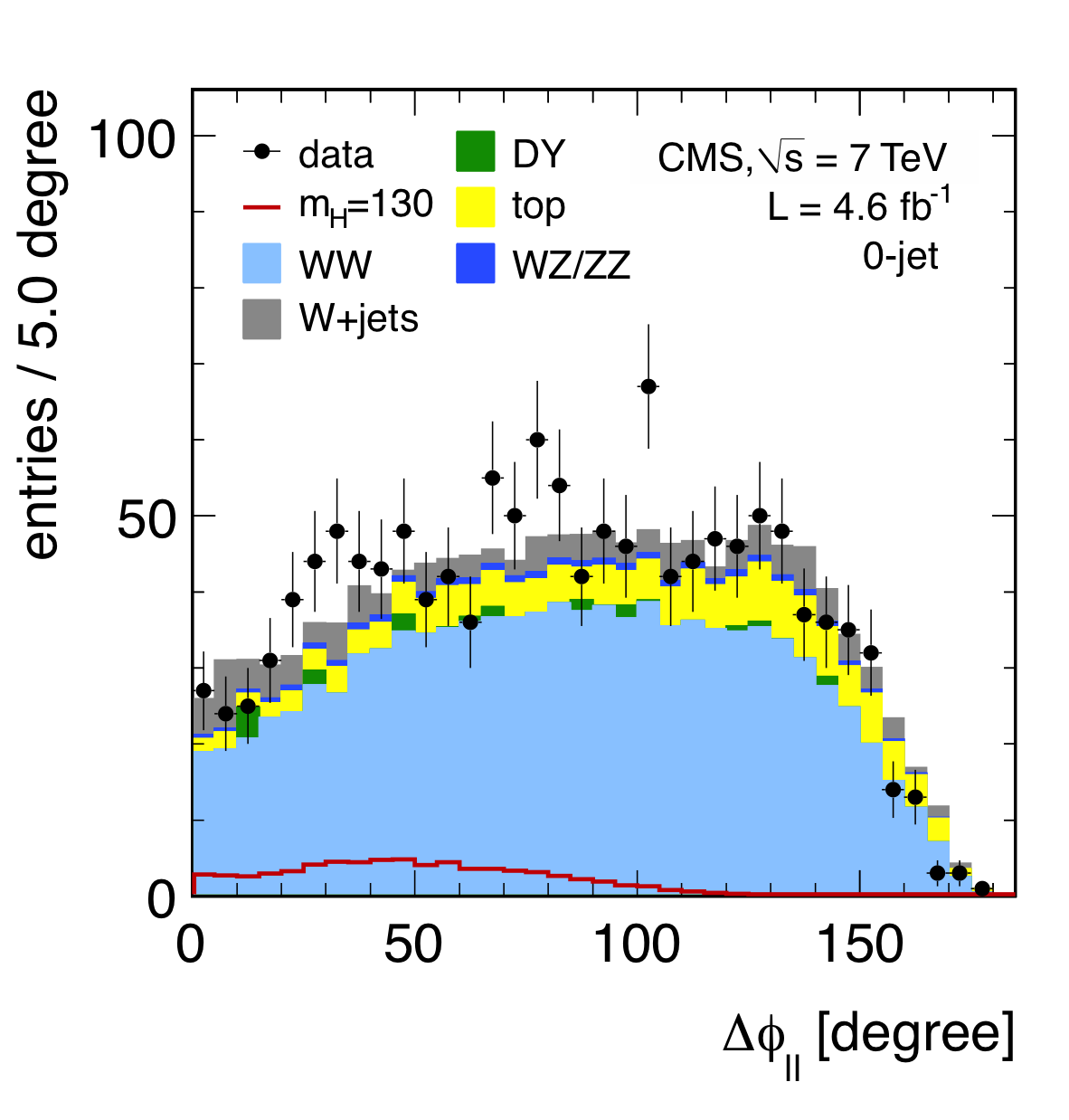}
\includegraphics[width=.50\textwidth]{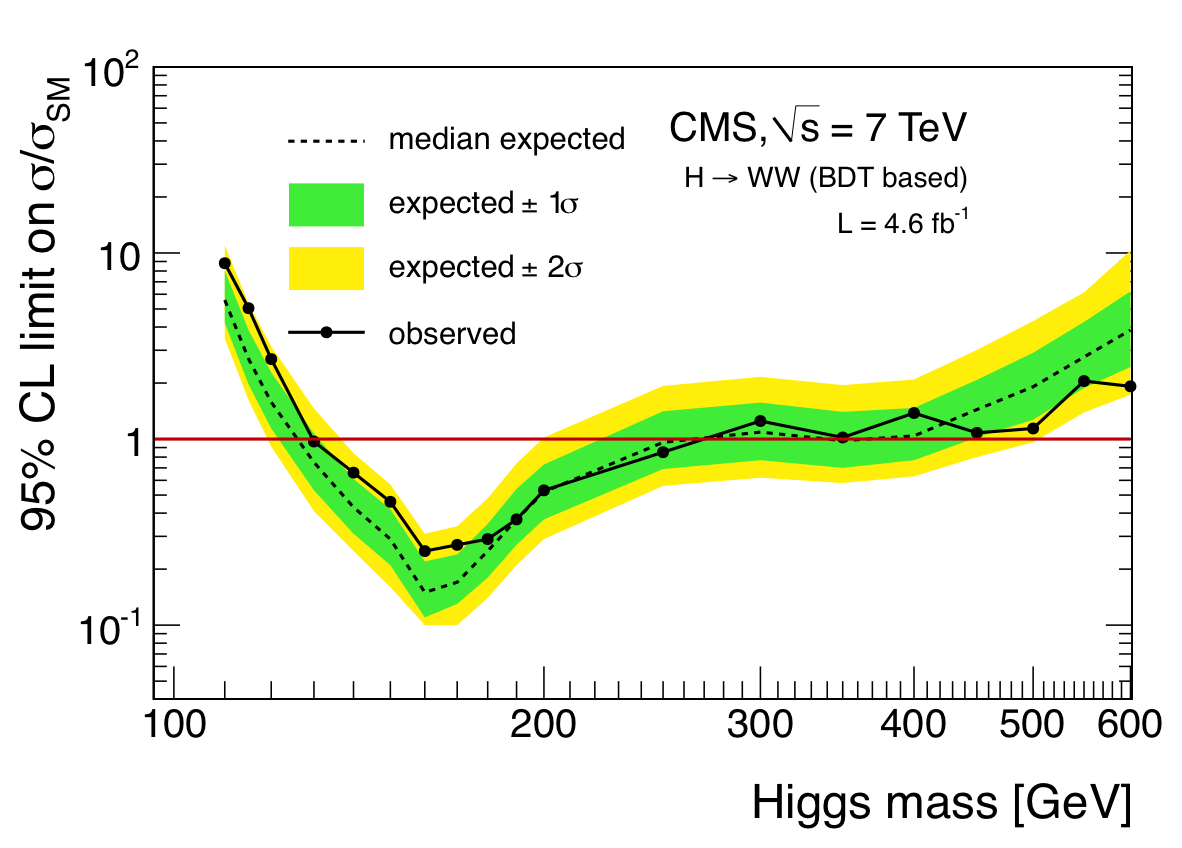}
\end{center}
\caption{\small \it
(Left): The distribution of the azimuthal angle separation $\Delta \phi_{\ell\ell}$
in the H+0 jet channel of the CMS analysis.
The expected signal for a Standard Model Higgs boson with $m_H$ = 130~\GeV\ is superimposed.
(Right): Expected (dashed) and observed (solid) 95\%~C.L. upper limits on the cross section, normalized
to the Standard Model cross section, as a function of $m_H$. The results at neighbouring mass points are highly correlated due to the limited
mass resolution in this final state (from Ref.~\protect\cite{CMS:1202.1489}).
}
\label{f:higgs_WW_CMS}
\end{figure}
As another example, Fig.~\ref{f:higgs_WW_CMS} (left) shows the distribution
of the azimuthal angle difference
($\Delta \phi_{\ell \ell}$) between the two selected leptons for events in the 0-jet
category in the CMS experiment. Also this distribution is well described by the expected
background processes.
To enhance the sensitivity, the CMS experiment exploits
two different analysis strategies for the 0-jet and 1-jet categories, the first one using a cut-based approach and the second one using a multivariate technique \cite{CMS:1202.1489}.

Since no significant excesses of events are found in any of the event categories in both
the ATLAS and CMS experiments, upper limits on the production
cross section are set, as shown in Figs.~\ref{f:higgs_WW_ATLAS} and
\ref{f:higgs_WW_CMS}.
The ATLAS experiment excludes the existence of a Standard Model Higgs boson
over a mass range from 130 - 260~\GeV, while the expected exclusion, in case no Higgs boson
is present,
is 127 $\leq m_H \leq$ 234~\GeV. The CMS experiments excludes a mass range from
129 - 270~\GeV, with an expected range from 127 - 270~\GeV.

\subsubsection{Search for \Htau\ and \Hbb\ decays}

In addition to the searches described above, the search for the Higgs boson has also been
performed in the \Htau\ \cite{ATLAS-CONF-2012-014,Chatrchyan:1425096}
and \Hbb\ final states \cite{ATLAS-CONF-2012-015,Chatrchyan2012284}.
These searches do not yet reach the sensitivity of the others described above.
They are, however, included in the overall combination of the results of the two
collaborations \cite{ATLAS-CONF-2012-019,Chatrchyan:1422382}. The observed and
expected cross section limits are included in Fig.~\ref{f:higgs_all_sigma95}.

\subsubsection{Search for the Higgs boson in the high mass region}

For higher Higgs boson masses ($\mH > 2  \mZ$) the decays \HWW\ and \HZZ\ dominate.
Due to the higher mass and improved signal-to-background conditions, also the
decays
$H \to  Z Z \to \ell \ell \ \nu \nu$ \cite{Chatrchyan:1424679,ATLAS-CONF-2012-016},
$H \to  Z Z \to \ell \ell \ q q $ \cite{ATLAS-CONF-2012-017,Chatrchyan:1422383},
$H \to  Z Z \to \ell \ell \ \tau \tau $ \cite{Chatrchyan:1424729} and
$H \to  W W \to \ell \nu \ q q $ \cite{ATLAS-CONF-2012-018}
provide additional sensitivity.

The $H \to ZZ \to \ell \ell \nu \nu $ is the most sensitive channel. The
selection of two leptons and large missing transverse energy gives rather
good signal-to-background conditions. The dominant backgrounds are from diboson
and \ttbar\ production. Also in this case the transverse mass $m_T$ of the
$\ell \ell - \ETmiss$ system is used as discriminating variable. The distributions
are shown in Fig.~\ref{f:higgs_ZZllnn_mass} for the ATLAS (left) and CMS (right)
experiments together with expected signals at 400~\GeV.
No indications for excesses
are seen and upper limits on the Higgs boson production cross sections are set.
They are included as well in Fig.~\ref{f:higgs_all_sigma95}.

\begin{figure}[hbtn]
\begin{center}
\includegraphics[width=.50\textwidth]{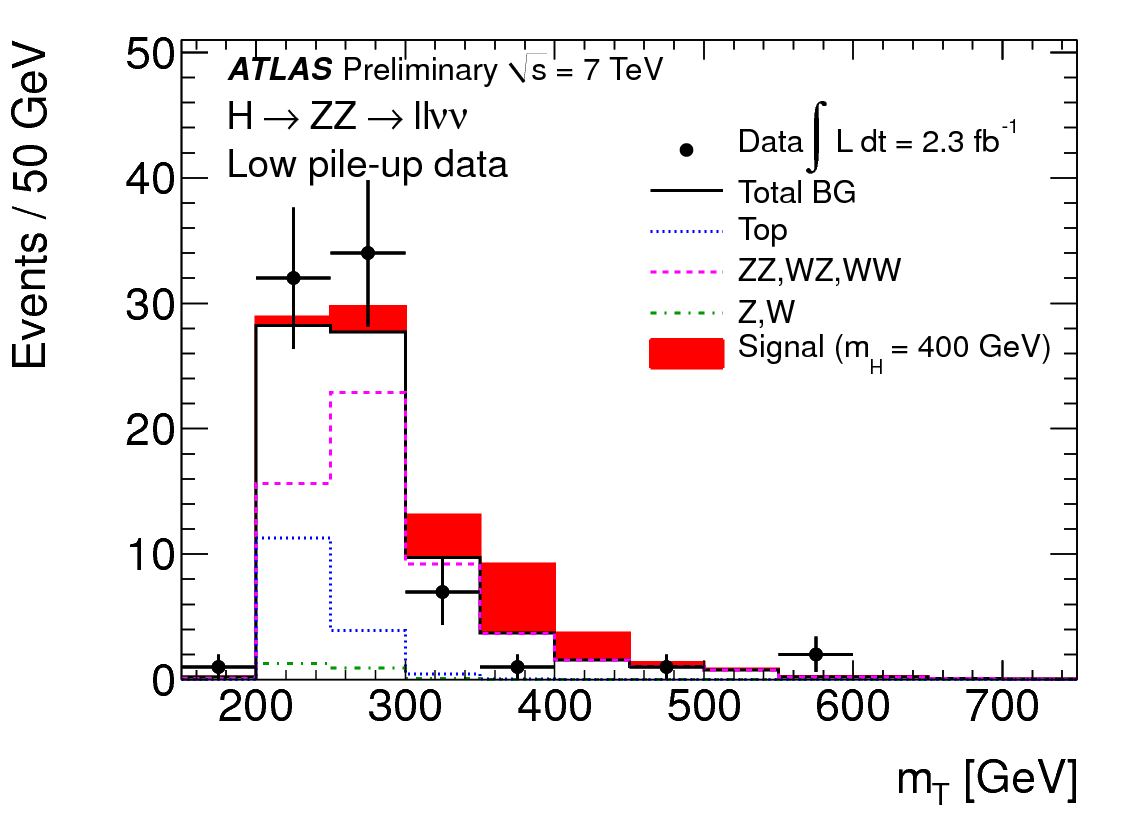}
\includegraphics[width=.36\textwidth]{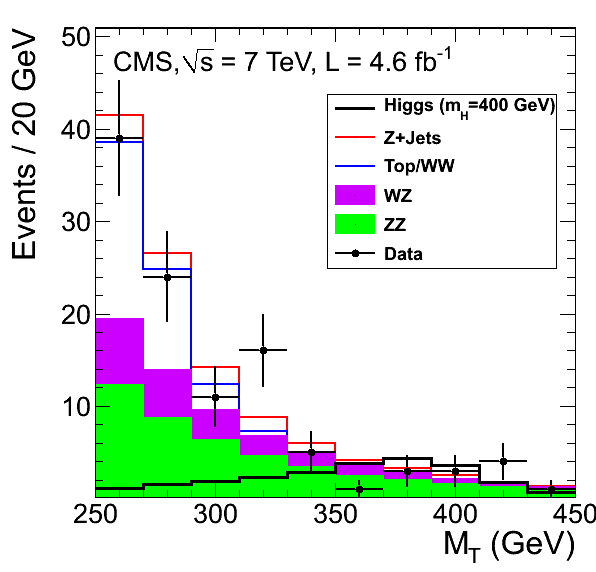}
\end{center}
\caption{\small \it
The distributions of the transverse mass of $H \to  ZZ \to \ell \ell \nu \nu$ candidates
in the ATLAS (left) and CMS (right) experiments. Expected signals for a Higgs boson
with a mass of 400~\GeV\ are superimposed
 (from Refs.~\protect\cite{ATLAS-CONF-2012-016, Chatrchyan:1424679}).
}
\label{f:higgs_ZZllnn_mass}
\end{figure}

\subsection{Combination results of searches for the Standard Model Higgs boson}

\subsubsection{Excluded mass ranges}

The ATLAS and CMS experiments have combined their respective search results on the
Standard Model Higgs boson \cite{ATLAS-CONF-2012-019, Chatrchyan:1422382}.
The combination procedure is based on the profile likelihood ratio test statistic
$\lambda({\mu})$~\cite{asym}, which extracts the information on the
signal strength $\mu = \sigma / \sigma_{\rm{SM}}$
from a full likelihood including all the parameters
describing the systematic uncertainties and their correlations.
More details on the statistical procedure used are described in Ref.~\cite{HCG}.

\begin{figure}[htb]
  \begin{center}
    \includegraphics[width=0.59\textwidth]{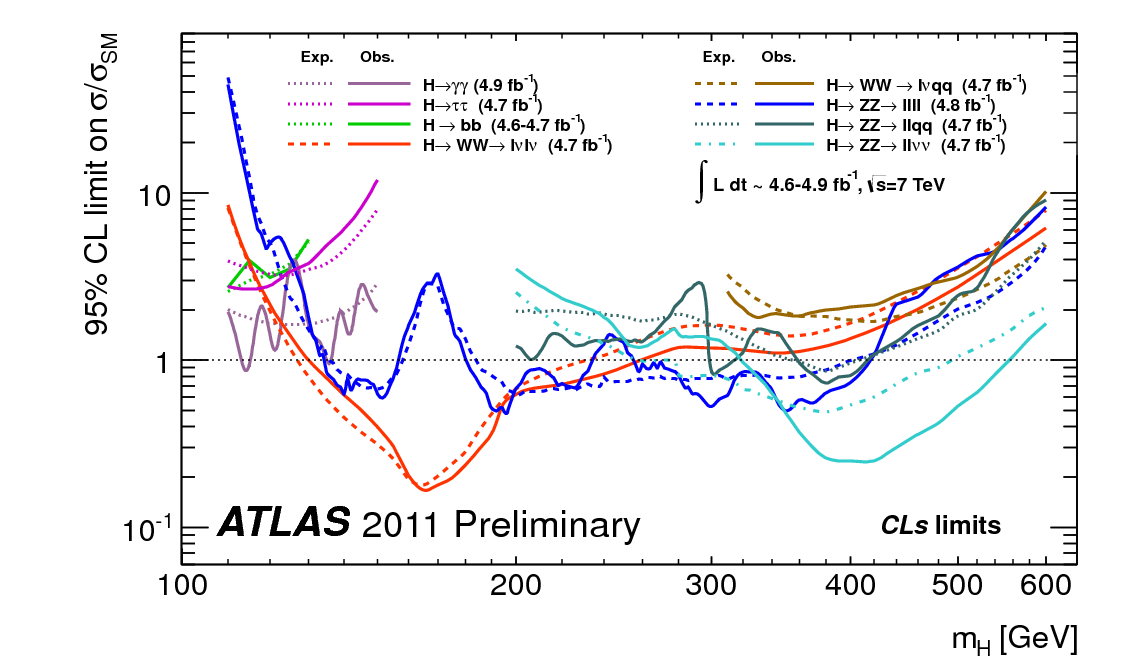}
    \includegraphics[width=0.40\textwidth]{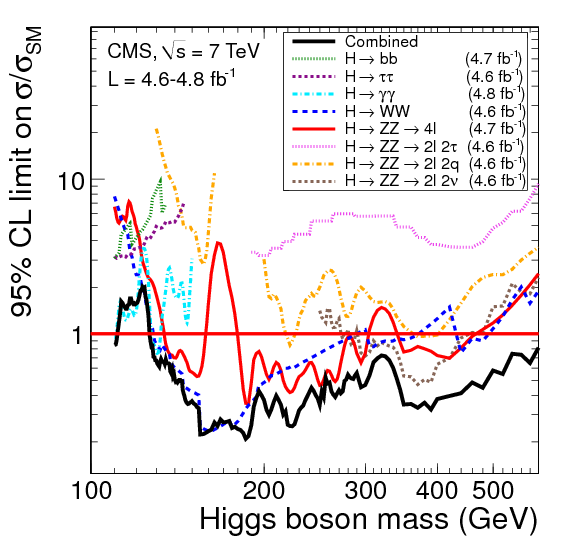}
  \end{center}
\vspace*{-0.5cm}
\caption{ The observed (solid) and expected (dashed) 95\%~C.L. cross
  section upper limits for the individual search channels in the ATLAS (left) and
CMS (right) experiments, normalized
  to the Standard Model Higgs boson production cross section, as a function of the
  Higgs boson mass. The expected limits are those for the
  background-only hypothesis, i.e. in the absence of a Higgs boson
  signal (from Refs.~\protect\cite{ATLAS-CONF-2012-019, Chatrchyan:1422382}).
}
  \label{f:higgs_all_sigma95}
\end{figure}

In Figure~\ref{f:higgs_all_sigma95} the expected and observed 95\% C.L. limits are shown from
the individual channels entering this combination,  separately for the ATLAS and CMS experiments.
The combined 95\%~C.L. exclusion limits are shown in Fig.~\ref{fig:CLs} as a
function of $\mH$ for the full mass range and for the low mass range.
The combined expected 95\%~C.L. exclusion regions for the two experiments are very similar and
cover the \mH\ range from 120 to 555~\GeV\ for the ATLAS and from
118 to 543~\GeV\ for the CMS experiment. Based on the observed limit,
the ATLAS experiment excludes at the 95\% C.L.
the Standard Model Higgs boson in three mass ranges:
from 110.0 - 117.5~\GeV,  from 118.5 to 122.5~\GeV\ and from 129 to 539~\GeV.
The 95\% C.L. CMS exclusion covers the range 127 - 600~\GeV.
The observed exclusion covers a
large part of the expected exclusion range, with the exception of the
low mass region where an excess of events above the expected
background is observed. It is striking that both
experiments are not able to cover the mass window from about 118 to 129~\GeV,
despite their sensitivity in this range.

\begin{figure}[hbtn]
\begin{center}
    \includegraphics[width=0.52\textwidth]{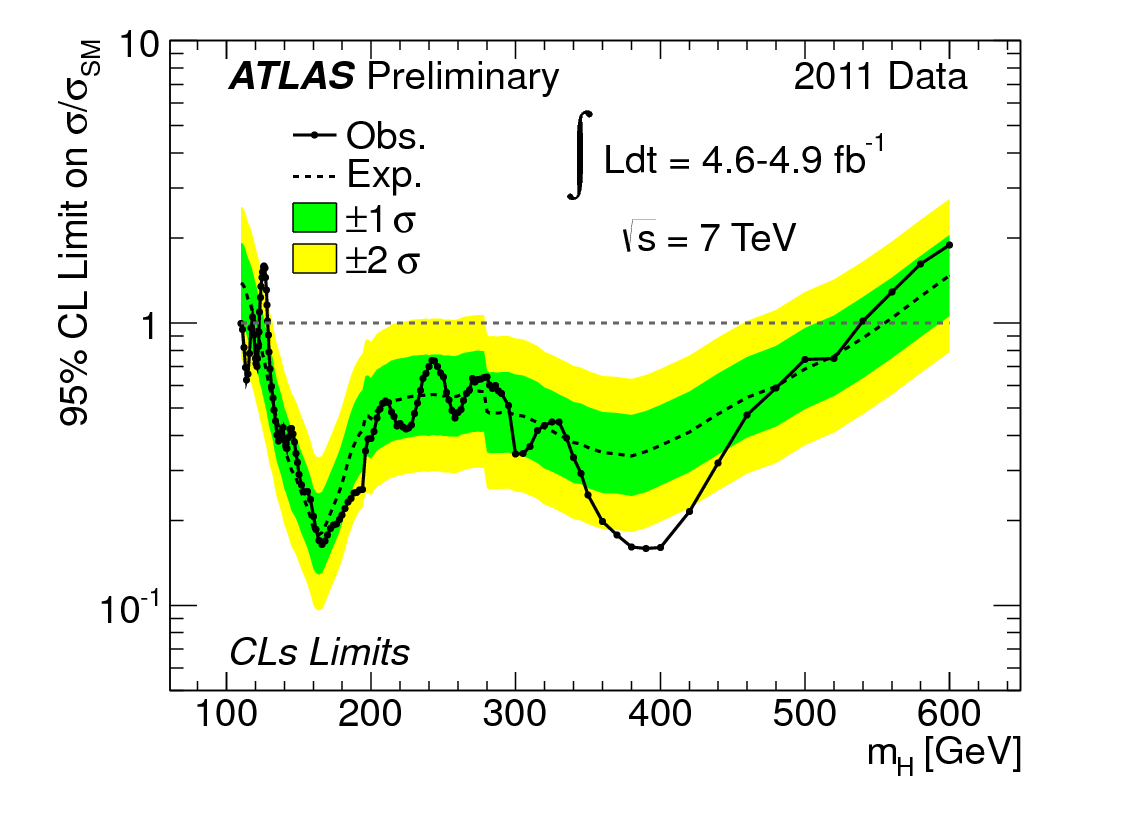}
    \includegraphics[width=0.40\textwidth]{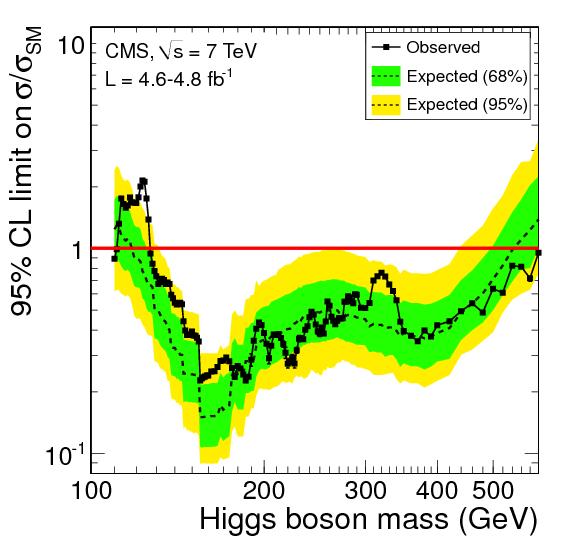}
    \includegraphics[width=0.52\textwidth]{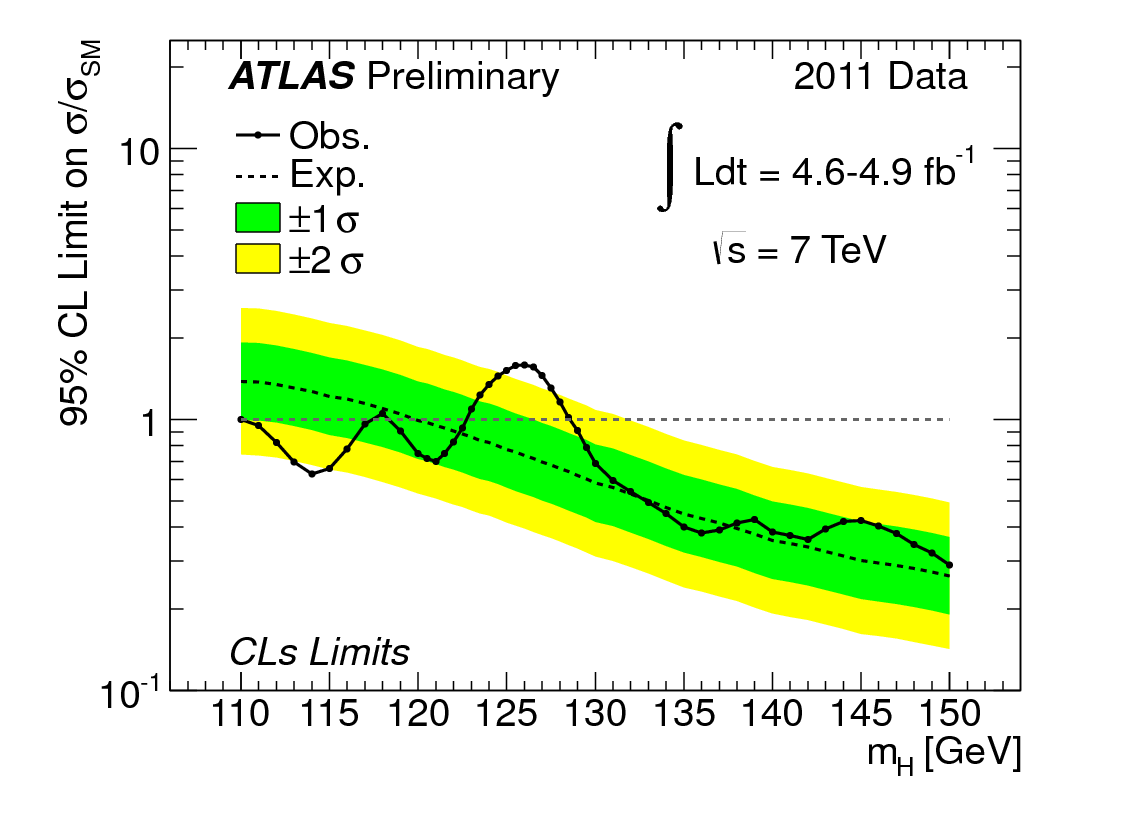}
    \includegraphics[width=0.40\textwidth]{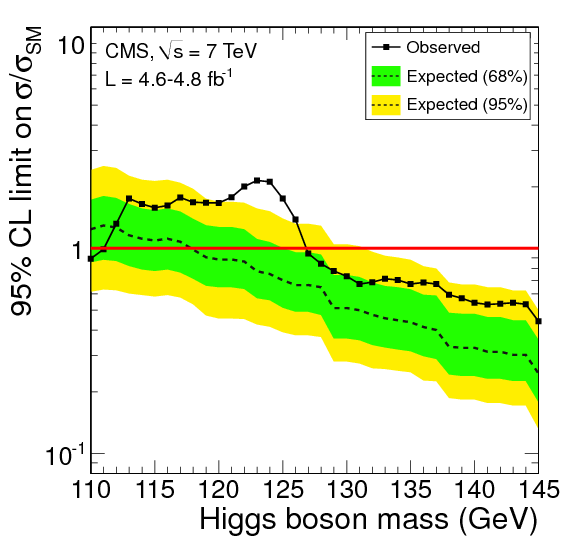}
\end{center}
  \caption{The observed (full line) and expected (dashed line) 95\%~C.L.
    combined upper limits on the Standard Model Higgs boson production cross
    section divided by the Standard Model expectation as a function of
    \mH\ in the full mass range considered in the analyses (top) and in
    the low mass range (bottom) for the ATLAS (left) and CMS (right) experiments.
    The dotted curves show the median
    expected limit in the absence of a signal and the green and yellow
    bands indicate the corresponding 68\%\ and 95\%\ intervals
 (from Refs.~\protect\cite{ATLAS-CONF-2012-019, Chatrchyan:1422382}).
}
  \label{fig:CLs}
\end{figure}

\subsubsection{Compatibility with the background-only hypothesis}

\begin{figure}[htbn]
  \begin{center}
    \includegraphics[width=0.46\textwidth]{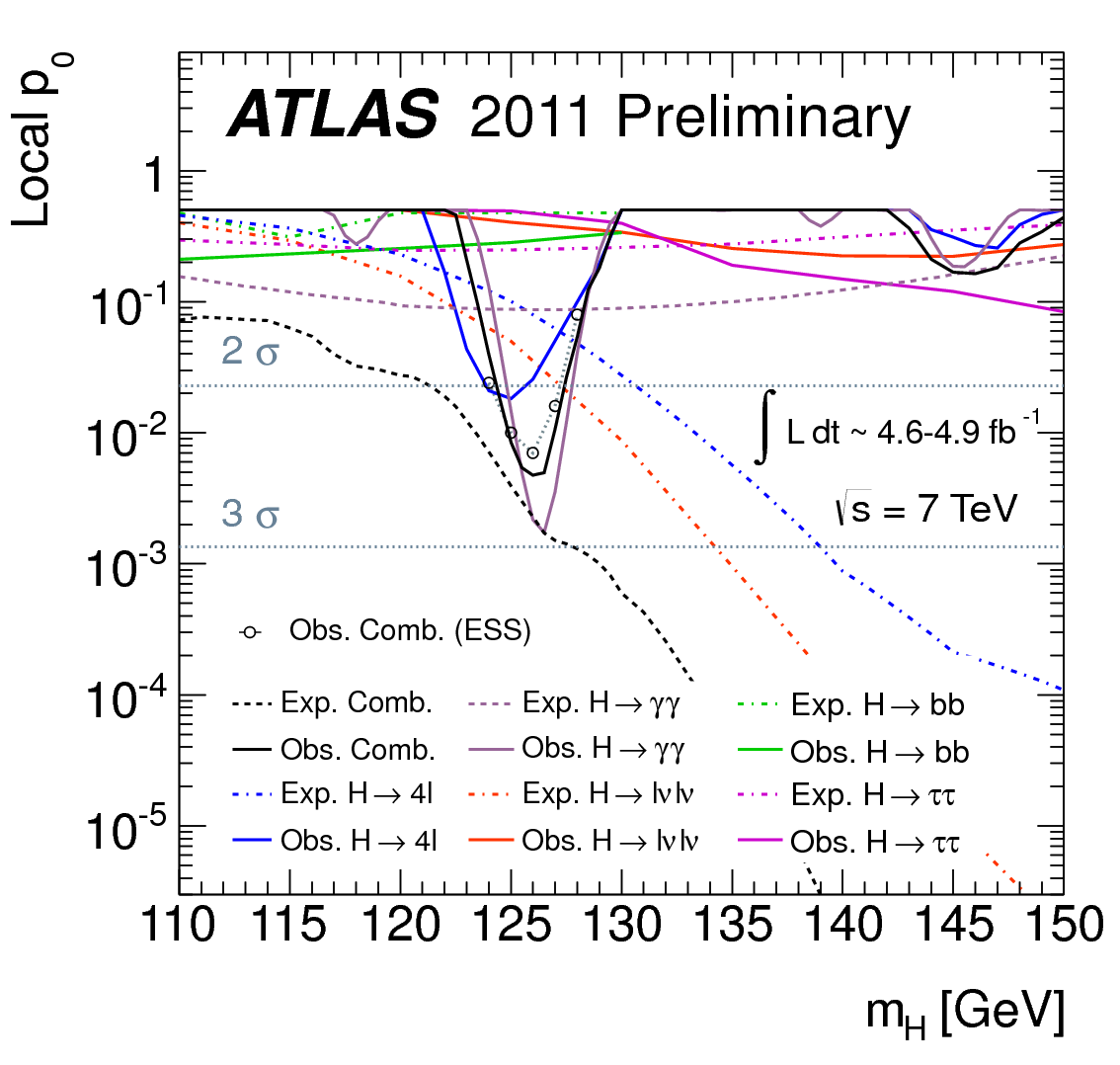}
    \includegraphics[width=0.46\textwidth]{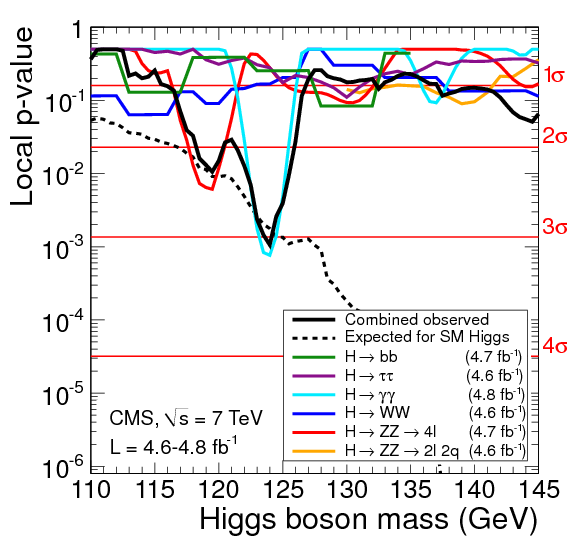}
  \end{center}
  \caption{The local probability $p_0$ for a background-only
    experiment to be more signal-like than the observation
    in the low mass range
    as a
    function of \mH\ for the ATLAS (left) and CMS (right) experiments.
    The $p_0$ values are shown for individual channels as well as for the
    combination.
    The dashed curves show the median expected local
    $p_0$ under the hypothesis of a Standard Model Higgs boson
    production signal at that mass. The horizontal dashed lines
    indicate the $p$ values corresponding to significances of
    1$\sigma$ to 5$\sigma$
(from Refs.~\protect\cite{ATLAS-CONF-2012-019, Chatrchyan:1422382}).
}
  \label{fig:CLb}
\end{figure}

Excesses of events are observed in the ATLAS experiment near 126~\GeV\
in the \Hgg\ and \Hllll\ channels, both of which provide fully reconstructed
candidates with high-resolution in invariant mass. The CMS experiment
observes two localized excesses, one at 119.5~\GeV\ associated with three
$Z \to 4\ell $ events and the other one at 124~\GeV, arising mainly from
the $\gamma \gamma$ channel. In addition, a broad
offset of about one standard deviation is seen for the
low resolution channels $H \to WW, H \to \tau \tau$ and $H \to b \bar b$.
The observed local $p_0$ values, calculated using the asymptotic
approximation, as a function of \mH\ and the expected value in the
presence of a Standard Model Higgs boson signal are shown in
Fig.~\ref{fig:CLb} in the low mass region for the two experiments.

In the ATLAS data the local significance for the combined result
reaches 2.6$\sigma$ for \mH=126~\GeV\ with an expected value
in the presence of a signal at that mass of 2.9$\sigma$.
The local significance for the combination of the CMS channels
at $m_H$ = 124~\GeV\ amounts to 3.1$\sigma$

The significance of the excesses is mildly sensitive to energy scale
systematic (ESS) uncertainties and the resolution for photons and
electrons. The observed effect of the ESS uncertainty is small and reduces
the maximum local significance in the ATLAS experiment from 2.6$\sigma$
to 2.5$\sigma$.

The global $p_0$ for local excesses depends on the range of \mH\ and
the channels considered.
 The global probability for an excess as large as the one observed in the ATLAS
combination at 126~\GeV\
to occur anywhere in the mass range  110--600~\GeV\
is estimated to be approximately 30\%, decreasing to 10\%\ in the
range 110--146~\GeV, which is not excluded at the 99\%\ confidence
level by the LHC combined Standard Model Higgs boson search~\cite{lhcCombination}.
The global significance for the CMS excess is estimated to be 1.5$\sigma$
for the full search range from 110--600~\GeV\ and 2.1$\sigma$
for the restricted search range from  110--145~\GeV.

The best-fit value of $\mu$, denoted $\hat{\mu}$, is displayed for the
combination of all channels for the two experiments in Fig.~\ref{fig:muhat}.
The bands around $\hat{\mu}$ illustrate the
$\mu$ interval corresponding to $-2\ln \lambda(\mu)<1$ and represent
an approximate $\pm 1\sigma$ variation. The excess observed for
$\mH=126~\GeV$~in the ATLAS experiment corresponds to $\hat{\mu}$ of approximately
$0.9^{+0.4}_{-0.3}$, which is compatible
with the signal strength expected from a Standard Model Higgs boson at that mass
($\mu=1$). Also for the CMS experiment the  $\hat{\mu}$ values are within one
sigma of unity in the mass range from 117--126~\GeV.

\begin{figure}[p!]
  \begin{center}
    \includegraphics[width=0.55\textwidth]{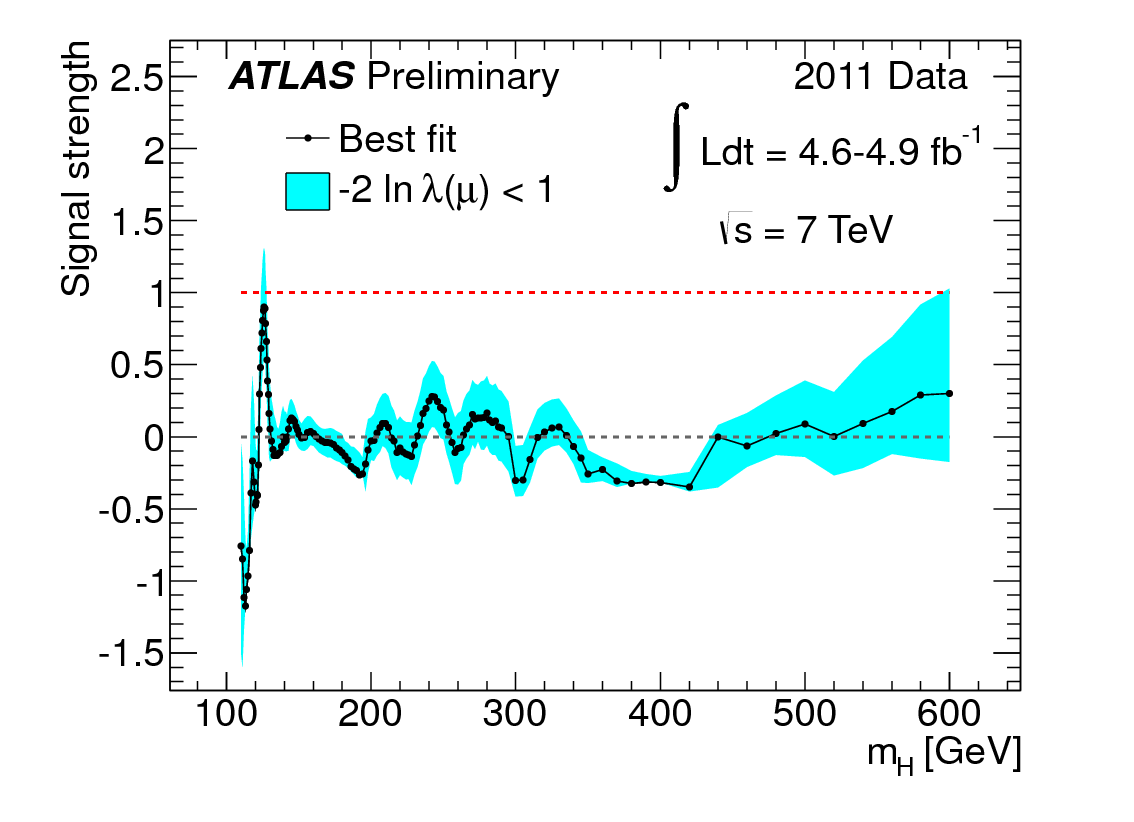}
    \includegraphics[width=0.42\textwidth]{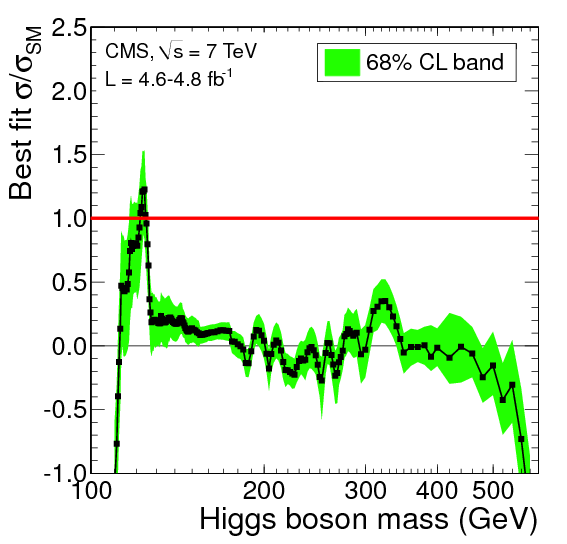}
\end{center}
  \caption{The best-fit signal strength $\hat \mu$ as
    a function of the Higgs boson mass hypothesis in the full
    mass range for the combination of the ATLAS (left) and CMS (right)
    analyses.
    The
    $\mu$ value indicates by what factor the Standard Model Higgs boson
    cross section would have to be scaled to best match the observed
    data. The band
    shows the interval around $\hat{\mu}$ corresponding to a variation
    of $-2\ln \lambda(\mu)<1$
(from Refs.~\protect\cite{ATLAS-CONF-2012-019, Chatrchyan:1422382}).
}
  \label{fig:muhat}
\end{figure}

\clearpage

\section{Search for Supersymmetric Particles}

\noindent Due to the high centre-of-mass energy of 7~\TeV, the LHC has a large discovery potential for new heavy particles beyond the
Tevatron limits. This holds in particular for particles with colour charge, such as squarks
and gluinos in supersymmetry (SUSY) \cite{SUSY-refs1,*SUSY-refs2}.
However, due to the excellent luminosity performance of the LHC in 2011,
sensitivity also exists for electroweak production of charginos and neutralinos, the supersymmetric partners of the
electroweak gauge bosons and the Higgs boson. In the following a few results of the searches by the ATLAS and CMS
collaborations for supersymmetry
with up to 2 fb$^{-1}$ of LHC $pp$ data at $\sqrt{s}$ = 7~\TeV\ are summarized. Since none of the analyses have observed
any excess above the Standard Model expectations, limits on SUSY parameters or masses of SUSY particles
are set. The discussion presented here follows largely the  review of Ref.\cite{SUSY-DeJong}
on the results from the ATLAS collaboration.

\subsection{Searches with jets and missing momentum}

\noindent Assuming conservation of R-parity, the lightest supersymmetric particle (LSP) is stable and weakly interacting, and
will typically escape detection. If the primary produced particles are squarks or gluinos (and assuming a negligible
lifetime of these particles), they will decay to final states with energetic jets and significant missing transverse momentum.
This final state can be produced in a large number of R-parity conserving models \cite{susy:Fayet,*susy:Fayet1},
in which squarks, $\squark$, and gluinos,  $\gluino$, can be produced in pairs as  $\gluino \gluino$, $\gluino \squark$,
or $\squark \squark$. They can decay via $\squark \to q \ninoone$ and $\gluino \to q \bar{q} \ninoone$ to weakly interacting
neutralinos, $\ninoone$. However, also charginos or heavier neutralinos might appear in the decay cascade and these particles
may produce high transverse momentum leptons in their decays into the LSP.

The ATLAS and CMS collaborations have carried out analyses with a lepton veto \cite{Aad:2011ib, CMS-1109.2352},
requiring one isolated lepton \cite{ATLAS:2011ad, CMS-PAS-SUS-11-015},
or requiring two or more leptons \cite{ATLAS_1110.6189, CMS-PAS-SUS-11-013}.
In addition, a dedicated search was performed for events with high jet
multiplicity with six or more jets \cite{ATLAS_JHEP11-2011-99}.
Data samples corresponding
to integrated luminosities between 1.0 and 1.3 fb$^{-1}$ were used. Events are triggered either on the presence of a jet
plus large missing momentum, or on the presence of at least one high-$p_T$ lepton. Backgrounds to the searches arise
from Standard Model processes such as vector boson production plus jets ($W$ + jets, $Z$ + jets), top quark pair production
and single top production, QCD multijet production, and diboson production. They are estimated in a semi-data-driven way,
using control regions in combination with a transfer factor obtained from simulation.
The results are interpreted in the MSUGRA/CMSSM model~\cite{msugra_cmssm}, and in particular as limits in the plane spanned by
the common scalar mass parameter at the GUT scale $m_0$ and the common gaugino mass parameter at the GUT scale
$m_{1/2}$, for values of the common trilinear coupling parameter $A_0$ = 0, Higgs mixing parameter $\mu$ > 0, and ratio of
the vacuum expectation values of the two Higgs doublets  $\tan \beta$ = 10. Figure~\ref{f:susy_01}~(left)
shows the results for the analyses of the ATLAS collaboration with
$\geq 2, \geq 3$ or $\geq 4$ jets plus missing transverse momentum, and the multijets plus missing momentum analysis. For a choice
of parameters leading to equal squark and gluino masses, squark and gluino masses below approximately 1~\TeV\ are
excluded. The 1-lepton and 2-lepton results are less constraining in MSUGRA/CMSSM for this choice of parameters,
but these analyses are complementary, and therefore no less important. The exclusion contours obtained by the CMS
collaboration in different final states, including the lepton channels, are shown in Fig.~\ref{f:susy_02}.

\begin{figure}[htp]
  \begin{center}
\includegraphics[width=0.48\textwidth]{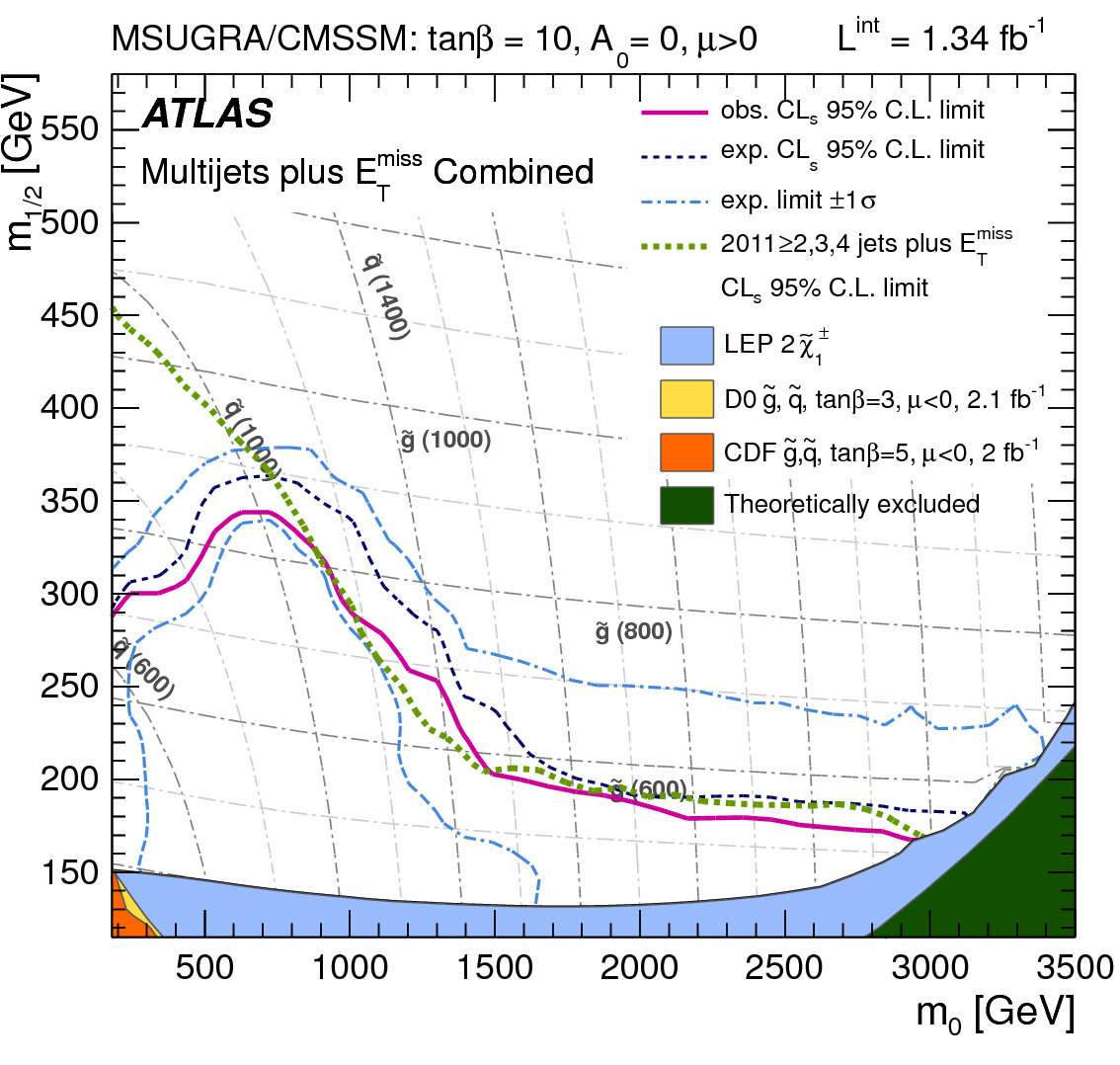}
\includegraphics[width=0.48\textwidth]{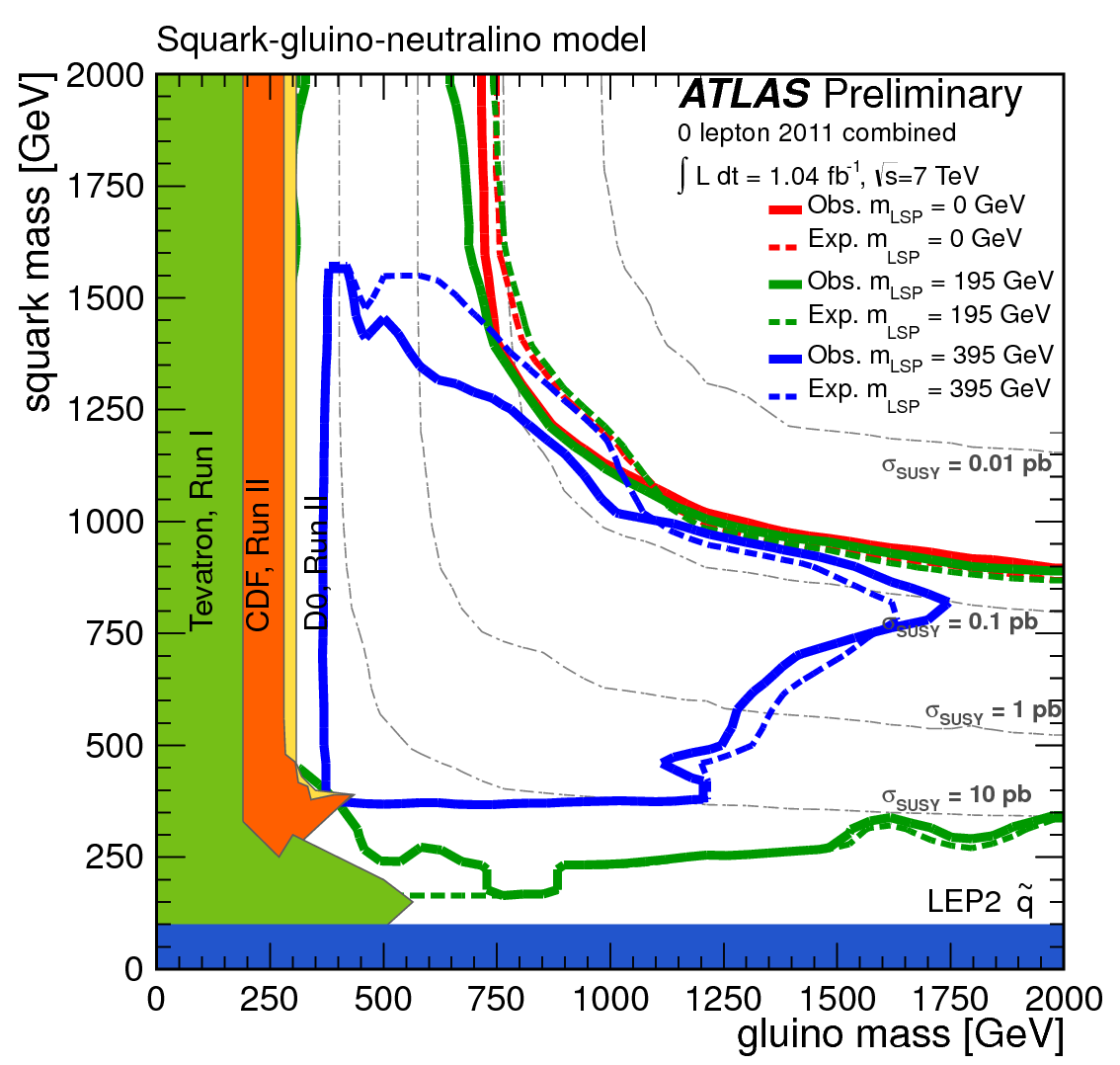}
\end{center}
\caption{\it\small
(Left): Exclusion contours in the MSUGRA/CMSSM ($m_0-m_{1/2}$)-plane for $A_0$ = 0, $\tan \beta$ = 10 and $\mu$ > 0, arising from the
analysis of the ATLAS collaboration with $\geq 2, \geq 3$ or $\geq 4$ jets plus missing transverse momentum, and the multijets
plus missing momentum analysis (from Ref.~\protect\cite{ATLAS_JHEP11-2011-99}).
(Right): Exclusion contours from the ATLAS analyses in the squark-gluino mass plane for three values of the LSP mass using
the simplified model description (see text) (from Ref.~\protect\cite{ATLAS-CONF-2011-155}).
}
\label{f:susy_01}
\end{figure}

\subsection{Simplified model interpretation}

\noindent The various analyses have also been interpreted in simplified models
assuming specific production and decay modes. The constraints implied by the MSUGRA/CMSSM models~\cite{msugra_cmssm}
are relaxed, leaving more freedom for the variation of particle masses and decay
modes. Interpretations in simplified models thus show better the limitations of the analyses as a function of the relevant
kinematic variables.

Inclusive search results with jets and missing momentum are interpreted using simplified models with either pair
production of squarks or of gluinos, or production of squark-gluino pairs. Direct squark decays
($\tilde{q} \rightarrow q\tilde{\chi}^0_1$) or direct gluino decays
($\tilde{g}\rightarrow q\tilde{q}\tilde{\chi}^0_1$) are dominant if all other particle masses have multi-\TeV\
values, so that those do not play a role. Using these assumptions, the excluded mass regions are sensitive to the
mass of the LSP ($\tilde{\chi}^0_1$). Figure~\ref{f:susy_01} (right) shows the ATLAS results interpreted in terms of limits on
(first and second generation) squark and gluino
masses, for three values of the LSP ($\tilde{\chi}^0_1$) mass, and assuming that all other SUSY particles
are very massive \cite{ATLAS-CONF-2011-155}.
Further interpretations are also done in terms of limits on gluino mass versus LSP mass assuming high squark masses,
or in terms of limits on squark mass vs LSP mass assuming large gluino
masses \cite{ATLAS:2011ad,ATLAS-CONF-2011-155}.

\begin{figure}[hbtn]
\begin{center}
\includegraphics[width=0.70\textwidth,angle=0]{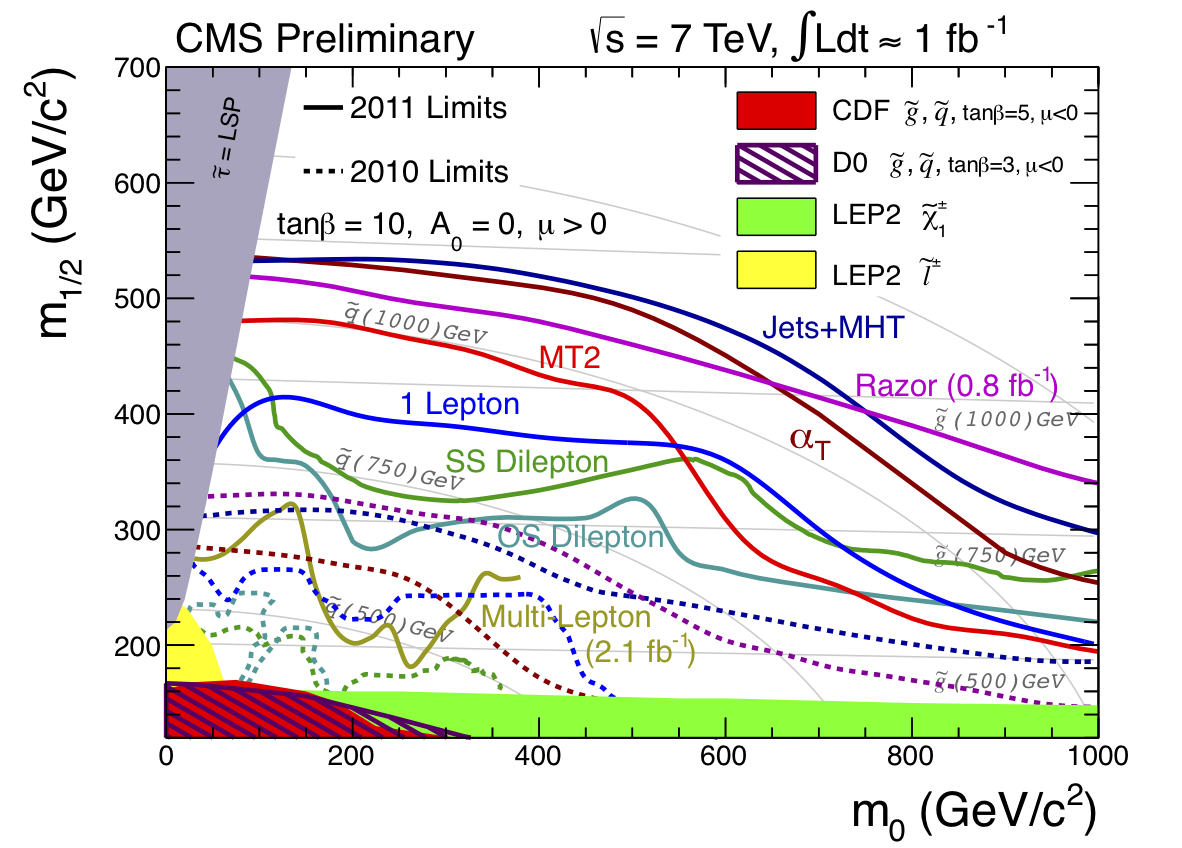}
\end{center}
\caption{\small \it
Summary of exclusion contours in the MSUGRA/CMSSM ($m_0-m_{1/2}$)-plane for the
parameters $A_0$ = 0, $\tan \beta$ = 10 and $\mu$ > 0 for various analyses and different final states from the CMS
collaboration (from Ref.~\protect\cite{CMS-PAS-SUS-11-015}).
}
\label{f:susy_02}
\end{figure}

The results of the inclusive jets plus missing momentum searches, interpreted in these simplified models, indicate that masses
of first and second generation squarks and of gluinos must be above approximately 750~\GeV. An important
caveat in this interpretation is the fact that this is only true for neutralino LSP masses below approximately
250~\GeV\ (as in MSUGRA/CMSSM~\cite{msugra_cmssm} for values of $m_{1/2}$ below $\sim$600~\GeV). For higher LSP masses, the squark and
gluino mass limits are significantly less restricting. It will be a challenge for further analyses to extend the sensitivity
of inclusive squark and gluino searches to the case of heavy neutralinos. If the LSP is heavy, events are characterized
by less energetic jets and less missing transverse momentum. This will be more difficult to trigger on, and
lead to higher Standard Model backgrounds in the analysis.

\subsection{Search for stop and sbottom production}

Important motivations for electroweak-scale supersymmetry are the facts that SUSY might provide a natural solution
to the hierarchy problem by preventing \lq unnatural\rq\ fine-tuning of the Higgs sector, and that the lightest stable
SUSY particle is an excellent dark matter candidate. It is instructive to consider what such a motivation really requires
from SUSY: a relatively light top quark partner (the stop, $\stop$ and an associated sbottom-left, $\sbottomL$), a gluino
not much heavier than about 1.5~\TeV\ to keep the stop light, given that it receives radiative corrections from loops
like $\stop \to \gluino t \to \stop$, and electroweak gauginos below the \TeV\
scale \cite{Barbieri-HCP}.
There are no strong constraints on first and second generation squarks and sleptons; in fact heavy squarks
and sleptons make it easier for SUSY to satisfy the strong constraints from flavour physics.

Motivated by these considerations, the ATLAS and CMS collaborations have also carried out a number of searches for supersymmetry
with $b$-tagged jets, which are sensitive to sbottom and stop production, either in direct production or in
production via gluino decays. Jets are tagged as originating from $b$-quarks by an algorithm that exploits both track
impact parameter and secondary vertex information.

Direct sbottom pair production is searched for in a data sample corresponding to an integrated luminosity of
2 fb$^{-1}$ by requiring two $b$-tagged
jets with $p_T$ > 130, 50~\GeV\ and significant missing transverse momentum of more than 130~\GeV\ \cite{ATLAS-1112.3832}.
The final discriminant in the ATLAS analysis is the boost-corrected contransverse mass $m_{CT}$ \cite{contransverse-mass},
and signal regions with $m_{CT}$ > 100, 150, 200~\GeV\ are considered.
No excesses are observed above the expected backgrounds from top, $W$+heavy flavour and $Z$+heavy flavour production.
Figure~\ref{f:susy_sbottom} (left) shows the resulting limits in the sbottom-neutralino mass plane, assuming sbottom pair production
and sbottom decays into a $b$-quark plus a neutralino (LSP) with a 100\% branching fraction. Under
these assumptions, sbottom masses up to 390~\GeV\ are excluded for neutralino masses below 60~\GeV.

The ATLAS collaboration has searched for stop quark production in gluino decays \cite{ATLAS-CONF-2011-130}
using an analysis requiring at least four
high-$p_T$ jets of which at least one should be $b$-tagged, one isolated lepton, and significant missing transverse momentum.
Since the number of observed events agrees with the expectations from Standard Model processes,
limits are set in the gluino-stop mass plane, assuming the gluino to decay as $\tilde{g}\rightarrow \tilde{t}t$,
and the stop quark to decay as $\tilde{t} \rightarrow b\tilde{\chi}^{\pm}_1$. The obtained mass limits are shown in
Fig.~\ref{f:susy_sbottom} (right).

\begin{figure}[hbtn]
\begin{center}
\includegraphics[width=0.49\textwidth,angle=0]{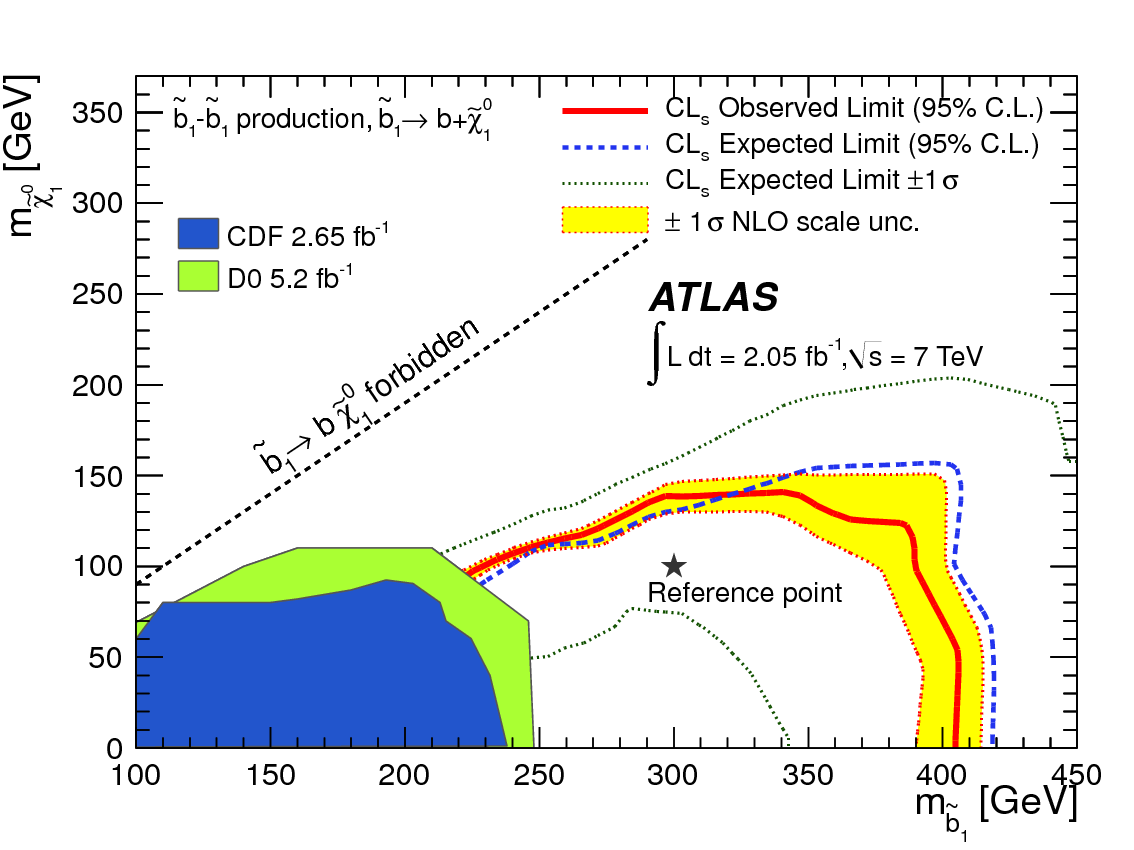}
\includegraphics[width=0.49\textwidth,angle=0]{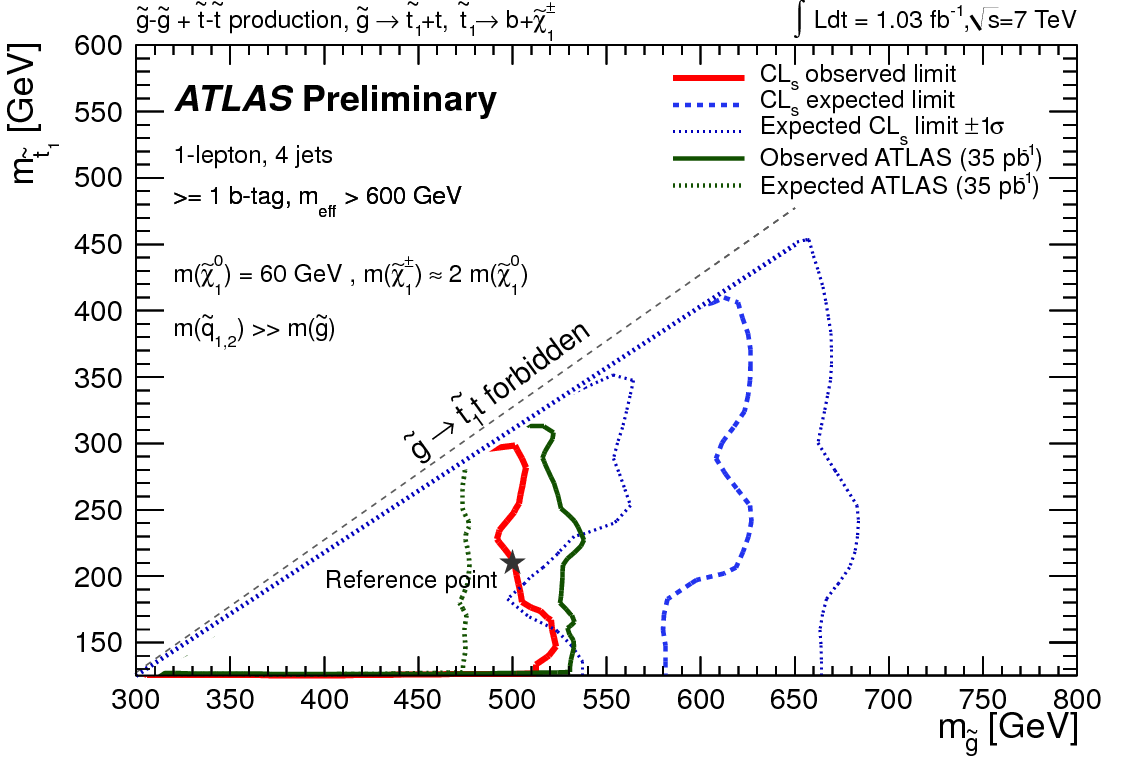}
\end{center}
\caption{\small \it
(Left): Exclusion contours from the ATLAS analyses in the sbottom-neutralino mass plane resulting from the analysis searching for sbottom
pair production assuming $\tilde{b}_1 \to b\tilde{\chi}^{0}_1$ decays (from Ref.~\protect\cite{ATLAS-1112.3832}).
(Right): Exclusion contours from the ATLAS analyses in the gluino-stop mass plane resulting from the analysis searching for stop
production via gluino decays. The assumptions made to derive the limits are given in the figure
(from Ref.~\protect\cite{ATLAS-CONF-2011-130}).
}\label{f:susy_sbottom}
\end{figure}

Further searches for direct stop pair production are in progress. These searches are challenging due to the
similarity with the top-quark pair-production final state for stop masses similar to the top mass, and due to the
low cross section for the production of stops with high mass.
The ATLAS collaboration has searched for signs of new phenomena in events passing
a top-quark pair selection with large missing transverse momentum \cite{ATLAS-1109.4725}. Such an analysis
is sensitive to pair production of massive partners of the top quark, decaying to a top quark and a long-lived
undetected neutral particle. No excess above background was observed, and limits on the cross section for pair production
of top quark partners are set. These limits constrain fermionic exotic fourth generation quarks, but not
yet scalar partners of the top quark, such as the stop quark \cite{ATLAS-1109.4725}.

\subsection{Search for supersymmetry in multilepton final states}

The search for final states with several leptons and missing transverse momentum are sensitive to the production
of charginos and/or heavier neutralinos (other than the LSP), decaying
leptonically into the LSP. These analyses comprise the golden search modes at the Tevatron, but are also rapidly
gaining relevance at the LHC and both the ATLAS and CMS collaborations have performed corresponding analyses
\cite{ATLAS_1110.6189, CMS-PAS-SUS-11-013}. The ATLAS collaboration has published results of various analyses searching for dilepton
events plus missing momentum in data corresponding to an integrated luminosity of 1.0 $\ifb$ \cite{ATLAS_1110.6189}.
Three searches are performed for new phenomena in final states with opposite-sign and same-sign
dileptons and missing transverse momentum. These searches also include signal regions that place requirements
on the number and  $\pt$ of energetic jets in the events. For all signal regions good agreement is found between
the numbers of observed events and the predictions of expected events from Standard Model processes.
Additionally, in opposite-sign events, a search is made for an excess of same-flavour over different-flavour
lepton pairs. Effective production cross sections in excess of 9.9 fb for opposite-sign events
with missing transverse momentum greater than 250~\GeV\ are excluded at 95\% C.L. For same-sign events
with missing transverse momentum greater than 100~\GeV, effective production cross sections in excess of 14.8 fb are
excluded at 95\% C.L. The latter limit is interpreted in a simplified electroweak gaugino production model excluding
chargino masses up to 200~\GeV, under the assumption that slepton decays are dominant \cite{ATLAS_1110.6189}.

The CMS collaboration has presented preliminary results, based on data corresponding to an integrated luminosity of
2.1 $\ifb$, on the search for supersymmetric particles in three- and four-lepton
final states, including hadronic decays of $\tau$ leptons \cite{CMS-PAS-SUS-11-013}. The backgrounds from Standard Model processes
are suppressed by requiring missing transverse energy, Z-mass vetos of the invariant dilepton mass
or high jet activity. Control samples in data are used to obtain reliable background estimates.
Within the statistical and systematic uncertainties the numbers of observed events are consistent with the expectations from
Standard Model processes. These results are used to exclude previously unexplored regions
of the supersymmetric parameter space assuming R-parity conservation with the lightest
supersymmetric particle being  a neutralino. The corresponding exclusion contours in the MSUGRA/CMSSM~\cite{msugra_cmssm}
interpretation are shown in Fig.~\ref{f:susy_multilepton} in the ($m_0-m_{1/2}$) plane for $A_0$ = 0, $\mu$ > 0, and
for $\tan \beta$ values of 3 and  10.
They extend significantly the regions excluded by the CDF \cite{CDF-SUSY-multileptons}
and \dzero \cite{D0-SUSY-multileptons} experiments and those excluded with previous searches at the
LHC \cite{CMS:1106.0933}.

\begin{figure}[hbtn]
\begin{center}
\includegraphics[width=0.48\textwidth,angle=0]{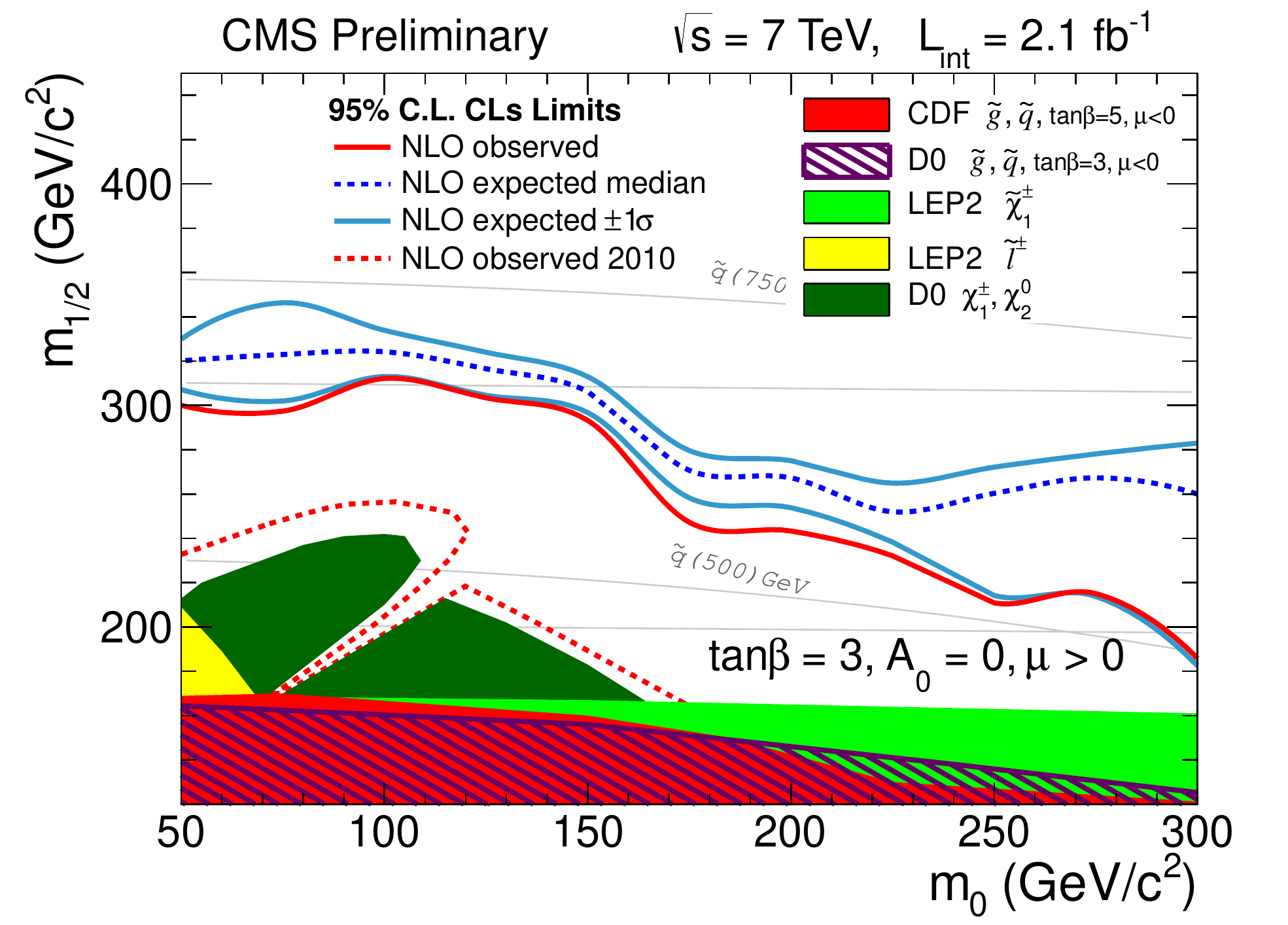}
\includegraphics[width=0.48\textwidth,angle=0]{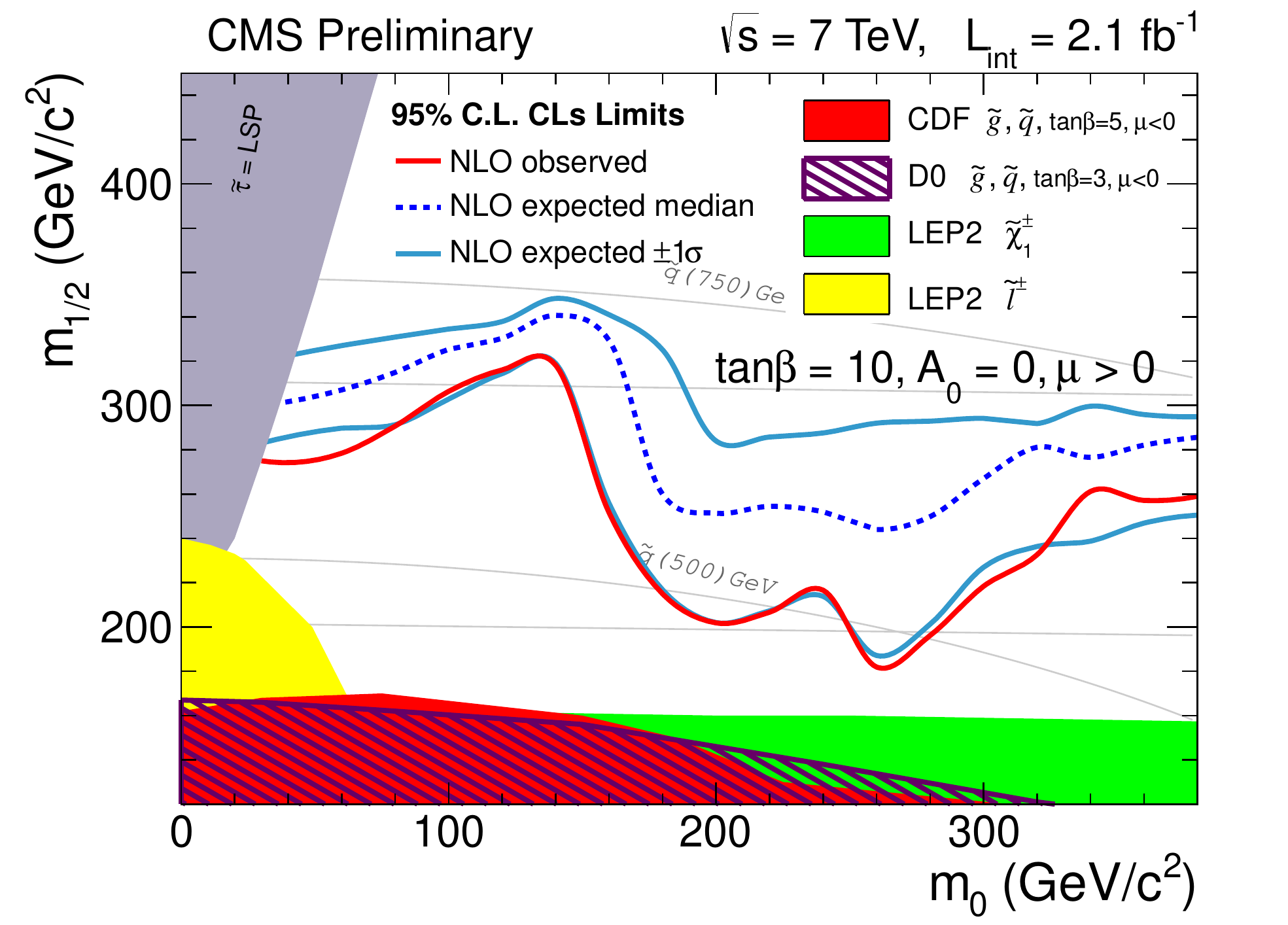}
\end{center}
\caption{\small \it
Exclusion contours in the MSUGRA/CMSSM ($m_0-m_{1/2}$)-plane for the
parameters $A_0$ = 0, $\mu$ > 0 and $\tan \beta$ = 3 (left) and $\tan \beta$ = 10 (right) obtained from searches for
SUSY production in final states with multileptons by the CMS collaboration (from Ref.~\protect\cite{CMS-PAS-SUS-11-013}).
}
\label{f:susy_multilepton}
\end{figure}

\subsection{Summary and outlook on SUSY searches}

\noindent Many different searches for the production of supersymmetric
particles have been performed in a large variety of final states by the ATLAS and CMS collaborations at the LHC.
Data corresponding to integrated luminosites in the range between 1.0 and 4.7 $\ifb$ taken during the year 2011
have been analyzed. In all channels, the number of observed events is in agreement with the expectations from Standard
Model processes and no evidence for the production of supersymmetric particles has been found so far. The data
have been used to set already rather strong limits on the masses of possible supersymmetric particles. A summary of the
most important mass limits is given in Fig.~\ref{f:susy_summary}.

In addition to the analyses summarized here, many other analyses have been performed and many different
final states have been explored. There are investigations of SUSY searches in gauge mediated supersymmetry breaking
models, by using final states with photons or multileptons. In addition, in many models (split SUSY, R-hadrons,
anomaly-mediated SUSY breaking and in certain parts of the phase space of gauge-mediated SUSY breaking scenarios)
SUSY particles may be long-lived either because their decay is kinematically suppressed or due to very small
couplings, e.g. in R-parity violating models. Many of these scenarii have already been explored and the reader is referred
to the corresponding publications of the ATLAS and CMS collaborations. Also in all these searches for more exotic
SUSY scenarios the number of observed events in in agreement with the expectations from background from Standard Model
processes.

\begin{figure}[hbtn]
\begin{center}
\includegraphics[width=0.57\textwidth,angle=0]{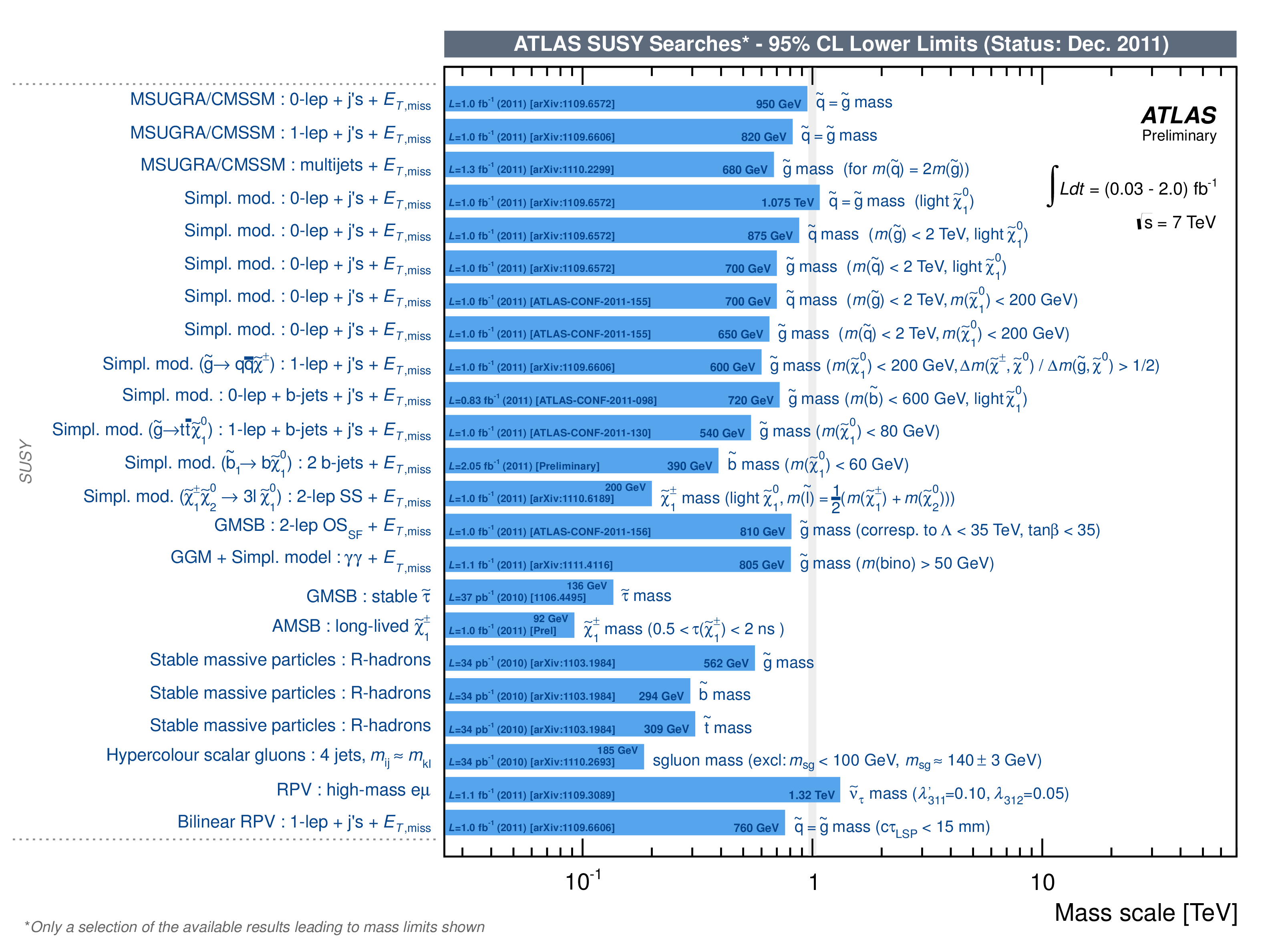}
\includegraphics[width=0.41\textwidth,angle=0]{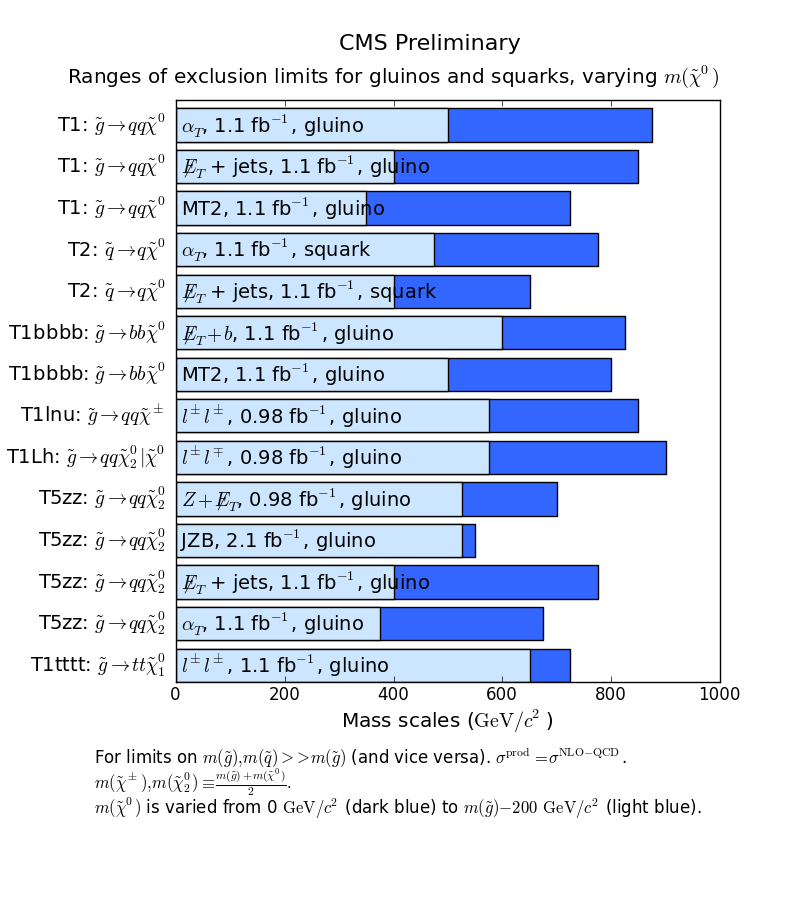}
\end{center}
\caption{\small \it
Summary of excluded mass ranges from a variety of searches for the production of supersymmetric particles from the ATLAS (left) and
CMS (right) collaborations.
Only a representative selection of available results is shown. The CMS results indicate the change of the limits under variation of
the neutralino mass from 0~to~200~\GeV.
}
\label{f:susy_summary}
\end{figure}

Although no signs of supersymmetry have been found so far, it is important to realize that actual tests
of \lq natural\rq~supersymmetry are only just beginning.
In this respect, the LHC run of 2012, with an expected luminosity of more than 10 fb$^{-1}$,
possibly at $\sqrt{s}$ = 8~\TeV, will be very important. However, experimentally there will be considerable challenges
in triggering and in dealing with high pile-up conditions. In the longer term, increasing the LHC beam energy to > 6~\TeV\
will again enable the crossing of kinematical barriers and open the way for multi-\TeV\ SUSY searches.

\section{Search for other Physics Scenarios Beyond the Standard Model}

As already mentioned in Section 2, the Standard Model is an
extremely successful effective theory which has been extensively
tested over the past forty years. However, a number of fundamental questions are left unanswered.
Many models for physics Beyond the Standard Model (BSM) have been proposed and the
ATLAS and CMS experiments have used the data collected in 2010 and 2011 to search for
indications of new physics. An impressive list of analyses has been performed. So far, no
indications for deviations from the Standard Model have been found. The event numbers
and kinematical distributions in all final states
considered agree with the expectations from Standard Model processes. Therefore, these
analyses have been used to constrain the parameter space of many BSM models.

Since it is impossible to present and discuss all analyses in such a summary paper, a few
benchmark processes are selected and the search results are presented
in the following. This concerns the search for
new vector bosons, or more general the search for heavy dilepton resonances, the search
for compositeness and the search for dijet resonances.
Finally the results from other searches are briefly summarized.

\subsection{Search for heavy dilepton resonances}

The ATLAS and CMS collaborations have performed searches for
narrow high-mass neutral and charged resonances decaying into $e^+e^-$ or $\mu^+\mu^-$ pairs
or $e  \nu$ or $\mu \nu$, respectively.
In several extensions of the Standard Model new heavy
spin-1 neutral gauge bosons such as $Z'$~\cite{RevModPhys.81.1199,Erler:2009jh,PhysRevD.34.1530},
technimesons~\cite{Lane1989274,PhysRevD.67.115011,PhysRevD.79.035006}, as well as spin-2 Randall-Sundrum
gravitons, $G^*$,~\cite{PhysRevLett.83.3370} are predicted. Additional heavy charged gauge bosons appear e.g.
in left-right-symmetric models~\cite{lrsm}.

The benchmark models considered in the analyses for the $Z'$ are the
Sequential Standard Model~\cite{RevModPhys.81.1199}, with the same couplings to fermions as the $Z$ boson, and the $E_6$ grand unified
symmetry group~\cite{PhysRevD.34.1530}, broken into $SU (5)$ and two additional $U(1)$ groups, leading to new neutral gauge fields
$\psi$ and $\chi$. The particles associated with the additional fields can mix in a linear combination to form the $Z'$
candidate: $ Z'(\theta_{E_6})=Z_{\psi}'\cos\theta_{E_6} + Z_{\chi}^{'} \sin\theta_{E_6}$ , where $\theta_{E_6}$ is the mixing angle
between the two gauge bosons. The pattern of spontaneous symmetry breaking and the value of $\theta_{E_6}$ determine the $Z'$
couplings to fermions.

Other models predict additional spatial dimensions as a possible explanation for the gap between the electroweak
symmetry breaking scale and the gravitational energy scale. The Randall-Sundrum (RS) model~\cite{PhysRevLett.83.3370} predicts
excited Kaluza-Klein modes of the graviton, which appear as spin-2 resonances. These modes have a narrow
intrinsic width when $k/\MPl$ < 0.1, where $k$ is the spacetime curvature in the extra dimension, and
$\MPl = M_{\rm{Pl}}/\sqrt{8\pi}$ is the reduced Planck scale.

The search performed by the ATLAS experiment \cite{ATLAS:1108.1582} is based on a dataset corresponding to an integrated luminosity of
up to 1.2 $\ifb$. The observed invariant mass spectrum is shown in Fig.~\ref{f:Z-prime_ATLAS} (left) for the $e^+e^-$
final state after final selections.
The backgrounds from Drell-Yan, \ttbar, diboson and $W$+jets production are determined from Monte
Carlo simulation after normalization to the respective (N)NLO cross sections. The background
from QCD multijet production is estimated using data-driven methods with the inversion of lepton identification
criteria. The simulated backgrounds are rescaled so that the total sum of the backgrounds
matches the observed number of events observed in data in the 70-110~\GeV\ mass interval.
The scaling factor is within 1\% of unity. The advantage of this approach is that the uncertainty
on the luminosity and any mass independent uncertainties on efficiencies,
cancel between the $Z'$ ($G^*$) and the $Z$ boson. The dilepton invariant mass distributions
are well described by the prediction from Standard Model processes. Figure~\ref{f:Z-prime_ATLAS} (left)
also displays the expected $Z'$ signals in the Sequential Standard Model for three mass hypotheses.

\begin{figure}[hbtn]
\begin{center}
\includegraphics[width=0.51\textwidth,angle=0]{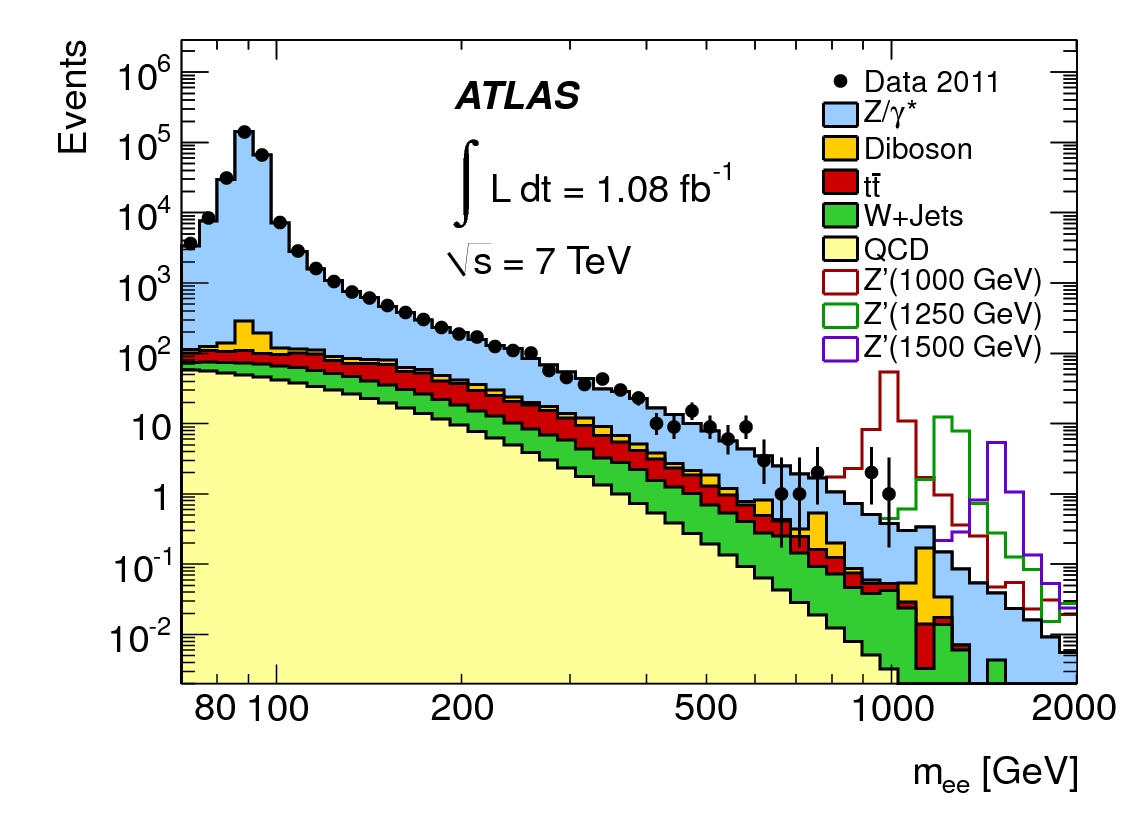}
\includegraphics[width=0.38\textwidth,angle=0]{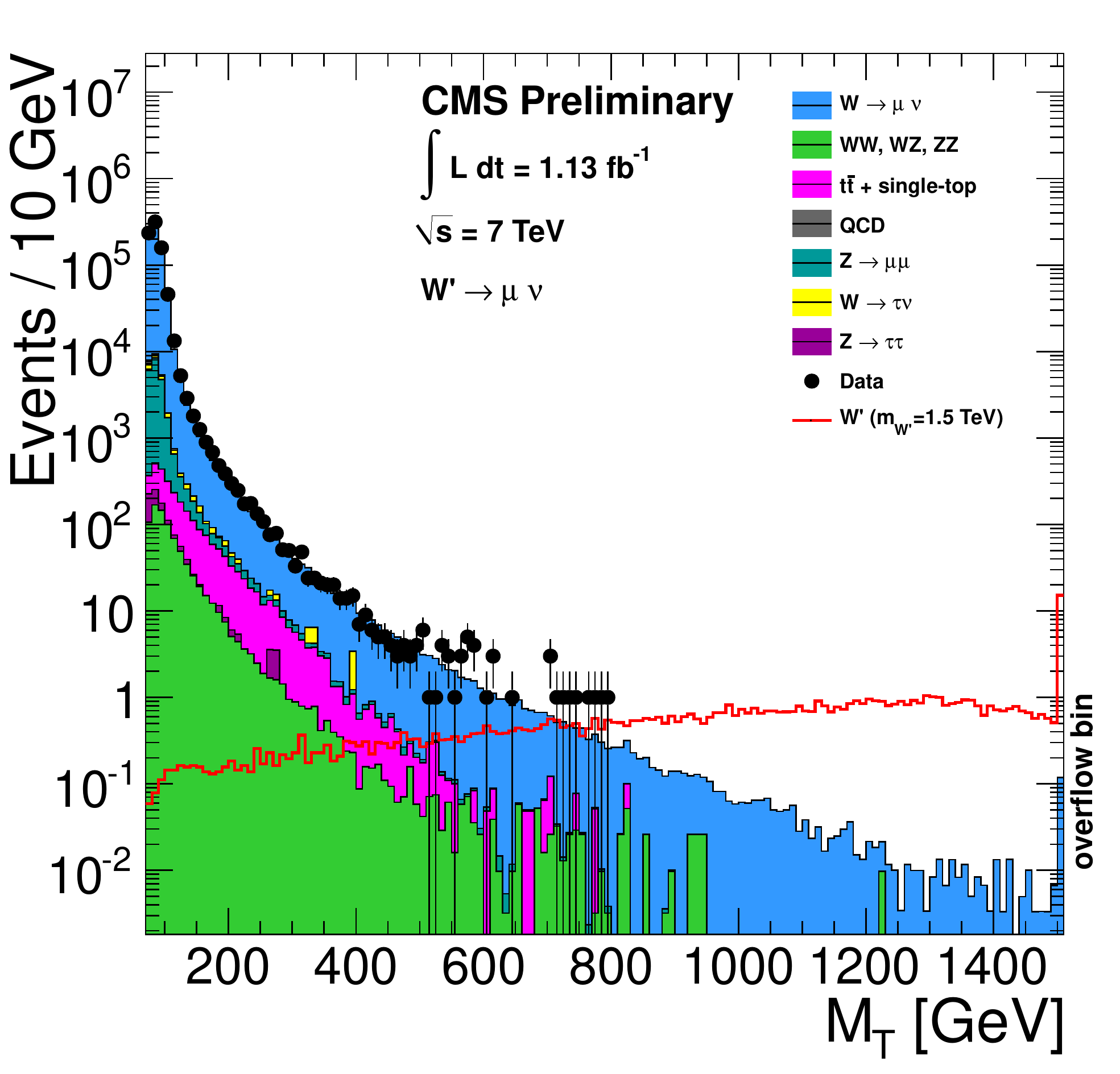}
\end{center}
\caption{\small \it
(Left): Distribution of the dielectron invariant mass after final selections in the ATLAS experiment, compared to the
stacked sum of all expected backgrounds, with three example
$Z_{\rm{SSM}}^{'}$ signals overlaid. The bin width is constant in $\log m_{\ell \ell}$
(from Ref.~\protect\cite{ATLAS:1108.1582}).
(Right): Distribution of the $\mu-\ETmiss$ transverse mass after final selections in the CMS experiment. The expected
signal from a hypothetical $W'$ boson with a mass of 1.5~\TeV\ is superimposed
(from Ref.~\protect\cite{CMS-PAS-EXO-11-024}).
}
\label{f:Z-prime_ATLAS}
\end{figure}

Given the good agreement between the data and the Standard Model expectations,
limits are set on the cross section times branching ratio for the different $Z'$ models. The resulting mass limits are
1.83~\TeV\ for the Sequential Standard Model $Z'$ boson, 1.49-1.64~\TeV\ for various $E_6$-motivated $Z'$ bosons,
and 0.71-1.63~\TeV\ for a Randall-Sundrum graviton with couplings ($k/\MPl$)
in the range 0.01-0.1. Similar analyses have been performed by the CMS collaboration \cite{CMS-PAS-EXO-11-019}
and comparable limits have been extracted. They are included in the summary of results from different experiments
for various physics models in Table~\ref{t:Z-prime}.

The benchmark model considered in the search for the $W'$ is the
Sequential Standard Model \cite{RevModPhys.81.1199}, with the same couplings to fermions as the $W$ boson. In this case the
transverse mass of the lepton and $\ETmiss$ system is used as discriminating variable. As an example,
the measured transverse mass distribution
in the muon final state in the CMS experiment \cite{CMS-PAS-EXO-11-024}
is shown in Fig.~\ref{f:Z-prime_ATLAS} (right). The expectation for a $W'$ signal with a mass of 1.5~\TeV\ is superimposed.
Also the transverse mass distributions measured by the LHC experiments are well described by the prediction from
Standard Model processes and the data
allow to exclude heavy $W'$ bosons with masses below 2.15~\TeV\  (ATLAS)~\cite{ATLAS:1108.1316}
and 2.25~\TeV\  (CMS)~\cite{CMS-PAS-EXO-11-024} at the 95\% C.L.

The mass limits obtained at the LHC are the most stringent to date, including indirect limits set by LEP2. It is striking to see
how fast the LHC experiments have superseded the limits obtained with much higher luminosity at the Tevatron. The
analyses based on the data from 2010 ($L_{\rm{int}}$ = 36~\ipb) resulted in comparable limits to those obtained at the Tevatron
based on an integrated luminosity of 5.5~\ifb\ (see Table~\ref{t:Z-prime}).

\begin{table}
\begin{small}
\begin{center}
\caption{\small \it
Observed 95\% C.L. mass lower limits on $Z'$, $G^*$ gravitons
and $W'$ resonances obtained for various models in the ATLAS and CMS experiments. The results from searches at the
Tevatron are included for comparison.}
\begin{tabular}{llc|ccc|l}
            \hline
Model       & \rule[-3mm]{0mm}{8mm}{Experiment} & $L_{\rm{int}}$ & \multicolumn{3}{c|} {95\% C.L. limits} & Ref.  \\
              &  &  & $e^{+}e^{-}$ & $\mu^{+}\mu^{-}$ & $\ll$  \\
                        &    &  (\ifb)       & (\TeV)      & (\TeV)       &  (\TeV) &  \\
\hline
              $Z^{'}_{SSM}$ & CDF/$\dzero$    & 5.5      &          &       &  1.07  & \protect\cite{dzerotab1,*cdftab1} \\
                    & ATLAS/CMS  & 0.036     &  0.96    & 0.83  &  1.05/1.14   &     \protect\cite{ATLAStab1,*CMStab1}    \\
                     & ATLAS  &1.1 / 1.2 &  1.70    & 1.61  &  1.83  &   \protect\cite{ATLAS:1108.1582}     \\
                     & CMS    & 1.1      &          &       &  1.94  &   \protect\cite{CMS-PAS-EXO-11-019}  \\
\hline
             $Z^{'}$ $E_6$ models    & ATLAS  & 1.1 / 1.2 &     &   &  1.49 - 1.64 &    \protect\cite{ATLAS:1108.1582}   \\
                                     &CMS    &  1.1       &     &   &   1.62       &     \protect\cite{CMS-PAS-EXO-11-041}  \\
\hline
              $G^*$  $ k/\MPl$ = 0.01   &ATLAS  & 1.1 / 1.2   &     &     &  0.71  &   \protect\cite{ATLAS:1108.1582}   \\
                     $G^*$  $ k/\MPl$ = 0.03  & & 1.1 / 1.2   &     &     &  1.03  &     \\
                     $G^*$  $ k/\MPl$ = 0.05  & & 1.1 / 1.2   &     &     &  1.33  &     \\
                     $G^*$  $ k/\MPl$ = 0.10 &  & 1.1 / 1.2   &     &     &  1.63  &     \\
              $G^*$  $ k/\MPl$ = 0.05   &CMS    & 1.1         &     &     &  1.45  &   \protect\cite{CMS-PAS-EXO-11-019}  \\
                    $G^*$  $ k/\MPl$ = 0.10   & & 1.1         &     &     &  1.78  &    \\
\hline
              $W^{'}_{SSM}$       &ATLAS  & 1.04   &  2.08    & 1.98  &  2.15  &   \protect\cite{ATLAS:1108.1316}   \\
                                &CMS    &  1.1    &          &       &  2.27  &   \protect\cite{CMS-PAS-EXO-11-024}   \\
\hline
\end{tabular}
\label{t:Z-prime}
\end{center}
\end{small}
\end{table}


\subsection{Limits on new physics from jet production}

The measurements on inclusive and dijet production, as discussed in Section \ref{s:QCD_jets},
can also be used to constrain contributions from new physics that would
modify the expected QCD behaviour in the jet production cross sections. Two examples are discussed in the following.

\subsubsection{Substructure of quarks}

Both collaborations have searched for quark compositeness by investigating the angular
distribution of jet events \cite{ATLAS:1103.3864, CMS-compositeness}.
At small scattering angles in the centre-of-mass system of the two partons, the angular
distribution is expected to be proportional to the Rutherford cross section,
$d \hat{\sigma}/d \cos \theta^* \sim 1 / (1-\cos \theta^*)^2$.
For the scattering of massless partons, which are assumed to be collinear with the beam protons,
the longitudinal boost of the parton-parton centre-of-mass frame with respect to the proton-proton
centre-of-mass frame, $y_{\rm{boost}}$, and $\theta^*$ are obtained from the rapidities
$y_1$ and $y_2$ of the jets from the two
scattered partons by $y_{\rm{boost}} = \frac{1}{2} (y_1 + y_2)$
 and $| \cos \theta^*| = \tanh  y^*$, where $y^* = \frac{1}{2}  | y_1 - y_2 |$
and where $\pm y^*$ are the rapidities of the two jets in the parton-parton centre-of-mass
frame.
The variable $\chi_{\rm{dijet}} = e^{2y^*}$ is used to measure the dijet
angular distribution, which for collinear massless-parton scattering takes the form
$\chi_{\rm{dijet}} = ( 1 + | \cos \theta^* |) / ( 1 - | \cos \theta^* |)$.
This choice of $\chi_{\rm{dijet}}$, rather than $\theta^*$, is motivated by the fact that
$d {\sigma_{\rm{dijet}}} /d \chi_{\rm{dijet}}$ is flat for Rutherford scattering.

\begin{figure}[hbtn]
\begin{center}
\includegraphics[width=0.49\textwidth,angle=0]{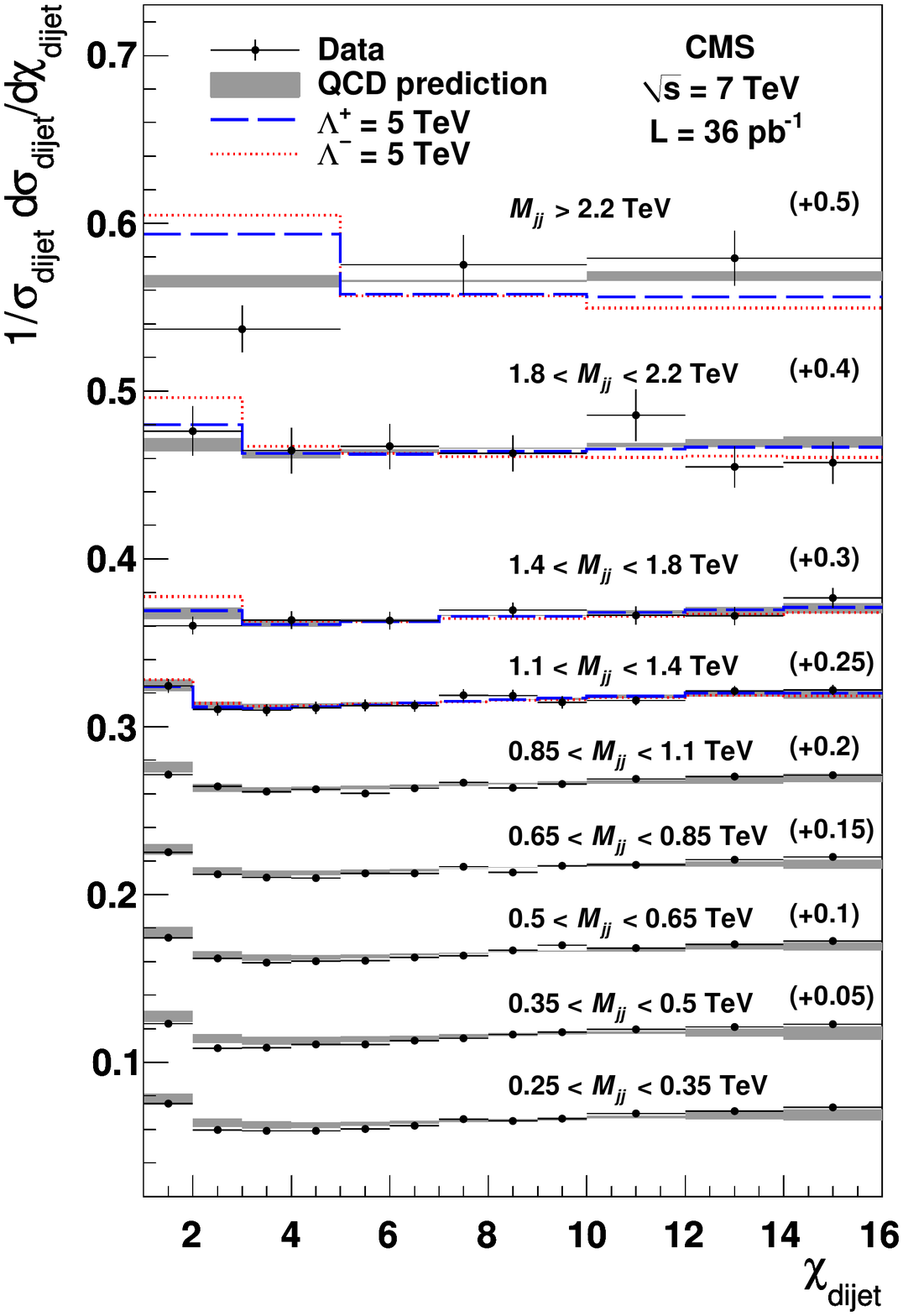}
\includegraphics[width=0.49\textwidth,angle=0]{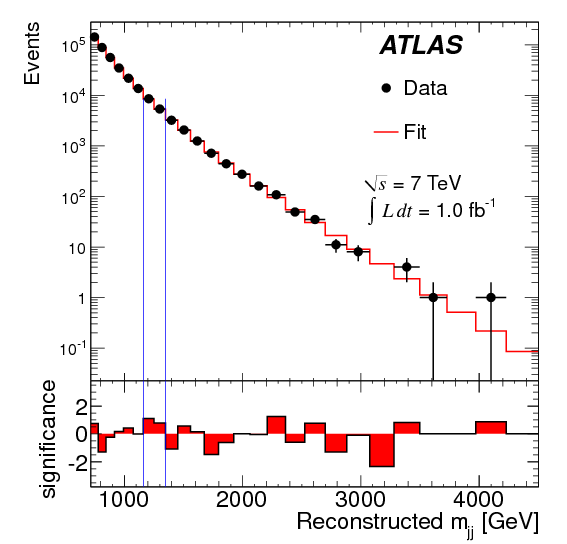}
\end{center}
\caption{\small \it
(Left):
Normalized dijet angular distributions in several ranges of the dijet mass as measured by the CMS collaboration. The data points include statistical and systematic uncertainties. The results are compared with the predictions of pQCD at NLO (shaded bands) and with the predictions including a contact interaction term of compositeness scale $\Lambda^{+}$ = 5~\TeV\ (dashed histogram) and $\Lambda^{-}$ = 5~\TeV\ (dotted histogram). The shaded bands show the effect on the NLO pQCD predictions due to $\mu_R$ and $\mu_F$ scale variations and PDF uncertainties, as well as the uncertainties from the non-perturbative corrections added in quadrature (from Ref.~\protect\cite{CMS-compositeness}).
(Right): The reconstructed dijet mass distribution (filled points) measured by the ATLAS collaboration fitted with a smooth functional form describing the QCD background. The bin-by-bin significance of the data-background difference is shown in the lower panel. Vertical lines show the most significant excess found (from Ref.~\protect\cite{ATLAS-compositeness}).
}
\label{f:exotics_dijet}
\end{figure}

The differential dijet angular distributions for different $m_{\rm{jj}}$ ranges,
and corrected for detector effects as measured by the CMS experiment using the 2010 data ($L_{\rm{int}}$ = 36~\ipb)
are shown in Fig.~\ref{f:exotics_dijet} (left). The data are found to be in good agreement with
pQCD predictions at NLO calculated with NLOJET++ \cite{NLO_jets,NLO_jets2}, which are superimposed on the figure.
The measured dijet angular distributions can be used to set limits on quark compositeness
parametrized by a four-fermion contact interaction term in addition to the QCD Lagrangian.
The value of the mass scale $\Lambda$ characterizes the strengths of the quark substructure binding
interactions and the physical size of the composite states. A color- and isospin-singlet contact
interaction (CI) of left-handed quarks gives rise to an effective Lagrangian
term \cite{eichten1, eichten2}
\begin{equation}
  L_{qq} = \eta_0 \frac{2\pi}{\Lambda^2} (\bar{q}_L\gamma^{\mu}q_L)(\bar{q}_L\gamma_{\mu}q_L),
\end{equation}
where $\eta_0$ = +1 corresponds to destructive interference
between the QCD and the new physics term, and $\eta_0$ = -1 to constructive interference.
From the measured $\chi_{\rm{dijet}}$ distribution, lower limits on the contact interaction scale
of $\Lambda^+$ = 5.6~\TeV\ and $\Lambda^-$ = 6.7~\TeV\ for destructive and constructive interference, respectively,
have been set by the CMS collaboration at the 95\% C.L.~\cite{CMS-compositeness}.
The expected limits in case of no substructure are
5.0~\TeV\ and 5.8~\TeV, respectively.
The ATLAS collaboration has performed a similar analysis and excludes at the 95\% C.L. quark contact interactions
with a scale $\Lambda <$ 9.5~$\TeV$~\cite{ATLAS:1103.3864}. However, it should be noted that this observed limit is
significantly above the expected limit of 5.7~\TeV\ for the data sample corresponding to an integrated luminosity
of 36~\ipb.
Very recently, the CMS collaboration has published the results of an updated analysis based on data
corresponding to an integrated luminosity of 2.2~$\ifb$~\cite{CMS:1202.5535} and taking NLO calculations
for the QCD predictions into account.
Also this larger data set has been found to be in good agreement with the QCD expectations.
For the contact interaction model described above, 95\% C.L. limits of $\Lambda^+$ = 7.5~\TeV\
and $\Lambda^-$ = 10.5~\TeV\ have been set. The expected limits are 7.0~\TeV\ and 9.7~\TeV, respectively.

\subsubsection{Dijet resonances}

The ATLAS and CMS collaborations have also examined the dijet mass spectrum for resonances due to new phenomena localised
near a given mass, employing data-driven background estimates that do not rely on detailed
QCD calculations \cite{ATLAS-compositeness,CMS:1107.4771}.
The searches are based on data corresponding to an integrated luminosity of
1.0~fb$^{-1}$. As an example, the observed dijet mass distribution measured in the ATLAS experiment,
which extend up to masses of $\sim4$~\TeV\ is displayed in Fig.~\ref{f:exotics_dijet} (right).
It is found to be in good agreement with
a smooth function representing the Standard Model expectation. Since no
evidence for the production of new resonances is found,
95\% C.L. mass limits  have been set in the context of several models of new physics:
excited quarks (q$^*$)~\cite{excitedquarks,excitedquarks1}, axigluons~\cite{axigluon,axigluon1,axigluon2},
scalar colour octet states~\cite{coloroctet} and scalar diquarks predicted in Grand Unified Theories based on
the $E_6$ gauge group~\cite{GUT-E6}.
The results are summarized in
Table~\ref{t:dijet-resonances}. Also these limits are the most stringent ones to date.

\begin{table}
\begin{small}
\caption{ \small \it    The 95\% C.L. mass lower limits on dijet resonance models.}
\begin{center}
\begin{tabular}[h]{lllccl}
  \hline
Model       & Experiment & $L_{\rm{int}}$ & \multicolumn{2}{c} {95\% C.L. limits} & Ref.  \\
            &            &   ($\ifb$)      & Expected (\TeV)  & Observed (\TeV)    & \\ \hline
Excited quark $q*$ & ATLAS     & 1.0  & 2.81 & 2.99 &  \protect\cite{ATLAS-compositeness} \\
                   & CMS       & 1.0  & 2.68 & 2.49 &  \protect\cite{CMS:1107.4771} \\
Axigluon           & ATLAS     & 1.0  & 3.07 & 3.32 &  \protect\cite{ATLAS-compositeness} \\
                   & CMS       & 1.0  & 2.66 & 2.47 &  \protect\cite{CMS:1107.4771} \\
Colour Octet Scalar & ATLAS    & 1.0  & 1.77 & 1.92 &  \protect\cite{ATLAS-compositeness} \\
$E_6$ diquarks      & CMS      & 1.0  & 3.28 & 3.52 &  \protect\cite{CMS:1107.4771} \\
\hline
\end{tabular}
\label{t:dijet-resonances}
\end{center}
\end{small}
\end{table}

\subsection{Summary of results on other searches}

\noindent Many different searches for the Beyond the Standard Model processes have been performed
by the ATLAS and CMS collaborations at the LHC.
Data corresponding to integrated luminosities in the range between 1.0 and 4.7 $\ifb$ taken during the year 2011
have been analyzed and many different final states have been investigated.  So far, no
indications for deviations from the Standard Model have been found. The event numbers
and kinematical distributions in all final states
considered agree with the expectations from Standard Model processes. Therefore, these
analyses have been used to constrain the parameter space of many BSM models. A summary of the
most important limits from the ATLAS collaboration is given in Fig.~\ref{f:exotics_summary}.
Comparable limits have been set by the CMS collaboration.

\begin{figure}[hbtn]
\begin{center}
\includegraphics[width=0.98\textwidth,angle=0]{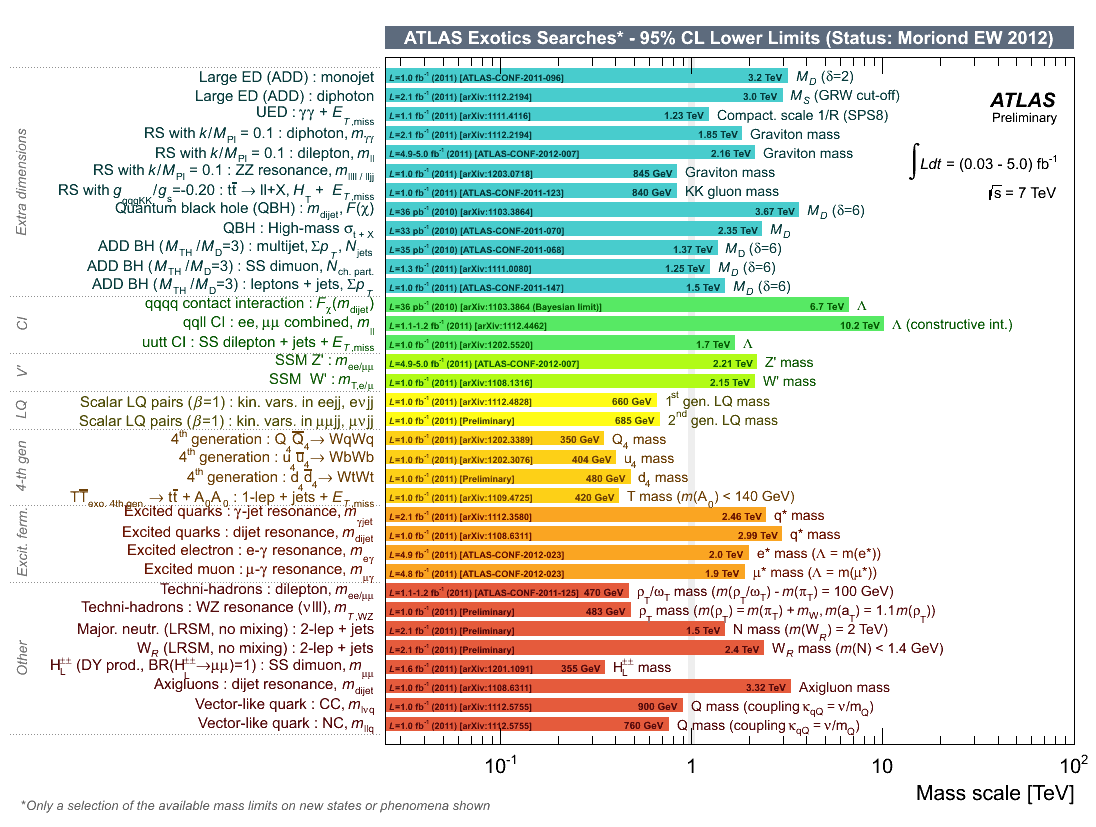}
\end{center}
\caption{\small \it
Summary of excluded mass ranges from a variety of searches from the ATLAS experiment for Beyond the Standard Model physics processes.
Only a representative selection of available results is shown.
}
\label{f:exotics_summary}
\end{figure}

\section{Conclusions}

With the start-up of the operation of the LHC at high energies particle physics has entered
a new era. Both the accelerator and the detectors have worked magnificently. Until the end
of 2011 data corresponding to an integrated luminosity of 5.5 \ifb\ have been recorded
with high efficiency by the LHC experiments. Based on these data, many tests of the predictions
of the Standard Model and searches for physics Beyond the Standard Model have been
performed in the new energy regime. So far, all measurements have been found to
be in good agreement with the predictions from the Standard Model.
Towards the end of 2011, the experiments have reached sensitivity for
the Standard Model Higgs boson. A large fraction of the possible Higgs boson mass range has already been
excluded by the ATLAS and CMS experiments with a confidence level of 95\%.
However, it is striking that both experiments
are not able to exclude the existence of the Higgs boson in the mass range from 118 - 129~\GeV, despite their sensitivity in this range.
In addition, tantalizing hints for a Higgs boson signal have been seen by both experiments
in the two high resolution channels \Hgg\ and \Hllll. However, the statistical significance is
not sufficient to claim evidence. More data are needed to
clarify the situation. With a successful run of the LHC in 2012 a final conclusion on the existence
of the Standard Model Higgs boson might be reached and the year 2012 might enter as
the ``Year of the Higgs Boson'' into the history of Physics.

%

\providecommand{\href}[2]{#2}\begingroup\raggedright\endgroup

\end{document}